\documentclass[twocolumn,superscriptaddress,showpacs,prd,aps,amsmath,amssymb,nofootinbib]{revtex4-1}
\usepackage{graphicx}
\usepackage{amssymb}
\usepackage{bm} 
\usepackage{mathrsfs} 

\usepackage{color}

\newcommand{\Madm}{M_{\rm ADM}}
\newcommand{\MK}{M_{\rm K}}
\newcommand{\MP}{M_{\rm P}}

\newcommand{\Msol}{M_\odot}

\newcommand{\beq}{\begin{equation}} 
\newcommand{\eeq}{\end{equation}} 
\newcommand{\beqn}{\begin{eqnarray}} 
\newcommand{\eeqn}{\end{eqnarray}} 
\newcommand{\pa}{\partial}
\newcommand{\na}{\nabla}
\newcommand{\gab}{g^\alpha\!_\beta}
\newcommand{\gabu}{g^{\alpha\beta}}
\newcommand{\gabd}{g_{\alpha\beta}}

\newcommand{\gmabu}{\gamma^{ab}}
\newcommand{\gmabd}{\gamma_{ab}}
\newcommand{\tgmabu}{\tilde\gamma^{ab}}
\newcommand{\tgmabd}{\tilde\gamma_{ab}}
\newcommand{\tgamma}{\tilde\gamma}
\newcommand{\habu}{h^{ab}}
\newcommand{\habd}{h_{ab}}

\newcommand{\fabu}{f^{ab}}
\newcommand{\fabd}{f_{ab}}

\newcommand{\tbeta}{\tilde{\beta}}
\newcommand{\Aabu}{A^{ab}}
\newcommand{\Aabd}{A_{ab}}

\newcommand{\tAabd}{\tilde{A}_{ab}}

\newcommand{\albe}{{\alpha\beta}}

\newcommand{\Tabd}{T_{\alpha\beta}}
\newcommand{\Tabu}{T^{\alpha\beta}}
\newcommand{\Tab}{T^\alpha{}\!_\beta}

\newcommand{\Gabd}{G_{\alpha\beta}}

\newcommand{\TabMd}{T^{M}_{\alpha\beta}}
\newcommand{\TabMu}{T_{M}^{\alpha\beta}}

\newcommand{\TabFd}{T^{F}_{\alpha\beta}}
\newcommand{\TabFu}{T_{F}^{\alpha\beta}}

\newcommand{\Fabu}{F^{\alpha\beta}}
\newcommand{\Fabd}{F_{\alpha\beta}}

\newcommand{\Fcdu}{F^{\gamma\delta}}
\newcommand{\Fcdd}{F_{\gamma\delta}}

\newcommand{\ttR}{{}^{3}\!\tilde R}

\newcommand{\tD}{\tilde D}
\newcommand{\zD}{{\raise1.0ex\hbox{${}^{\ \circ}$}}\!\!\!\!\!D}
\newcommand{\alone}{{\raise0.5ex\hbox{${}^{\ 1}$}}\!\!\!\!\alpha}
\newcommand{\Od}{{O}}

\newcommand{\tS}{\tilde S}

\newcommand{\dSab}{dS_{\alpha\beta}}
\newcommand{\dSa}{dS_{\alpha}}

\newcommand{\dl}{\delta}

\newcommand{\Dl}{\Delta}

\newcommand{\Lie}{\mbox{\pounds}}

\newcommand{\nalam}{\mathrel{\raise0.9ex\hbox{$^\lambda$}\mkern-14mu
\lower0.0ex\hbox{$\nabla$}}}

\newcommand{\dis}{\displaystyle}
\newcommand{\gmaa}{\gamma^\alpha{}_a}
\newcommand{\gmbb}{\gamma^\beta{}_b}
\newcommand{\gmaal}{\gamma_a{}^\alpha}

\newcommand{\tgmcdu}{\tilde\gamma^{cd}}
\newcommand{\tgmcdd}{\tilde\gamma_{cd}}

\newcommand{\tphi}{\tilde\phi}

\newcommand{\rhoH}{\rho_{\rm H}}

\newcommand{\Nrf}{{N_r^{\rm f}}}
\newcommand{\Nrm}{{N_r^{\rm m}}}

\newcommand{\tA}{\tilde A}

\newcommand{\Kabd}{K_{ab}}
\newcommand{\tKabd}{\tilde K_{ab}}
\newcommand{\zeroD}{{\raise1.0ex\hbox{${}^{\ \circ}$}}\!\!\!\!\!D}

\newcommand{\zLap}{{\raise1.0ex\hbox{${}^{\ \circ}$}}\!\!\!\!\Delta}
\newcommand{\zna}{{\raise1.0ex\hbox{${}^{\ \circ}$}}\!\!\!\!\!\nabla}
\newcommand{\zS}{{\raise1.0ex\hbox{${}^{\ \circ}$}}\!\!\!\!\!S}
\newcommand{\hu}{h{\underbar u}}

\newcommand{\cocal}{{\sc cocal}}

\newcommand{\Bpolmax}{{B^{\rm max}_{\rm pol}}}
\newcommand{\Btormax}{{B^{\rm max}_{\rm tor}}}

\newcommand{\Bpol}{{B_{\rm pol}}}
\newcommand{\Btor}{{B_{\rm tor}}}

\newcommand{\rhoc}{\rho_{\rm c}}
\newcommand{\prhoc}{(p/\rho)_{\rm c}}
\newcommand{\Omegac}{\Omega_{\rm c}}

\begin{document}
\title{New code for equilibriums and quasiequilibrium initial data of compact objects. IV. 
Rotating relativistic stars with mixed poloidal and toroidal magnetic fields}

\author{K\=oji Ury\=u}
\email{uryu@sci.u-ryukyu.ac.jp}
\affiliation{
Department of Physics, University of the Ryukyus, Senbaru 1, 
Nishihara, Okinawa 903-0213, Japan}
\author{Shijun Yoshida}
\email{yoshida@astr.tohoku.ac.jp}
\affiliation{
Astronomical Institute, Tohoku University, Aramaki-Aoba, Aoba, Sendai 980-8578, Japan}
\author{Eric Gourgoulhon}
\email{eric.gourgoulhon@obspm.fr}
\affiliation{
Laboratoire Univers et Th\'eories, UMR 8102 du CNRS,
Observatoire de Paris, Universit\'e Paris Diderot, F-92190 Meudon, France}
\author{Charalampos Markakis} 
\email{c.markakis@damtp.cam.ac.uk}
\affiliation{
DAMTP, Centre for Mathematical Sciences, University of Cambridge, Cambridge, CB3 0WA, UK}
\affiliation{
NCSA, University of Illinois at Urbana-Champaign, Urbana, IL 61801, USA}
\affiliation{
School of Mathematical Sciences, Queen Mary University of London, London, E1 4NS, UK}
\author{Kotaro Fujisawa}
\email{fujisawa@resceu.s.u-tokyo.ac.jp}
\affiliation{
Research Center for the Early universe, Graduate School of Science, University of Tokyo, 
Hongo 7-3-1, Bunkyo, Tokyo 113-0033, Japan}
\author{Antonios Tsokaros}
\email{tsokaros@illinois.edu}
\affiliation{
Department of Physics, University of Illinois at Urbana-Champaign, Urbana, IL 61801}
\author{Keisuke Taniguchi}
\email{ktngc@sci.u-ryukyu.ac.jp}
\affiliation{
Department of Physics, University of the Ryukyus, Senbaru 1, 
Nishihara, Okinawa 903-0213, Japan}
\author{Yoshiharu Eriguchi}
\email{eriguchi@ea.c.u-tokyo.ac.jp}
\affiliation{
Department of Earth Science and Astronomy, Graduate School of Arts and Sciences, University of Tokyo, 
Komaba 3-8-1, Meguro, 153-8902 Tokyo, Japan}
%

\date{\today}  

\begin{abstract} 
A new code for computing fully general relativistic solutions of strongly 
magnetized rapidly rotating compact stars is developed as a part of the \cocal\ 
(Compact Object CALculator) code.  The full set of Einstein's equations, 
Maxwell's equations and magnetohydrodynamic equations are consistently solved 
assuming perfect conductivity, stationarity, and axisymmetry, and 
strongly magnetized solutions associated with mixed poloidal and toroidal 
components of magnetic fields are successfully obtained in generic (non-circular) 
spacetimes.  
We introduce the formulation of the problem and the numerical method in detail, 
then present examples of extremely magnetized compact star solutions and 
their convergence tests.  It is found that, in extremely 
magnetized stars, the stellar matter can be expelled from the region of 
strongest toroidal fields.  
Hence we conjecture that a toroidal electro-vacuum region may appear inside of 
the extremely magnetized compact stars, which may seem like the neutron star 
becoming the strongest toroidal solenoid coil in the universe.  
\end{abstract} 


\maketitle
 
\section{Introduction}

A magnetar, a neutron star (NS) associated with very strong 
surface magnetic fields around $10^{14}-10^{15}$G, has become a widely 
accepted model for Soft Gamma Repeaters (SGR) and Anomalous 
X-ray pulsars (AXP) \cite{Magnetar}.  
Although electromagnetic fields of observed magnetars 
are very strong, their electromagnetic energy may not be 
expected to dominate over internal or gravitational energies.  
Therefore in most theoretical models of magnetars, 
the electromagnetic fields are treated separately from the hydrostatic 
equilibrium of the compact stars as, for example, in \cite{Oron:2002gs}, 
or they are treated as perturbations.  With the perturbative techniques,
general relativistic stars having purely poloidal magnetic fields and 
both toroidal and poloidal magnetic fields were calculated in 
\cite{Konno:1999} and \cite{PolTorMS}, respectively.  Effects of 
stable stratification on structures of stars with mixed 
poloidal-toroidal magnetic fields were included in \cite{SSMS}.  

However, the electromagnetic fields of newly born magnetars could be 
strong enough to have a comparable amount of energy, or could be 
highly concentrated and distributed anisotropically, so that the fields 
may largely alter the hydrostatic equilibrium of stars globally, 
or locally, respectively.  
From a theoretical viewpoint, it is also interesting to compute 
extreme solutions such as compact stars associated with 
the electromagnetic fields in their strongest limit, 
and to investigate their impact onto the hydrostatic as well as 
the spacetime structure \cite{RNS}.  

Several numerical methods have been developed in the last three decades 
for computing such stationary and axisymmetric equilibriums of 
relativistic compact stars, which are largely deformed due to 
strong electromagnetic fields and rapid rotation.  
The first success was achieved by the Meudon group (LUTH) for computing 
compact stars associated with poloidal magnetic fields 
\cite{Bocquet:1995je}.  Those associated with purely toroidal magnetic 
fields were solved by Kiuchi and Yoshida \cite{Kiuchi:2008ch}, 
and later by Frieben and Rezzolla \cite{Frieben:2012dz}.  
More recently, the Florence group published a series of 
articles for computing magnetized compact stars 
with purely poloidal, toroidal, as well as mixed magnetic fields
\cite{Firenze}.  In their computations, simplified formulations 
for the gravitational fields have been used, which enabled 
systematic computations of solutions in a wide region of parameter 
space.    

In our previous paper \cite{Uryu:2014tda}, we presented 
preliminary results for stationary and axisymmetric equilibriums of 
relativistic rotating stars associated with strong electromagnetic 
fields, in particular, with mixed toroidal and poloidal magnetic fields.  
Following \cite{Uryu:2014tda}, we detail below the formulations and 
a numerical methods for computing such equilibriums, including 
improvements on our earlier work \cite{Uryu:2014tda}.  
We then present a few examples of solutions 
associated with extremely strong electromagnetic fields and results 
of convergence tests.  In the newly calculated solutions, it is found 
that the toroidal magnetic fields concentrate near, but well below, 
the equatorial surface, and that the fields expel the matter when 
their strength becomes of order $10^{17}$G or higher for typical 
neutron stars.  From this finding we can conjecture that a neutron 
star associated with such extremely strong toroidal magnetic fields 
may have a toroidal magneto-vacuum tunnel in it, that is, 
such a neutron star may become a toroidal solenoid itself.

This paper is organized as follows.  The formulation for stationary 
and axisymmetric equilibriums of relativistic stars associated with 
electromagnetic fields is described in Sec.~\ref{sec:Formulation}
with emphasis on the 3+1 decomposition of Maxwell's equations 
and the derivation of a system of first integrals and 
integrability conditions for ideal magnetohydrodynamic (MHD) flows.  
In Sec.~\ref{sec:Numerical}, the derived formulation is further 
modified into the form implemented in the present numerical code, the 
\cocal\ code, and then the numerical method used in the code is briefly 
described.  In Sec.~\ref{sec:Results}, three new numerical solutions 
calculated from the latest version of the \cocal\ code for magnetized 
rotating equilibriums are presented, and their convergence test with 
respect to resolution and number of multipoles included in 
the Poisson solver are presented.

\section{Formulation}
\label{sec:Formulation}

\subsection{Summary for formulation}

In the following, relativistic rotating stars associated with 
electromagnetic fields are modeled in the framework of 
a stationary and axisymmetric Einstein-Maxwell charged and magnetized 
perfect fluid spacetime.  We assume that the equilibriums of 
magnetized stars satisfy the ideal MHD condition.  
Because of the nature of mixed poloidal and toroidal components 
of magnetic fields as well as a possible existence of meridional 
flows of matter, the spacetime is no longer circular: it is not 
invariant under a simultaneous inversion of $t\rightarrow -t$ and 
$\phi \rightarrow -\phi$ \cite{circular}.
To incorporate all metric components that describe 
such non-circular spacetimes, we apply the waveless formulation 
which is developed for solving initial data sets for numerical relativity 
simulations \cite{Shibata:2004qz,MeudonWL,WLBNS}.  
The waveless formulation is based on a 3+1 decomposition and conformal 
decomposition of the spatial metric, which are commonly used 
in numerical relativity.  
%
%
Under appropriate gauge conditions, and time and rotational symmetries, 
the metric components are obtained by solving a system of elliptic 
partial differential equations (PDEs) on an asymptotically flat 
spacelike slice $\Sigma$.  

An analogous formulation is also applied to recast 
Maxwell's equations into 3+1 form, with the electromagnetic 
1-form obtained by solving elliptic PDEs.  
The formulation for the electromagnetic fields is detailed below, 
which differs from the standard formulation from which the well-known 
Grad-Shafranov equation is derived.  

A formulation for a system of ideal MHD equations has been 
discussed in our previous paper \cite{Gourgoulhon:2011gz}.  
In \cite{Gourgoulhon:2011gz}, integrability conditions 
to guarantee consistency of the stationary and axisymmetric 
system and associated set of first integrals have been derived.  
The basic idea of the formulation used in \cite{Uryu:2014tda} as well as 
in the present paper is essentially the same as that of 
\cite{Gourgoulhon:2011gz}, but an alternative choice of variables results 
in somewhat different set of equations to be solved.  In the formulation of 
\cite{Gourgoulhon:2011gz}, the electromagnetic 2-form $F=dA$ and its 
Hodge dual $\star F$ are decomposed covariantly using the 1-form basis 
dual to symmetry vectors $t^\alpha$ and $\phi^\alpha$, and three scalar 
fields which are the same in both decompositions of $F$ and $\star F$ 
(see e.g., Eqs.~(2.35) and (2.36) in \cite{Gourgoulhon:2011gz}).  
An analogous decomposition is applied to the vorticity 2-form $d(\hu)$, 
then, after careful algebraic manipulations, the relativistic 
transfield equation (a generalized form of the Grad-Shafranov equation 
with meridional flows) is derived.  

In the following formulation, unlike \cite{Gourgoulhon:2011gz}, 
we use the contravariant tensor $\Fabu$ instead of $\star F$, 
and an orthogonal basis of a reference flat metric defined 
in Sec.~\ref{sec:basis} to decompose the set of equations.  
This choice is probably more common in formulations of 
numerical relativity, and hence results in a more familiar 
form of the equations, although redundant components 
remain in the equations.  Another difference is that we 
do not reduce the number of variables by imposing axial 
symmetry in our formalism.  This allows enough generality in 
the new part of the code that will enable easy extension for 
computing, for example, non-axisymmetric configurations of 
electromagnetic fields, electromagnetic standing 
waves, or a magnetic dipole field misaligned with the rotation 
axis, in the future.  This also minimizes the effort to develop 
and debug a new code for such a rather complex problem, 
as computing tools having already implemented in \cocal, 
such as its multipole moment elliptic PDE solver, can be utilized.

The 3+1 decomposition and the waveless formulation for Einstein's equations 
are briefly summarized below, whose details are found, for example, in 
\cite{RNS,NR,NRBS,Gourg2012}, and in our previous papers 
\cite{Shibata:2004qz,WLBNS,Uryu:2016dqr}, respectively.   
The derivations of the formulations for Maxwell's equations and the first integrals 
of the ideal MHD equations are presented in full detail in the following subsections.  
In this paper, we use abstract index notation for tensors; 
the Greek letters $\alpha, \beta, \gamma, ...$ are for abstract 4D indices, 
the Latin lowercase letters $a, b, c, ...$ for 3D indices, and 
the Latin uppercase letters $A, B, C, ...$ for 2D indices.  

In the above, we expressed the 2-forms, $F, \star F, dA$, and $d(\hu)$ 
omitting indices.  Such index-free notation may also be used with 
caution, in particular, when calculations involve forms and vectors.  
A dot denotes an inner product, that is, a contraction between adjacent 
indices. For example, a vector $v$ and a $p$-form $\omega$ have inner product 
\beq
v\cdot \omega = v^\gamma \omega_{\gamma\alpha\dots\beta}, \quad 
\omega\cdot v = \omega_{\alpha\dots\beta\gamma}v^\gamma .
\eeq
In particular, the Cartan identity for a $p$-form $\omega$ 
in index-free notation is written, 
\beq
\Lie_v \omega \,=\, v\cdot d\omega \,+\, d(v\cdot \omega).
\label{eq:Cartan}
\eeq
Certain relations in index-free notation are summarized in Appendix 
\ref{secApp:forms}.

\subsection{Framework and notations}
\subsubsection{3+1 decomposition of spacetime}
\label{sec:3+1}

We consider globally hyperbolic spacetimes $({\cal M}, \gabd)$, 
${\cal M}={\mathbb R}\times \Sigma$, admitting two 
symmetries: stationarity associated with a timelike Killing vector $t^\alpha$ and 
axisymmetry associated with a spacelike rotational Killing vector $\phi^\alpha$.  
The spacetime is foliated by spacelike hypersurfaces $\Sigma_t=\chi_t(\Sigma_0)$ 
parametrized by a time coordinate $t$, where $\chi_t$ is a diffeomorphism 
generated by $t^\alpha$ and $\Sigma_0$ is an initial slice.  
Because of the time-translation symmetry, $\Sigma_t$ are identical for any $t$.  
The spacelike vector $\phi^\alpha$ generates a congruence of circles in $\Sigma_t$ 
parametrized by $\phi$ whose length is $2\pi$.  Those parameters $t$ and $\phi$ 
are chosen as coordinates.

The future pointing normal 1-form $n_\alpha$ is defined by 
$n_\alpha =-\alpha \na_\alpha t$, and it is related to $t^\alpha$ as 
\beq
t^\alpha \,=\, \alpha n^\alpha +\beta^\alpha,
\eeq 
where $\alpha$ and $\beta^\alpha$ 
are the lapse function and the shift vector, respectively, and 
the shift is spacelike, $\beta^\alpha n_\alpha=0$.  
The projection tensor to a slice $\Sigma_t$ is defined, 
\beq
\gamma_\albe \,=\, \gabd \,+\, n_\alpha n_\beta, 
\eeq
and its pullback to $\Sigma_t$ is written $\gmabd$.  
Then, the metric $\gabd$ in a chart $\{x^\alpha\}$ 
is split into 3+1 form in a chart $\{t,x^a \}$, 
\beqn
ds^2 &=& \gabd dx^\alpha dx^\beta \nonumber \\
&=& -\alpha^2 dt^2 \,+\, \gmabd(dx^a+\beta^a dt)(dx^b+\beta^b dt).
\eeqn

For the spatial metric $\gmabd$, a conformal decomposition is 
introduced as 
\beq
\gmabd = \psi^4 \tgmabd, 
\label{eq:Conformal}
\eeq 
where $\psi$ is the conformal factor, 
and $\tgmabd$ the spatial conformal metric.  This decomposition 
is specified through a condition $\tgamma=f$, where $\tgamma$ is the 
determinant of $\tgmabd$ and $f$ the determinant of the flat metric $\fabd$ 
in a chart $\{x^a\}$.  
The differences between the spatial conformal metric and the flat metric, 
$\habd$ and $\habu$, are defined by 
\beq
\tgmabd \,=\, \fabd \,+\, \habd, 
\ \ \mbox{and} \ \ 
\tgmabu \,=\, \fabu \,+\, \habu, 
\eeq
where $\tgmabu$ and $\fabu$ are the inverses of the corresponding metrics.  
Because of the conformal decomposition, the weight of the Levi-Civita 
tensor becomes 
\beq
\sqrt{-g} \,=\, \alpha\psi^6\sqrt{\tgamma} \,=\, \alpha\psi^6\sqrt{f}.
\label{eq:detg}
\eeq
We denote spatial derivative operators $D_a$, $\tD_a$, and $\zD_a$ 
which are compatible with spatial metrics $\gmabd$, $\tgmabd$, and 
$\fabd$, respectively, and a spacetime derivative operator compatible 
with the metric $\gabd$ by $\na_\alpha$.

Since we write down all field equations, equations of motion, and 
other associated relations, including coordinate conditions, using 
the flat derivative operator $\zD_a$, we have freedom to choose 
$\{x^a\}$, a coordinate system of the reference frame associated 
with $\fabd$, without changing the spacetime geometry.  In this paper, 
as in elementary vector analysis, we only choose one of Cartesian, 
cylindrical, or spherical coordinates associated with a set of 
orthogonal bases for $\{x^a\}$ (see, Appendix \ref{secApp:basis}).  
Under a choice of orthogonal basis, a difference in the weight 
(\ref{eq:detg}) arises from $f$ which may or may not be included in $\psi$ 
depending on whether one chooses a coordinate or non-coordinate basis.  

The extrinsic curvature of $\Sigma_t$ is defined by 
\beq
\Kabd \,=\, - \frac12 \gmaa \gmbb \Lie_n \gamma_\albe 
\,=\, -\frac1{2\alpha}\left(\pa_t\gmabd\,-\,\Lie_\beta \gmabd\right),
\label{eq:Kab}
\eeq
where the Lie derivatives $\Lie$ are defined on either $\cal M$ or 
$\Sigma_t$ depending on the vector to derive along, and $\pa_t$ is 
the pull back of $\Lie_t$ defined on $\cal M$ to $\Sigma_t$.  
The trace of $\Kabd$ is written $K$, and the trace free part of 
$\Kabd$ is defined by 
\beq
\Aabd \,=\, \Kabd-\frac13\gmabd K.
\label{eq:Aab}
\eeq
Substitution of Eq.~(\ref{eq:Kab}) to the trace free part (\ref{eq:Aab}) 
results in conformal Killing operators with respect to 
$t^\alpha$ and $\beta^\alpha$, 
\beqn
\Aabd &=& - \frac1{2\alpha}
\left(\, \gmaa \gmbb \Lie_{\alpha n} \gamma_\albe 
-\frac13\gmabd \gamma^\albe \Lie_{\alpha n} \gamma_\albe \,\right)
\nonumber\\ 
&=& \frac{\psi^4}{2\alpha}
\left(\, \Lie_\beta \tgmabd - \frac13 \tgmabd \tgmcdu \Lie_\beta \tgmcdd \right)
\nonumber\\
&-& \frac{\psi^4}{2\alpha}
\left(\, \pa_t \tgmabd - \frac13 \tgmabd \tgmcdu \pa_t \tgmcdd \right), 
\label{eq:AabCFK}
\eeqn
while the trace of Eq.~(\ref{eq:Kab}) becomes 
\beqn
K\,=\,\frac6{\alpha\psi}\left(\, \Lie_\beta \psi \,-\, \pa_t \psi \,\right)
+\frac1{2\alpha\tgamma}\left(\, \Lie_\beta \tgamma \,-\, \pa_t \tgamma \,\right) .
\label{eq:K}
\eeqn
As a result, under the condition $\tgamma=f$ with an assumption 
$\pa_t \tgamma=\pa_t f=0$, 
the time derivatives of $\tgmabd$ and $\psi$ are separately 
related to $\Aabd$ and $K$, respectively.  

In actual computations, we introduce conformally rescaled $\Kabd$ and $\Aabd$
defined by $\tKabd = \psi^{-4} \Kabd$ and $\tAabd = \psi^{-4} \Aabd$, 
for which their indices are raised (lowered) with $\tgmabu$ ($\tgmabd$).

\subsubsection{Stress energy tensors for the perfect fluid and electromagnetic fields}

A strongly magnetized (and possibly charged) compact star is described 
by Einstein-Maxwell, charged and magnetized, perfect fluid spacetime.  
The stress-energy tensor $\Tabu$ is the sum of the perfect-fluid 
stress-energy tensor $\TabMu$ and the electromagnetic stress-energy tensor 
$\TabFu$,  
\beq
\Tabu \,=\, \TabMu \,+\, \TabFu, 
\eeq
where $\TabMu$ is defined by 
\beq 
\TabMu = \epsilon u^\alpha u^\beta + p q^{\alpha\beta}, 
\label{eq:TabMu}
\eeq
and $\TabFu$ by 
\beq
\TabFu\,=\,
\frac1{4\pi}\left(F^{\alpha\gamma}F^\beta{}_\gamma
-\frac14\gabu\Fcdd\Fcdu\right).  
\label{eq:TabFu}
\eeq

In the definition of $\TabMu$, $u^\alpha$ is the fluid 
4-velocity, $p$ the pressure, $\epsilon$ the energy density, and  
\beq  
q^{\alpha\beta} = g^{\alpha\beta} + u^\alpha u^\beta 
\eeq 
is the projection tensor onto a surface orthogonal to $u^\alpha$. 
Here, we assume that the fluids satisfy equations of state (EOS) 
of the form 
\beq  
p = p(\rho,s),\ \ \  \epsilon = \epsilon(\rho,s),   
\label{eq:EOS2param}
\eeq   
where $\rho$ is the baryon-mass density,\footnote{That is, 
$\rho := m_B n$, with $n$ the number density of 
baryons and $m_B$ the mean baryon mass.} 
and $s$ the entropy per unit baryon mass, 
although later we assume a simpler one-parameter EOS.

In the definition of $\TabFu$, 
the electromagnetic field 2-form $\Fabd$ is related to 
the electromagnetic potential 1-form $A_\alpha$ by 
\beq
\Fabd = (dA)_\albe := 
\na_\alpha A_\beta-\na_\beta A_\alpha.  
\label{eq:Faraday}
\eeq
In this Eq.~(\ref{eq:Faraday}), $(dA)_\albe$ 
is the exterior derivative of the 1-form $A_\alpha$.  
Since $F_\albe$ is a closed 2-form,  
\beq
(dF)_{\albe\gamma} := 3\,\na_{[\alpha} F_{\beta\gamma]} 
= \na_\alpha F_{\beta\gamma}
+ \na_\beta F_{\gamma\alpha} 
+ \na_\gamma  F_{\alpha\beta}=0.
\eeq

\subsubsection{Stationarity and axisymmetry}
\label{sec:sym}

For stationary and axisymmetric systems, 
the Lie derivatives of field and matter variables, 
$\{\gabd, A_\alpha, u^\alpha, h, s\}$, along 
the time and axial symmetry vectors, asymptotically 
timelike vector $t^\alpha$, and a spacelike rotation 
vector $\phi^\alpha$, vanish: 
\beqn
&
\Lie_\eta \gabd = 0, \quad\ 
\Lie_\eta A_\alpha = 0, 
&
\nonumber\\
&
\Lie_\eta u_\alpha = 0, \quad
\Lie_\eta h = 0, \quad
\Lie_\eta s = 0, 
&
\label{eq:symmetry}
\eeqn
where $\eta^\alpha = t^\alpha$ or $\phi^\alpha$, and 
$h$ is the relativistic enthalpy defined by $h=(\epsilon+p)/\rho$.

As mentioned earlier, we use the same set of field equations for 
the gravity as the waveless formulation derived and used in 
\cite{Shibata:2004qz,MeudonWL,Uryu:2016dqr,WLBNS}.  
In this formulation, we do not impose $\phi$-symmetry explicitly 
onto the field equations.  The waveless condition becomes a part of 
the time symmetry conditions imposed on the time derivatives 
of field variables in the inertial frame.  Consequently, 
our formalism for solving the field equations may also be applicable 
for computing quasi-equilibrium solutions without axial symmetry.

\subsection{Formulation for gravitational fields}

\subsubsection{Summary of the waveless formulation}

As in the common formulations of numerical relativity, we decompose 
Einstein's equations, $\Gabd=8\pi\Tabd$, into normal and 
transverse components with respect to the hypersurface $\Sigma_t$ 
\cite{NR,NRBS,Gourg2012}.  In our equilibrium (or quasiequilibrium) 
initial data formalism for numerical relativity, we choose 
the following combinations of components, 
\beqn
&&\left(\Gabd - 8\pi\Tabd \right)\,n^\alpha n^\beta\,=\,0
\label{eq:Ham}
\\
&&\left(\Gabd - 8\pi\Tabd \right)\, \gmaa n_\beta\,=\,0
\label{eq:Mom}
\\
&&\left(\Gabd - 8\pi\Tabd \right)\left(\gamma^\albe+\frac12 n^\alpha n^\beta\right)\,=\,0
\label{eq:trG}
\\
&&\left(\Gabd - 8\pi\Tabd \right)\left(\gmaa \gmbb -\frac13 \gmabd \gamma^\albe \right)\,=\,0
\label{eq:trfreeG}
\eeqn
and {\it formally} recast them into a system of elliptic PDEs for the 3+1 variables 
$\{\psi,\tbeta_a,\alpha\psi,\habd \}$, respectively.  In the present 
computations, we separate the flat Laplacians, $\zLap:=\zD_a\zD^a$ operating 
on these variables, represented by $\Phi$, and moving other terms 
to the source terms, ${\cal S}$,\footnote{
The manner of determining $\Kabd$ in our formulation is analogous to 
the one used in an initial data formulation often referred 
to as the conformal thin-sandwich formalism 
\cite{York:1998hy,Cook:2000vr,NR,NRBS,Gourg2012}. }
\beq
\zLap \Phi \,=\, {\cal S}.  
\label{eq:GRfieldeq}
\eeq
The source terms ${\cal S}$ of Eqs.~(\ref{eq:trG}) and (\ref{eq:trfreeG}) 
contain a time derivative of the trace and trace-free part of extrinsic curvature 
$\pa_t K$ and $\pa_t \Aabd$, respectively, and, as mentioned earlier, 
$K$ and $\Aabd$ contain $\pa_t \psi$ and $\pa_t \tgmabd$, respectively, as 
Eqs.~(\ref{eq:K}) and (\ref{eq:AabCFK}).  

In most of formulations for numerical relativity simulations, gauge conditions 
are dynamically imposed through the so called $\alpha$-driver and $\Gamma$-driver 
that determine the lapse $\alpha$ and shift $\beta^a$ \cite{NR,NRBS,Gourg2012}.  
In our initial data formulation, $\alpha$ and $\beta^a$ are part of the metric 
potentials to be determined, and the gauge conditions are introduced by prescribing 
the values of the trace $K$ and the divergence $\zD_b\tgmabu$.  
We normally choose maximal slicing and Dirac gauge conditions, 
\beq
K=0 \ \ \ \mbox{and} \ \ \  \zD_b\tgmabu\,=\,0, 
\label{eq:gauge}
\eeq
for the four coordinate conditions.  
A method to impose these conditions has been described in 
\cite{WLBNS,Uryu:2016dqr}, and is repeated in the next subsection.  

In paper \cite{Shibata:2004qz}, a waveless condition is derived for 
the gravitational fields, which results in all metric potentials, 
including $\habd$, having Coulomb type fall off in the asymptotics under 
the gauge (\ref{eq:gauge}). 
Such waveless condition is to impose an asymptotic behavior on 
the time derivative of spatial conformal metric, 
\beq
\pa_t \tgmabu=\Od (r^{-3}).
\label{eq:waveless}
\eeq
In \cite{WLBNS,Uryu:2016dqr} as well as the present calculations, 
we impose a stronger condition 
\beq
\pa_t \tgmabu=0.
\label{eq:waveless0}
\eeq

As mentioned above, the time derivative terms in the gravitational field 
equations are $\{\pa_t \psi, \pa_t \tgmabd, \pa_t K, \pa_t \Aabd\}$.
Since the value of $K$ is fixed by the gauge condition 
(\ref{eq:gauge}), the time derivative of the conformal factor, 
$\pa_t \psi$, as seen in Eq.~(\ref{eq:K}) does not appear in the 
field equations.\footnote{It may appear in a gauge condition (see below).}   
The maximal slicing condition, $K=0$, is assumed to be satisfied 
not only instantaneously on the initial hypersurface, but 
also on the neighboring slices, hence $\pa_t K=0$.  
The waveless condition (\ref{eq:waveless0}) fixes the value of 
$\pa_t \tgmabd$.  The remaining $\pa_t \Aabd$, and other time derivative 
terms appearing in the equations of motion for the matter, may be prescribed 
by stationarity as in Eq.~(\ref{eq:symmetry}).\footnote{
The discussion above is to elucidate that the formulation can be 
applicable for quasi-stationary non-axisymmetric data.  One can 
assume global time symmetry and discard all time derivative terms 
from the beginning, which results in the same set of equations 
used in the following computations.}  

The waveless formulation has been successfully applied for computing 
equilibriums of single rotating stars, as well as quasi-equilibrium 
initial data of non-axisymmetric rotating stars and 
binary neutron stars \cite{Uryu:2016dqr,WLBNS} by replacing time 
symmetry vector with that in the rotating frame (the helical Killing 
vector $k^\alpha =t^\alpha+\Omega\phi^\alpha$ \cite{BD92D94,Friedman:2001pf}) 
except for $\pa_t\tgmabu$ on which the waveless condition (\ref{eq:waveless}) 
is imposed.\footnote{Under helical symmetry, 
non-axisymmetric initial data associated with standing gravitational waves 
may be calculated by imposing the symmetry also to the time derivative 
$\pa_t\tgmabu$, then rearranging the field equations to separate 
the Helmholtz operator for $\tgmabd$ in the left hand side 
\cite{BD92D94,Friedman:2001pf,Klein:2004be,PSWconsortium,YBRUF06}.  
}
The concrete form of Eq.~(\ref{eq:GRfieldeq}) for each metric potential 
$\{\psi,\tbeta_a,\alpha\psi,\habd \}$ is presented in 
\cite{Shibata:2004qz,WLBNS,Uryu:2016dqr}.

\subsubsection{Imposition of the gauge conditions}
\label{sec:gauge}

Recently we have developed a novel formulation for imposing arbitrary 
gauge conditions on the waveless initial data, and successfully computed 
a black hole toroid system in Kerr-Schild coordinates \cite{Tsokaros:2018zlf}.  
The maximal slice and Dirac conditions (\ref{eq:gauge}) are replaced by 
generalized gauge conditions 
\beq
K=K_{\cal G}  \ \ \ \mbox{and} \ \ \  \zD_b\tgmabu\,=\,{\cal G}^a, 
\label{eq:gauge0}
\eeq
where $K_{\cal G}$ and ${\cal G}^a$ are, respectively, 
a function and a vector given arbitrarily.  
Our computation for the strongly magnetized rotating star is therefore 
not limited to the choice (\ref{eq:gauge}), that is, $K_{\cal G}=0$ and 
${\cal G}^a=0$.  Since the asymptotic flatness may be imposed in most of 
applications for computing astrophysical compact objects, it will be 
convenient to choose gauge conditions which become the maximal and Dirac 
gauges except for the vicinity of the sources.  

Taking into account the gauge invariance of the linearized metric under 
transformations 
\beqn
\dl \beta^a &\rightarrow & \dl \beta^a \,+\, \zD^a \xi
\label{eq:gauge_transf_shift}
\\
\dl \gmabu &\rightarrow & \dl \gmabu \,-\, \zD^a \xi^b \,-\, \zD^b \xi^a, 
\label{eq:gauge_transf_gamma}
\eeqn
we introduce a gauge potential $\xi$ and gauge vector potentials $\xi^a$ 
to adjust $\beta^a$ and $\habu$ as 
\beqn
\beta^a{}' &=&  \beta^a \,+\, \frac1{\sqrt{\gamma}}\zD^a \xi, 
\label{eq:gauge_shift}
\\
\habu{}' &=&  \habu \,-\, \zD^a\xi^b \,-\, \zD^b\xi^a \,+\, \frac23\fabu \zD_c\xi^c.  
\label{eq:gauge_hab}
\eeqn
To impose gauge conditions (\ref{eq:gauge0}), 
we solve $K'=K_{\cal G}$ and $\zD_b \habu{}'={\cal G}^a$ for these gauge 
potentials $\xi$ and $\xi^a$, respectively, where $K'$ is Eq.~(\ref{eq:K}) 
in which $\beta^a{}'$ is substituted in place of $\beta^a$.  Then, the (primed) 
new variables are reconstructed accordingly with Eqs.~(\ref{eq:gauge_shift}) 
and (\ref{eq:gauge_hab}), and the original variables are replaced by 
the new ones, $\beta^a{}' \rightarrow \beta^a$ and $\habu{}' \rightarrow \habu$, 
during iterations for solving the field equations.  
Substituting Eqs.~(\ref{eq:gauge_shift}) and (\ref{eq:gauge_hab}) to these 
conditions, the gauge vector potentials $\xi$ and $\xi^a$ are solved from 
elliptic equations, 
\beqn
&&\qquad\qquad\qquad\quad 
\zLap \xi \,=\, {\cal S}_{\rm K} ,
\label{eq:gauge_eq_xi}
\\
&&
{\cal S}_{\rm K} \,=\, \sqrt{\gamma}
\left[\,\alpha K_{\cal G} - \zD_a\beta^a
- \frac{6}{\psi}\left(\Lie_\beta \psi-\pa_t\psi\right)\,\right],
\label{eq:gauge_source_xi}
\eeqn
and
\beqn
&&\qquad\qquad
\zLap \xi^a \,=\, {\cal S}_{\rm D}^a, 
\label{eq:gauge_eq_xia}
\\
&& 
{\cal S}_{\rm D}^a \,=\, -{\cal G}^a + \zD_b \habu -\frac13 \zD^a \zD_b\xi^b .
\label{eq:gauge_source_xia}
\eeqn

A time derivative term $\pa_t \psi$ in the source (\ref{eq:gauge_source_xi}) 
is prescribed in computing initial data on $\Sigma_t$, or may be absorbed 
in the gauge condition $K_{\cal G}$.  In the following, it is set 
$\pa_t\psi=0$.  The above system of elliptic equations 
(\ref{eq:gauge_eq_xi})--(\ref{eq:gauge_source_xia}) are solved 
simultaneously and iteratively together with the field equations 
\cite{WLBNS,Uryu:2016dqr,Tsokaros:2018zlf}.

\subsection{Maxwell's and relativistic ideal MHD equations}
\label{sec:MHDeqs}

Hereafter in this section , we describe the formulations for solving 
electromagnetic fields and equilibriums of magnetized matter in detail.  
Maxwell's equations are written 
\beqn
&&
(dF)_{\albe\gamma} = 0
\label{eq:EMBianchi}
\\[1mm]
&&
\na_\beta\Fabu = 4\pi j^\alpha,
\label{eq:Maxwell}
\eeqn
where $j^\alpha$ is the electric current density.  The converse of 
Poincar\'e lemma implies the existence of a potential 
1-form $A_\alpha$, such that $\Fabd = (dA)_\albe$.  
By construction, the current density is conserved, 
\beq
\na_\alpha j^\alpha = 0. 
\eeq

From the Bianchi identity 
\beq
\na_\beta\Tabu \,=\, \na_\beta\TabMu \,+\, \na_\beta\TabFu \,=\,0,
\eeq
and the rest mass conservation law, 
\beq
\na_\alpha (\rho u^\alpha) = 0, 
\label{eq:restmass}
\eeq
the relativistic MHD-Euler equations are derived; 
\beq
u^\beta (d(\hu))_{\beta\alpha} -T\na_\alpha s = \frac1{\rho} \Fabd j^\beta,
\label{eq:MHD-Euler}
\eeq
where $(d(\hu))_{\albe}:=\na_\alpha(hu_\beta)-\na_\beta(hu_\alpha)$ 
is the canonical vorticity 2-form.  
The system of equations for the matter is closed by adding 
an EOS for the thermodynamic variables and a relation for energy 
transport.  Since we introduce a one-parameter EOS for computing 
equilibriums, we do not need to consider the latter relation.  

Finally, we assume that the ideal MHD condition holds: 
\beq
F_\albe u^\beta = 0.  
\label{eq:idealMHD}
\eeq
This condition implies the conservation of 
the flux $\Fabd$ as recalled briefly in Appendix \ref{secApp:idealMHD}.

In our previous paper \cite{Gourgoulhon:2011gz}, we have shown that, 
under stationarity and axisymmetry, the above system of Maxwell's 
equations and ideal MHD equations (\ref{eq:EMBianchi})--(\ref{eq:idealMHD}) 
can be recast in a system of a single elliptic PDE for a master potential, 
the relativistic master transfield equation, and first integrals, 
where the master potential may be related to a flux function, for example, 
the $\phi$-component of the potential $A_\alpha$.  
In the absence of a meridional flow field, the PDE becomes the well known 
Grad-Shafranov equation.  The formulation in \cite{Gourgoulhon:2011gz} 
is superior to the other formulations, since the single governing equation 
for the master potential is derived in a fully covariant form thanks to 
the use of exterior calculus and the orthogonal decomposition of a tangent 
space into subspaces spanned by the Killing vectors $(t^\alpha,\phi^\alpha)$ and 
a remaining ``meridional'' spacelike 2-surface.  It is also shown that 
the system of the transfield equation and associated first integrals is the most 
general form which contains all types of limiting cases including purely 
poloidal/toroidal magnetic fields, no-magnetic fields with meridional 
flows, and purely circular flows.  As mentioned earlier, however, we do 
{\it not} follow this style of formulation presented in 
\cite{Gourgoulhon:2011gz}, in particular we do not solve the master transfield 
equation, but solve Maxwell's equations to determine all (3+1 decomposed) 
components of $A_\alpha$.

\subsection{Formulation for the electromagnetic field}

In this subsection, we derive a 3+1 form of Maxwell's equations, 
which are recast in a set of elliptic PDEs for the components of the 
(3+1 decomposed) electromagnetic potential 1-form $A_\alpha$.  
Although these calculations are straightforward 
(see e.g., \cite{NRBS,Knapp:2002fm}), 
we present them in detail since the resulting equations are 
implemented in \cocal\ code for actual computations.

\subsubsection{3+1 decomposition of electromagnetic fields}

In this subsection, we introduce the 3+1 decomposed variables 
for the electromagnetic potential 1-form $A_\alpha$ and the Faraday 
tensor $\Fabu$, as well as a decomposition of its divergence 
$\na_\beta\Fabu$.  To avoid confusion, for a given 4D object, 
its projection to $\Sigma_t$ defined on $\cal M$ is denoted with 
a barred symbol, and the one defined on $\Sigma_t$ itself 
is denoted using the same symbol with 4D indices being 
replaced by 3D ones.  For example, the 4D object $A_\alpha$ 
is related to $\bar{A}_\alpha$ and the 3D object $A_a$ as follows;
\beqn
&& \bar{A}_\alpha \,=\, \gamma_{\alpha}{}^\beta A_\beta \\
&& A_a  \,=\, \gmaal A_\alpha 
\,=\, \gmaal \bar{A}_\alpha, 
\eeqn
As usual, we omit the bar on a projected 4D object when 
the 4D object is spatial.

We define the 3+1 variables of the electromagnetic potential 
1-form $A_\alpha$ and Faraday tensor $\Fabu$ by 
\beqn
&
\Phi_\Sigma \,=\, - A_\alpha n^\alpha, 
\quad &
A_a = \gmaal A_\alpha, 
\\[1mm]
&
F^a
\,=\, \gamma^a{}_\alpha \Fabu n_\beta , 
\quad &
F^{ab} \,=\, \gamma^a{}_\alpha\gamma^b{}_\beta \Fabu, 
\eeqn
and $\Fabu n_\alpha n_\beta = 0$ by anti-symmetry.  
Then, $A_\alpha$ and $\Fabu$ are related to their spatial 
components by 
\beqn
&& 
{A}_\alpha 
\,=\, \Phi_\Sigma\, n_\alpha + \bar{A}_\alpha 
\,=\, \Phi_\Sigma\, n_\alpha + \gamma_\alpha{}^a {A}_a ,
\label{eq:Aa4D}
\\[1mm]
&&
\Fabu \,=\, \bar{F}^\albe + n^\alpha \bar{F}^\beta - n^\beta\bar{F}^\alpha.
\label{eq:Fab4D}
\eeqn

As shown in Appendix \ref{sec:App:Fab}, the projected Faraday tensor,
$F_a$ and $F_{ab}$, 
and its divergence defined on $\Sigma$ become 
\beqn
&&\quad
F_a\,=\, -\, \Lie_n A_a
\,-\, \frac1{\alpha}D_a(\alpha\Phi_\Sigma).  
\label{eq:Fad}
\\[1mm]
&&\quad
F_{ab} \,=\, D_a  A_b \,-\,D_b  A_a,
\label{eq:Fab3D}
\\[1mm]
&&
n_\alpha \na_\beta \Fabu 
\,=\, -D_a F^a, 
\label{eq:projn_divF}
\\[1mm]
&&
\gamma^a{}_\alpha \na_\beta F^{\alpha\beta}
\,=\,
\frac1{\alpha}D_b (\alpha {F}^{ab})
\,-\, \Lie_n {F}^a
\,+\, K {F}^a.  
\label{eq:projSigma_divF}
\eeqn

\subsubsection{3+1 decomposition of Maxwell's equations}

Using Eq.~(\ref{eq:projn_divF}), the projection of 
Maxwell's equations along the hypersurface normal $n^\alpha$
is written 
\beq
(\na_\beta\Fabu - 4\pi j^\alpha)n_\alpha 
\,=\, -D_a F^a +4\pi\rho_\Sigma \,=\, 0, 
\label{eq:Maxwell_n}
\eeq
where $\rho_\Sigma$ is the projection of the current 
$j^\alpha$ along the normal $n^\alpha$ defined by 
\beq
\rho_\Sigma \,=\, -j^\alpha n_\alpha, 
\eeq
and $F^a$ is derived from Eq.~(\ref{eq:Fad}) by raising the index, 
\beq
F^a\,=\, -\, \Lie_n A^a
\,+\, 2K^a{}_b A^b
\,-\, \frac1{\alpha}D^a(\alpha\Phi_\Sigma).  
\label{eq:Fau}
\eeq
Note that the charge $\rho_\Sigma$ is related to the time 
component of the 4 current as 
\beq
\rho_\Sigma \,=\, -j^\alpha n_\alpha  
\,=\, -j^\alpha (-\alpha \na_\alpha t) 
\,=\, \alpha j^t.
\eeq

Using Eq.~(\ref{eq:projSigma_divF}), the projection of 
Maxwell's equations onto the hypersurface $\Sigma_t$
is written 
\beqn
&&(\na_\beta\Fabu - 4\pi j^\alpha)\gamma^a{}_\alpha 
\nonumber\\
&&\quad
\,=\, \frac1{\alpha}D_b (\alpha F^{ab})
\,-\, \Lie_n F^a 
\,+\, K F^a 
-4\pi j_\Sigma^a \,=\, 0, \quad
\label{eq:Maxwell3Du}
\eeqn
where $j_\Sigma^a$ is defined by 
\beq
j_\Sigma^a \,=\, \gamma^a{}_\alpha j^\alpha.  
\eeq
Note that the projected current $j_\Sigma^a$ is related 
to the components of the 4 current $j^\alpha$ as 
\footnote{For later convenience, we use $j^a$ as a spatial 
part of 4D current $j^\alpha$, hence $j^a \ne \gamma^a{}_\alpha j^\alpha$.}
\beq
j_\Sigma^a \,=\, (g^a{}_\alpha + n^a n_\alpha)j^\alpha 
\,=\,j^a + j^t \beta^a.  
\eeq
For later use, the dual of Eq.~(\ref{eq:Maxwell3Du}) is derived,   
\beqn
&&(\na_\beta\Fabu - 4\pi j^\alpha)\gamma_{a \alpha}
\nonumber\\
&&\ \ 
\,=\, 
\frac1{\alpha}D_b (\alpha F_a{}^b)
\,-\, \Lie_n F_a 
\,-\, 2K_a{}^b F_b 
\,+\, K F_a 
-4\pi j^\Sigma_a 
\nonumber\\
&&\ \ 
\,=\, 0, 
\label{eq:Maxwell_Sigma}
\eeqn
where the relation
$\gamma_{ab}\Lie_n F^b
= \Lie_n F_a + 2 K_{ab}\, F^b $ is used.

\subsubsection{Imposition of stationarity}

We assume that the electromagnetic potential 1-form $A_\alpha$ 
respects the time and rotational symmetry as discussed 
in Sec.\ref{sec:sym}.  In our formulation, we impose 
the stationary condition explicitly on the equations, 
$\Lie_t A_\alpha = 0$.  Since $t^\alpha = \alpha n^\alpha + \beta^\alpha$, 
we have 
\beqn
&& \Lie_n A_a \,=\, \frac1{\alpha}(\pa_t A_a - \Lie_\beta A_a) 
\,=\, -\frac1{\alpha}\Lie_\beta A_a ,
\\
&& \Lie_n F_a \,=\, \frac1{\alpha}(\pa_t F_a - \Lie_\beta F_a) 
\,=\, -\frac1{\alpha}\Lie_\beta F_a ,
\eeqn
and similarly for the duals $A^a$ and $F^a$.
With this symmetry, Maxwell's equations (\ref{eq:Maxwell_n}) 
and the relation (\ref{eq:Fad}) are rewritten 
\beqn
&&
(\na_\beta\Fabu - 4\pi j^\alpha)n_\alpha 
\,=\, -D_a F^a +4\pi\rho_\Sigma \,=\, 0, 
\label{eq:Maxwell_n_tsym}
\\ && \qquad\quad
F_a\,=\,  \frac1{\alpha}\Lie_\beta A_a
\,-\, \frac1{\alpha}D_a(\alpha\Phi_\Sigma),  
\label{eq:Fad_tsym}
\eeqn
and Eq.~(\ref{eq:Maxwell_Sigma}) becomes
\beqn
&&
(\na_\beta\Fabu - 4\pi j^\alpha)\gamma_{a \alpha}
\nonumber\\
&&\ 
\,=\, \frac1{\alpha}D_b (\alpha F_a{}^b)
\,+\, \frac1{\alpha}\Lie_\beta F_a 
\,-\, 2A_a{}^b F_b 
\,+\, \frac13 K F_a 
-4\pi j^\Sigma_a 
\nonumber\\
&&\ 
\,=\, 0, 
\label{eq:Maxwell_Sigma_tsym}
\eeqn
where $A_{ab}$ is the trace free part of the extrinsic 
curvature $\Kabd$ as defined in Eq.~(\ref{eq:Aab}).

\subsubsection{Conformal decomposition and equations for electromagnetic potentials}

To write down the final form of Maxwell's equations implemented in 
our actual numerical code, we introduce a conformal decomposition of 
the spatial metric Eq.~(\ref{eq:Conformal}) with the condition 
$\tgamma=f$ as explained in Sec.\ref{sec:3+1}.  
We introduce conformally rescaled quantities of 
the spatial electromagnetic potential 1-form and vector, 
\beq
\tA_a = A_a \quad \mbox{and}\quad  \tA^a = \tgmabu \tA_b = \psi^4 A^a,
\eeq
and for the spatial Faraday tensor
\beq
{\tilde F}_{ab} = F_{ab} \quad \mbox{and}\quad
{\tilde F}_{a} = F_{a}, 
\eeq
where tilded objects are rescaled quantities whose indices are 
raised or lowered by $\tgmabu$ or $\tgmabd$, respectively.

Details on the conformal decomposition of 3+1 form of Maxwell's equations 
(\ref{eq:Maxwell_n_tsym}) and (\ref{eq:Maxwell_Sigma_tsym}) are provided 
in Appendix \ref{sec:App:Maxwell}.  
The projection of Eq.~(\ref{eq:Maxwell_n_tsym}) along $n^\alpha$ results in 
an elliptic PDE for $\alpha \Phi_\Sigma$, 
\beq
\zLap(\alpha \Phi_\Sigma) \,=\, S, 
\label{eq:Maxwell_n_poisson}
\eeq
where the source $S$ is written
\beqn
S &=& 
-\,
h^{ab} \zD_a \zD_b(\alpha\Phi_\Sigma)
\,-\,
\zD_a \tgmabu \zD_b(\alpha\Phi_\Sigma)
\nonumber\\
&+& \tgmabu \frac{\alpha}{\psi^2}
\tD_a \Big(\frac{\psi^2}{\alpha}\Big)\,\alpha{ F}_b
\,+\,\zD_a \tgmabu\, \Lie_\beta A_b 
\nonumber\\
&+&\tgmabu \zD_a \Lie_\beta A_b
-4\pi\alpha\psi^4 \rho_\Sigma. \qquad 
\label{eq:Maxwell_n_source}
\eeqn

The projection of Eq.~(\ref{eq:Maxwell_Sigma_tsym}) to $\Sigma_t$ results in 
elliptic PDEs for $A_a$, 
\beq
\zLap A_a \,=\, S_a, 
\label{eq:Maxwell_vecpotA}
\eeq
where the source $S_a$ is written 
\beqn
S_a  
&:=&
\,-\,h^{bc}\zD_b\zD_c\tA_a
\,+\,\tgamma^{bc}\zD_b(C^d_{ca}\tA_d)
\,+\,\tgamma^{bc}C^d_{bc}\tD_d\tA_a 
\nonumber\\
&&
\,+\, \tgamma^{bc}C^d_{ba}\tD_c\tA_d
\,+\, \tD_a \tD_b \tA^b
\,+\, \ttR_{ab}\tA^b
\nonumber\\
&&
\,+\,{\tilde F}_a\!{}^b \frac{\psi^2}{\alpha}
\tD_b \left(\frac{\alpha}{\psi^2}\right)
\,+\, \frac{\psi^4}{\alpha}\Lie_\beta F_a 
\,-\, 2\psi^4 A_a{}^b F_b 
\nonumber\\
&&
\,+\, \frac13 \psi^4 K F_a 
\,-\,4\pi \psi^4 j^\Sigma_a .  
\label{eq:Maxwell_vecpotA_source}
\eeqn
To shorten the expression of the source 
(\ref{eq:Maxwell_vecpotA_source}), we keep $\tD_a$ instead of 
replacing it with $\zD_a$ and a connection $C^c_{ab}$ in some terms, 
where 
$C^c_{ab}:=\frac12\tgmcdu(\zD_a\tgamma_{db}+\zD_b\tgamma_{ad}-\zD_d\tgmabd)$.  
The fifth term in the right-hand side (RHS) of the source 
(\ref{eq:Maxwell_vecpotA_source}), 
$\tD_a \tD_b \tA^b$, may be expanded as follows, 
\beq
\tD_a\tD_b\tA^b 
= \zD_a \zD^b\tA_b
+ \zD_a(h^{bc}\zD_b\tA_c -\tgamma^{bc}C^d_{bc}\tA_d ), 
\label{eq:Aterms2}
\eeq
and then the Coulomb gauge condition $\zD^a A_a=0$ may be imposed explicitly, 
or a simpler expression of this term may be written 
applying the condition $\tgamma=f$ explicitly, 
\beq
\tD_a\tD_b\tA^b 
= \zD_a \zD_b \tA^b. 
\label{eq:Aterms3}
\eeq

\subsubsection{Imposition of Coulomb gauge}

As discussed in Appendix \ref{secApp:gauge}, we have freedom 
to choose the spatial gauge for the spatial part of the electromagnetic 
potentials.  We impose Coulomb gauge analogously to 
that for coordinate conditions discussed in Sec.~\ref{sec:gauge}: 
\beq
\zD^a A_a = 0.  
\label{eq:Coulomb}
\eeq 
Although we have not tested gauge conditions other than 
(\ref{eq:Coulomb}), we formulate the method to impose more 
general gauge conditions analogously to the imposition of 
coordinate conditions discussed in Sec.~\ref{sec:gauge}: 
\beq
\zD^a A_a = A_{\cal G}, 
\label{eq:CoulombAG}
\eeq 
where $A_{\cal G}$ is a given arbitrary function.  

Considering the following gauge transformation, 
\beq
A_a \rightarrow A_a + \zD_a f.
\eeq
we let $A'_a$ defined by 
\beq
A'_a = A_a + \zD_a f
\label{eq:Coulombgauge}
\eeq
satisfy the gauge condition $\zD^a A'_a = A_{\cal G}$
which leads to 
\beq
\zLap f = A_{\cal G} -\zD^a A_a
\label{eq:CoulombPDE}
\eeq
Same as the coordinate imposition 
Eqs.~(\ref{eq:gauge_eq_xi})--(\ref{eq:gauge_source_xia}), 
the above equation (\ref{eq:CoulombPDE}) is solved simultaneously 
and iteratively together with the field equations.  
Then, substituting the solution $f$ to Eq.~(\ref{eq:Coulombgauge}), 
$A'_a$ is calculated and $A_a$ is replaced by $A'_a\rightarrow A_a$.  

\subsection{First integrals of relativistic ideal MHD equations}
\label{sec:fint}

A key to a successful formulation for computing equilibriums of 
compact stars is to derive a set of first integrals of a system of 
hydrostationary equations.  For ideal MHD case, including ideal MHD 
condition (\ref{eq:idealMHD}), all equations for the magnetized matter 
have to be analytically and consistently integrated.  In the formulation 
in \cite{Gourgoulhon:2011gz}, those first integrals are thoroughly used 
to reduce the number of variables in deriving a single master transfield 
equation, in particular to eliminate the current $j^\alpha$ in Maxwell's 
equations.  In the present formulation we simply solve the system of 
the first integrals simultaneously with the field equations.  

In what follows, we analyze the rest mass conservation law 
(\ref{eq:restmass}), each of the $t$, $\phi$ and meridional $x^A$ 
components of the relativistic MHD-Euler equations (\ref{eq:MHD-Euler}), 
the ideal MHD condition (\ref{eq:idealMHD}), as well as Maxwell's 
equations in its original form (\ref{eq:Maxwell}), applying 
$t$ and $\phi$ symmetries.

\subsubsection{Coordinates and basis}
\label{sec:basis}

To begin with, we introduce an orthogonal basis, 
$\{t^\alpha,\phi^\alpha, e_A^\alpha\}$, associated with 
the coordinates $t$, $\phi$ and two other spatial coordinates
$x^A$, where the $t$ and $\phi$ coordinates are adapted to the 
spacetime symmetries generated by the two Killing vectors 
$t^\alpha$ and $\phi^\alpha$.  
The remaining two spatial `meridional' coordinates $x^A$, 
with indices denoted by uppercase Latin letters 
$A, B, \cdots = 1, 2$, may, for example, be $(\varpi, z)$ for cylindrical, or 
$(r,\theta)$ for spherical coordinates, for example.  
These bases and natural 1-form bases generated from these coordinates 
are normalized as 
\beq
t^\alpha \na_\alpha t = 1, \quad
\phi^\alpha \na_\alpha \phi = 1, \quad
e_A^\alpha \na_\alpha x^B = \delta_A{}^B, \quad
\eeq
where $\delta_A^B$ is the Kronecker delta, 
and otherwise orthogonal.  
In the following sections, we will use the 4D flat metric $\eta_\albe$ 
and 3D flat metric $f_{ab}$ associated with these bases, 
\beqn
&& 
\eta_\albe \,=\, 
-\,\na_\alpha t \na_\beta t \,+\, \na_\alpha \phi \na_\beta \phi 
\,+\, \dl_{AB}\na_\alpha x^A \na_\beta x^B,\quad
\\[1mm] && 
\eta^\albe \,=\, 
-\,t^\alpha t^\beta \,+\,  \phi^\alpha \phi^\beta 
\,+\, \dl^{AB}e_A^\alpha e_B^\beta,
\\[1mm] && 
f_{ab} \,=\, 
\na_a \phi \na_b \phi \,+\, \dl_{AB}\na_a x^A \na_b x^B,
\\[1mm] && 
f^{ab} \,=\, 
\phi^a \phi^b \,+\, \dl^{AB}e_A^a e_B^b.
\eeqn

Objects with contravariant indices are expanded using these bases, 
and are denoted, for example, as 
\beq
u^\alpha = u^t t^\alpha +  u^\phi \phi^\alpha +  u^A e_A^\alpha.  
\label{eq:4velcomp}
\eeq
for the 4-velocity $u^\alpha$.  
It is understood that the last term is summed over $A = 1, 2$.  
Similarly, objects with covariant 
indices (such as p-forms) are expanded with respect to the basis 
$\{\na_\alpha t,\na_\alpha \phi,\na_\alpha x^A\}$.

\subsubsection{Rest mass conservation equation}

The densitized rest mass conservation equation is written, 
\beqn
\na_\alpha(\rho u^\alpha) \sqrt{-g}
&=& 
\pa_\alpha\left[\,\rho(u^t t^\alpha 
+  u^\phi \phi^\alpha +  u^A e_A^\alpha)\sqrt{-g}\,\right]
\nonumber\\
&=& 
\pa_A(\rho u^A \sqrt{-g}) \,=\, 0, 
\eeqn
where we have applied the symmetries, 
\beq
\pa_\alpha (\rho u^t t^\alpha\sqrt{-g}) = 
\Lie_t (\rho u^\alpha \na_\alpha t\sqrt{-g}) =0, 
\eeq
and similarly to a term associated with $\phi^\alpha$.  
This suggests to introduce a stream function $\sqrt{-g}\Psi$ 
for the meridional flow fields $u^A$, 
\beq
u^A = \frac1{\rho\sqrt{-g}}\epsilon^{AB}\pa_B(\sqrt{-g}\Psi), 
\label{eq:merflowfn}
\eeq
where, $\epsilon^{AB}$ is an antisymmetric matrix
with a signature 
$\epsilon^{12} = -1$ 
\beq
\epsilon^{AB} = 
\left(
\begin{array}{rr}
0&-1\\
1&0
\end{array}
\right).
\eeq
The scalar function $\Psi$ is introduced to make explicit that 
the stream function $\sqrt{-g}\Psi$ is a densitized scalar.

\subsubsection{Components of electromagnetic 2-form 
and vorticity 2-form}

We write the components of $F = dA$, and impose symmetries.  
For economical reason, we omit spacetime indices $\alpha,\beta, \cdots$ 
in the following of this section and in the next section.
\\
$(t, \phi)$-component:
\beqn
F_{t\phi} 
&=& - F_{\phi t} 
= (t\cdot F) \cdot \phi
= (t \cdot dA)\cdot \phi
\nonumber\\
&=& \left[\,\Lie_t A - d(t\cdot A)\,\right]\cdot \phi
=-\Lie_\phi(t\cdot A)
=0.
\eeqn
$(t, x^A)$-components:
\beqn
F_{At} 
&=& - F_{tA} 
= - (t \cdot F)\cdot e_A
= - (t \cdot dA)\cdot e_A
\nonumber\\
&=& e_A\cdot d(t\cdot A)
=\pa_A A_t.
\eeqn
$(\phi, x^A)$ components:
\beqn
F_{A\phi} 
&=& - F_{\phi A} 
= - (\phi \cdot F)\cdot e_A
= - (\phi \cdot dA)\cdot e_A
\nonumber\\
&=& e_A\cdot d(\phi\cdot A)
=\pa_A A_\phi.
\eeqn
$(x^A, x^B)$ components are obviously,
\beq
F_{AB} = - F_{BA} 
= (dA)_{AB}  
= \pa_A A_B - \pa_B A_A, 
\eeq
or explicitly, 
\beqn
F_{AB} 
&=& - F_{BA} 
= (e_A\cdot F)\cdot e_B 
= (e_A\cdot dA)\cdot e_B 
\nonumber\\
&=& \left[\,\Lie_{e_A} A - d(e_A \cdot A)\,\right]\cdot e_B
= \pa_A A_B - \pa_B A_A. \ \ \ 
\eeqn

We introduce expressions for the spatial components of 
the 2-form $F$ and its dual as follows, 
\beqn
&&
F_{A\phi} \,=\, \pa_A A_\phi \,=\, - \epsilon_A{}^B B_B, 
\label{eq:F_Aphi}
\\
&&
F_{AB} \,=\, (dA)_{AB} \,=\, \epsilon_{AB} B_\phi, 
\label{eq:FABd}
\\
&&
F^{AB} \,=\, (dA)^{AB} \,=\, \epsilon^{AB} B, 
\label{eq:FABu}
\eeqn
where $\epsilon_{AB}$ and $\epsilon_A{}^B$ are also anti-symmetric 
matrices with signatures $\epsilon_{12}=-1$ and $\epsilon_1{}^2=-1$.  

Analogously, the components of the vorticity two-form 
$d(\hu)$ are written as follows:
\\
$(t, \phi)$ component:
\beq
d(\hu)_{\phi t} \,=\, - d(\hu)_{t\phi} \,=\, 0.
\eeq
$(t, x^A)$ components:
\beq
d(\hu)_{At} 
\,=\, - d(\hu)_{tA} 
\,=\, \pa_A (hu_t).
\eeq
$(\phi, x^A)$ components:
\beqn
d(\hu)_{A\phi} 
\,=\, - d(\hu)_{\phi A} 
\,=\, \pa_A (hu_\phi).
\eeqn
$(x^A, x^B)$ components:
\beq
d(\hu)_{AB} 
\,=\, - d(\hu)_{BA} 
\,=\, \pa_A (hu_B) - \pa_B (hu_A). 
\eeq

We introduce expressions for the spatial components of 
the 2-form $d(\hu)$ as follows,\footnote
{
The magnetic flux density $B$ and the vorticity $\omega$ 
are related to the Hodge dual of $F$ and $d(\hu)$ , 
respectively, as $B = u\cdot \star F$, and 
$\vec{\underline{\omega}}= \star d(\hu)\cdot u$.  
}
\beqn
&&
d(\hu)_{A\phi} \,=\, \pa_A (hu_\phi) 
\,=\, \epsilon_A{}^B \omega_B, 
\label{eq:dhu_Aphi}
\\
&&
d(\hu)_{AB} \,=\, - \epsilon_{AB} \,\omega_\phi. 
\label{eq:dhu_AB}
\eeqn

\subsubsection{Ideal MHD condition}

Substituting the 4-velocity in terms of a basis (\ref{eq:4velcomp}) 
to the ideal MHD condition $F\cdot u = 0$, and applying 
the symmetries to the 2-form $F = dA$ as discussed in 
the previous section, each component is written as follows: 
\\ 
$t$-component:
\beqn
t\cdot(F\cdot u) 
&=& t\cdot(F\cdot e_A)u^A 
= u^A F_{tA} 
\nonumber\\
&=& -u^A\pa_A A_t=0, 
\label{eq:idealMHD_t}
\eeqn
$\phi$-component:
\beqn
\phi\cdot(F\cdot u) 
&=& \phi\cdot(F\cdot e_A)u^A 
= u^A F_{\phi A} 
\nonumber\\
&=& -u^A\pa_A A_\phi=0, 
\label{eq:idealMHD_phi}
\eeqn
$x^A$-components:
\beqn
&&
e_A\cdot(F\cdot u) 
= e_A\cdot(F\cdot t)u^t + e_A\cdot(F\cdot \phi)u^\phi
\nonumber\\
&&\qquad\qquad\quad\ 
+ e_A\cdot(F\cdot e_B)u^B
\nonumber\\
&&\qquad\quad
= F_{At}u^t + F_{A\phi}u^\phi + F_{AB}u^B
\nonumber\\
&&\qquad\quad
= u^t\pa_A A_t + u^\phi\pa_A A_\phi  + (dA)_{AB}u^B =0.
\ \ 
\label{eq:idealMHD_xA}
\eeqn

Substituting the stream function $\sqrt{-g}\Psi$ defined by 
Eq.~(\ref{eq:merflowfn}) 
to each of the $t$ and $\phi$ components (\ref{eq:idealMHD_t}) and 
(\ref{eq:idealMHD_phi}), and then multiplying $\rho\sqrt{-g}$, we have, 
\beqn
&& 
\epsilon^{AB}\pa_A A_t \,\pa_B(\sqrt{-g}\Psi) =0, 
\label{eq:idealMHD_t_form1}
\\[1mm]
&& 
\epsilon^{AB}\pa_A A_\phi \,\pa_B(\sqrt{-g}\Psi) =0. 
\label{eq:idealMHD_phi_form1}
\eeqn
These relations imply that 
the constant surfaces of the scalars $A_t$, $A_\phi$, and the scalar density 
$\sqrt{-g}\Psi$ coincide.\footnote{
This means that the stream function $\sqrt{-g}\Psi$ depends on 
the choice of coordinate conditions.}  
Therefore, introducing the master potential 
$\Upsilon$ as an independent variable, they are written 
\beqn
&&
A_t = A_t(\Upsilon), 
\quad\ \ 
A_\phi = A_\phi(\Upsilon), 
\nonumber\\[1mm]
&&\!\!\!\!\!\!\!\!\!\!\!\!\!\!\!\!\!\!
\mbox{and}
\qquad\quad
\sqrt{-g}\Psi = [\,\sqrt{-g}\Psi\,](\Upsilon).  
\label{eq:def_fn_At_Psi}
\eeqn
These are part of the integrability conditions.

To obtain the first integral of the $x^A$-components (\ref{eq:idealMHD_xA}), 
we again substitute the definition of the stream function
(\ref{eq:merflowfn}), and $(dA)_{AB}=\epsilon_{AB}B_\phi$, 
to rewrite 
\beq
\rho u^t\sqrt{-g}\pa_A A_t 
\,+\, \rho u^\phi\sqrt{-g}\pa_A A_\phi
\,-\, B_{\phi}\pa_A[\sqrt{-g}\Psi] \,=\,0.
\label{eq:idealMHD_xA_form1}
\eeq
Because of the conditions (\ref{eq:def_fn_At_Psi}), it is rewritten 
\beq
\left(A'_t \rho u^t\sqrt{-g}
\,+\, A'_\phi \rho u^\phi\sqrt{-g}
\,-\,[\sqrt{-g}\Psi]'B_{\phi}
\,\right)\pa_A \Upsilon
=0, 
\eeq
where the primes $A'_t$, $A'_\phi$ and $[\sqrt{-g}\Psi]'$ 
stands for a derivative with respect to the master potential $\Upsilon$, 
\beqn
&&
A'_t:=\frac{d A_t(\Upsilon)}{d \Upsilon}, 
\quad\ \ 
A'_\phi:=\frac{d A_\phi(\Upsilon)}{d \Upsilon}, 
\nonumber\\[1mm]
&&\!\!\!\!\!\!\!\!\!\!\!\!\!\!\!\!\!\!
\mbox{and}
\qquad\qquad
[\sqrt{-g}\Psi]' := \frac{d [\sqrt{-g}\Psi](\Upsilon)}{d \Upsilon}.
\eeqn
Therefore, we have one of the first integrals, 
a consistency relation for components to be satisfied,
\beqn
A'_t \rho u^t\sqrt{-g}
\,+\, A'_\phi \rho u^\phi\sqrt{-g}
\,-\,[\sqrt{-g}\Psi]'B_{\phi}
=0.  
\quad
\label{eq:fint_idealMHD_xA}
\eeqn
In the absence of meridional 
flows, $[\sqrt{-g}\Psi]'(\Upsilon)=0$, 
Eq.~(\ref{eq:fint_idealMHD_xA}) implies 
a relativistic version of Ferraro's isorotation law \cite{Ferraro(1937)}, 
\beq
\frac{u^\phi}{u^t}\,:=\,\Omega(\Upsilon)\,=\,-\frac{A'_t}{A'_\phi}.  
\label{eq:rotlaw_noMC}
\eeq

\subsubsection{Maxwell's equations}

For any coordinate basis of the 1-form 
$\na_\alpha x$ $\ (x=t, \phi, x^A)$, 
the projections (components) of the divergence of the Faraday tensor 
$\na_\beta F^\albe$ become 
\beqn
\na_\alpha x \na_\beta F^\albe
&=& \na_\beta (F^\albe \na_\alpha x ) - 
F^\albe \na_\beta \na_\alpha x 
\,=\,
\na_\beta F^{x\beta}.
\nonumber\\
&=& 
\na_\beta (F^{xt}t^\beta+F^{x\phi}\phi^\beta+F^{xA}e_A^\beta)
\nonumber\\
&=& 
\Lie_{e_A}  F^{xA} + F^{xA} \frac1{\sqrt{-g}}\Lie _{e_A}\sqrt{-g} 
\nonumber\\
&=& 
\frac1{\sqrt{-g}}\Lie_{e_A}(F^{xA}\sqrt{-g}), 
\eeqn
where we have used $\na_\alpha t^\alpha=0$ and $\na_\alpha \phi^\alpha=0$ 
for Killing vectors $t^\alpha$ and $\phi^\alpha$.  

Then, the components of Maxwell's equations 
$\na_\beta\Fabu = 4\pi j^\alpha$ become as follows:
\\
$t$-component:
\beq
\Lie_{e_A}(F^{tA}\sqrt{-g})=4\pi j^t \sqrt{-g}, 
\eeq
$\phi$-component:
\beq
\Lie_{e_A}(F^{\phi A}\sqrt{-g})=4\pi j^\phi \sqrt{-g}, 
\eeq
$x^A$-component: 
\beq
\Lie_{e_B}(F^{AB}\sqrt{-g})=4\pi j^A \sqrt{-g}. 
\eeq
Since $F^{tA}$, $F^{\phi A}$, $F^{AB}$ are components, 
these equations are also written 
\beqn
4\pi j^t \sqrt{-g} &=& \pa_A(F^{t A}\sqrt{-g})
\\[1mm]
4\pi j^\phi \sqrt{-g} &=& \pa_A(F^{\phi A}\sqrt{-g})
\\[1mm]
4\pi j^A \sqrt{-g} &=& \pa_B(F^{AB}\sqrt{-g})
\nonumber\\[1mm]
&=& \epsilon^{AB}\pa_B(B\sqrt{-g}), 
\label{eq:mercurrent}
\eeqn 
where we substitute Eq.~(\ref{eq:FABu}) to $F^{AB}$ in Eq.~(\ref{eq:mercurrent}).

\subsubsection{MHD-Euler equations}

MHD-Euler equations (\ref{eq:MHD-Euler}) are also written 
in index free notation,
\beq
- d(\hu)\cdot u  - Tds - \frac1{\rho} dA\cdot j=0.  
\eeq
Substituting the current vector 
\beq
j^\alpha = j^t t^\alpha + j^\phi \phi^\alpha + j^A e_A^\alpha, 
\label{eq:current}
\eeq
and using the Cartan identity, and the $t, \phi$ symmetries,
each component becomes as follows:
\\
$t$-component: 
\beqn
&&
t\cdot\Big[\,- d(\hu)\cdot u  - Tds - \frac1{\rho} dA\cdot j\,\Big]
\nonumber\\
&&\quad
\,=\, u\cdot d\left[\,t\cdot(\hu)\,\right]
\,+\, \frac1{\rho}j\cdot d(t\cdot A)
\nonumber\\
&&\quad
\,=\, u^A\pa_A (hu_t)
\,+\, \frac1{\rho}j^A \pa_A A_t
\,=\,0,
\label{eq:MHD-Euler_t}
\eeqn
$\phi$-component: 
\beqn
&&
\phi\cdot\Big[\,- d(\hu)\cdot u  - Tds - \frac1{\rho} dA\cdot j\,\Big]
\nonumber\\
&&\quad
\,=\, u\cdot d\left[\,\phi\cdot(\hu)\,\right]
\,+\, \frac1{\rho}j\cdot d(\phi\cdot A)
\nonumber\\
&&\quad
\,=\, u^A\pa_A (hu_\phi)
\,+\, \frac1{\rho}j^A \pa_A A_\phi
\,=\,0,
\label{eq:MHD-Euler_phi}
\eeqn
$x^A$-component: 
\beqn
&&
-e_A \cdot\Big[\,-d(\hu)\cdot u  - Tds - \frac1{\rho} dA\cdot j\,\Big]
\nonumber\\
&&\quad
\,=\, -\,(u^t t + u^\phi \phi + u^B e_B)\cdot d(\hu)\cdot e_A
\nonumber\\
&&\quad\quad\ 
\,-\, \frac1{\rho}(j^t t + j^\phi \phi + j^B e_B)\cdot dA \cdot e_A
\,+\,T ds\cdot e_A
\nonumber\\
&& \quad
\,=\,u^t\pa_A (hu_t)
\,+\,u^\phi\pa_A (hu_\phi)
\,+\,u^B d(\hu)_{AB}
\nonumber\\
&& \quad\quad\ 
\,+\,\frac1{\rho}j^t\pa_A A_t
\,+\,\frac1{\rho}j^\phi\pa_A A_\phi
\,+\,\frac1{\rho}j^B (dA)_{AB}
\nonumber\\
&& \quad\quad\ 
\,+\,T \pa_A s
\,=\,0,  
\label{eq:MHD-Euler_xA_formA}
\eeqn

Analogously to the ideal-MHD conditions, after a somewhat 
lengthy calculation, integrability conditions and a set of 
first integrals of MHD-Euler equations are derived as follows:
\\
$t$-component:
\beq
-[\sqrt{-g}\Psi]' \, hu_t
\,+\,\frac1{4\pi}A'_t \,B\sqrt{-g}
\,=\,[\sqrt{-g}\Lambda_t](\Upsilon), 
\label{eq:fint_MHD-Euler_t}
\eeq
$\phi$-component:
\beq
-[\sqrt{-g}\Psi]' \, hu_\phi
\,+\,\frac1{4\pi}A'_\phi \,B\sqrt{-g}
\,=\,[\sqrt{-g}\Lambda_\phi](\Upsilon).
\label{eq:fint_MHD-Euler_phi}
\eeq
These are combined and written using another function of $\Upsilon$, 
$\Lambda(\Upsilon)$, 
\beq
A'_\phi\, hu_t \,-\, A'_t \, hu_\phi
\,=\, \Lambda(\Upsilon)
\,:=\,
\frac{A'_t[\sqrt{-g}\Lambda_\phi]\,-\,A'_\phi[\sqrt{-g}\Lambda_t]}
{[\sqrt{-g}\Psi]'}. 
\label{eq:fint_MHD-Euler_tphi}
\eeq
$x^A$-components:
\beqn
&&
-\,\frac12\left[
A'_\phi\left([\sqrt{-g}\Psi]''hu_t + [\sqrt{-g}\Lambda_t]' \right)
\right.
\nonumber\\
&&\quad
\,-\,A''_\phi \left([\sqrt{-g}\Psi]'hu_t + [\sqrt{-g}\Lambda_t] \right)
\nonumber\\
&&
\quad
\,+\, A'_t\left([\sqrt{-g}\Psi]'' hu_\phi + [\sqrt{-g}\Lambda_\phi]' \right)
\nonumber\\
&&
\left.\quad
\,-\, A''_t \left([\sqrt{-g}\Psi]' hu_\phi + [\sqrt{-g}\Lambda_\phi] \right)
\right]B_\phi
\nonumber\\
&&
\,+\,\frac12(A''_t\,hu_\phi - A''_\phi\,hu_t+\Lambda')
(A'_t u^t - A'_\phi u^\phi)\rho\sqrt{-g}
\nonumber\\
&&
\,+\, A'_t A'_\phi[\sqrt{-g}\Psi]'\omega_\phi
\,+\,A'_t  A'_\phi  s'   T   \rho \sqrt{-g}
\nonumber\\
&&
\,+\, (A'_t)^2 A'_\phi \, j^t\sqrt{-g}
\,+\,  A'_t (A'_\phi)^2\, j^\phi\sqrt{-g}
\,=\,0.
\label{eq:fint_MHD-Euler_xA}
\eeqn
Derivations of the above 
Eqs.~(\ref{eq:fint_MHD-Euler_t})-(\ref{eq:fint_MHD-Euler_xA})
are detailed in Appendix \ref{sec:App:fint_MHD-Euler}.  

For the case without meridional flow fields, $u^A=0$, a set of 
first integrals for the stationary and axisymmetric system can be 
derived from the above set of equations by taking a limit of 
the stream function, $[\sqrt{-g}\Psi](\Upsilon) \rightarrow$ constant.  
In this limit, the right hand side of the first integral 
(\ref{eq:fint_MHD-Euler_tphi}) becomes finite as shown in 
\cite{Gourgoulhon:2011gz}.  
In Appendix \ref{sec:App:purerotational}, we present a direct 
proof for the same case with pure rotational flows, since our formulation 
is slightly different from \cite{Gourgoulhon:2011gz}.  

\section{Formulation for numerical computation and numerical method}
\label{sec:Numerical}

In Sec.~\ref{sec:Formulation}, a set of elliptic PDEs
for computing gravitational fields 
$\{\psi, \tbeta_a, \alpha\psi, \habd\}$, 
and electromagnetic fields $\{\alpha\Phi_\Sigma, A_a \}$
of stationary and axisymmetric systems are derived from 
Einstein's and Maxwell's equations.  The number of variables 
for gravitational fields is 11, as it is augmented with 
the conformal factor $\psi$ and a condition $\tgamma=f$ is 
added to determine it.  The number of electromagnetic potentials are 
4, and 4 PDEs are derived from Maxwell's equations (\ref{eq:Maxwell}).   
The apparent number of variables and equations matches, though there are 
four additional coordinate conditions (\ref{eq:gauge}) to be imposed 
on the metric, and a gauge condition (\ref{eq:Coulomb}) 
on the electromagnetic potentials.  
For a set of matter and electric currents, 
$\{h, T, s, \rho, u^\alpha, j^\alpha\}$, 12 variables in total 
appear in a system of 10 equations in Sec.~\ref{sec:MHDeqs} which are 
MHD-Euler equations (\ref{eq:MHD-Euler}), ideal MHD condition 
(\ref{eq:idealMHD}) (3 components), the continuity equations for 
the rest mass conservation and the current conservation, 
and normalization of the 4-velocity $u\cdot u=-1$.  
Instead of solving the equation for local thermal energy conservation 
and the 2-parameter EOS (\ref{eq:EOS2param}) (that is, $h = h(\rho,s)$) 
simultaneously with the above system, we assume 1-parameter (barotropic) 
EOS.\footnote{The component along $u^\alpha$ of 
the ideal MHD condition $F\cdot u=0$ is trivial.   
The component along $u^\alpha$ of the relativistic MHD-Euler 
equation constrains the flow to be isentropic.}  
The apparent number of the variables and equations for the matter and 
current also match.  This is also the case for the derived system of 
algebraic equations for the first integrals and integrability conditions.

As shown in previous sections, for stationary and axisymmetric 
ideal MHD, the system of equations for matter and currents are integrable 
analytically when the $t$ and $\phi$ components of the electromagnetic 
potential 1-form and the densitized stream function are homologous.  
We rewrite these equations to be solved iteratively in our numerical 
code.

\subsection{Formulation for solving Maxwell's equations in ideal MHD}


Our formulation is to solve the 3+1 decomposed Maxwell's equations 
in the form of elliptic equations for the electromagnetic 1-form.  
Those are Eqs.~(\ref{eq:Maxwell_n_poisson}) and (\ref{eq:Maxwell_vecpotA}) 
and can be integrated by prescribing the current $j^\alpha$.  
The ideal MHD condition constrains the 4 components of $j^\alpha$.  
In particular, the $t$ and $\phi$ components appearing in 
Eq.~(\ref{eq:fint_MHD-Euler_xA}) are an inseparable combination, 
which is a consequence of the 
integrability conditions (\ref{eq:def_fn_At_Psi}) requiring the potentials 
$A_t$ and $A_\phi$ to be functions of a single master potential $\Upsilon$.  
There are several ways to rearrange the system of equations 
derived in Sec.~\ref{sec:fint} to compute $j^\alpha$.  In the rearrangement 
used in this paper, we choose that the master potential $\Upsilon$ 
to be equal to $A_\phi$.  This choice is general enough to generate 
interesting solutions for rotating compact stars associated with mixed 
toroidal and poloidal magnetic fields.

\subsubsection{Form of the currents}
\label{sec:current}

As shown in Eq.~(\ref{eq:mercurrent}), 
the Maxwell's equations relate $F^{AB} =\epsilon^{AB} B$ 
(\ref{eq:FABu}) with the $x^A$ components of the current $j^A$, 
\beq
j^A \sqrt{-g} 
\,=\, \frac1{4\pi}\epsilon^{AB}\pa_B(\sqrt{-g}B).  
\label{eq:current_def}
\eeq
Multiplying Eq.~(\ref{eq:current_def}) by $A'_t A'_\phi$ and 
using the first integral of the $t$ and $\phi$ components of MHD-Euler 
equations (\ref{eq:fint_MHD-Euler_t}) and (\ref{eq:fint_MHD-Euler_phi}), 
we have 
\beqn
&&\!\!\!\!\!\!\!\!\!\!\!\!\!\!\!\!\!\!\!\!\!\!
A'_t A'_\phi j^A \sqrt{-g} 
\nonumber\\ 
\,=\,
&-&\frac12\left[
A'_\phi\left([\sqrt{-g}\Psi]''hu_t + [\sqrt{-g}\Lambda_t]' \right)
\qquad
\right.
\nonumber\\
&&
\,-\,A''_\phi \left([\sqrt{-g}\Psi]'hu_t + [\sqrt{-g}\Lambda_t] \right)
\nonumber\\
&&
\,+\, A'_t\left([\sqrt{-g}\Psi]'' hu_\phi + [\sqrt{-g}\Lambda_\phi]' \right)
\nonumber\\
&&
\left. 
\,-\, A''_t \left([\sqrt{-g}\Psi]' hu_\phi + [\sqrt{-g}\Lambda_\phi] \right)
\right]\epsilon^{AB}\pa_B \Upsilon
\nonumber\\
&+&\frac12 A'_\phi[\sqrt{-g}\Psi]' \epsilon^{AB}\pa_B(hu_t)
\nonumber\\
&+&\frac12 A'_t[\sqrt{-g}\Psi]' \epsilon^{AB}\pa_B(hu_\phi). 
\eeqn

The combination of the $t$ and $\phi$-components of the current $j^\alpha$ 
has a similar form as above; from the first integral 
of the MHD-Euler equations (\ref{eq:fint_MHD-Euler_xA}), we have 
\beqn
&&
(A'_t)^2 A'_\phi \, j^t\sqrt{-g}
\,+\,  A'_t (A'_\phi)^2\, j^\phi\sqrt{-g}
\nonumber\\
&&\quad
\,=\,
\,\frac12\left\{
A'_\phi\left([\sqrt{-g}\Psi]''hu_t + [\sqrt{-g}\Lambda_t]' \right)
\right.
\nonumber\\
&&\quad\qquad
\,-\,A''_\phi \left([\sqrt{-g}\Psi]'hu_t + [\sqrt{-g}\Lambda_t] \right)
\nonumber\\
&&\quad\qquad
\,+\, A'_t\left([\sqrt{-g}\Psi]'' hu_\phi + [\sqrt{-g}\Lambda_\phi]' \right)
\nonumber\\
&&\quad\qquad
\left.
\,-\, A''_t \left([\sqrt{-g}\Psi]' hu_\phi + [\sqrt{-g}\Lambda_\phi] \right)
\right\}B_\phi
\nonumber\\
&&
\,-\,\frac12(A''_t\,hu_\phi - A''_\phi\,hu_t+\Lambda')
(A'_t u^t - A'_\phi u^\phi)\rho\sqrt{-g}
\nonumber\\
&&
\,-\, A'_t A'_\phi[\sqrt{-g}\Psi]'\omega_\phi
\,-\,A'_t  A'_\phi  s'   T   \rho \sqrt{-g} \ .
\label{eq:current_jphijt}
\eeqn

The above expressions for $j^\alpha$ are symmetric in $t$ and $\phi$, 
and they can be used for taking the limit as either $A_t$ or $A_\phi$ 
approaches to a constant.  
When $A_\phi$ is not constant, one can, without loss of generality, choose 
$\Upsilon = A_\phi$ because $A_\phi$ is an arbitrary function of 
the master potential $\Upsilon$.  
Since $A'_\phi = 1$ and $A''_\phi = 0$ for this case, 
we can derive simpler expressions for $j^\alpha$ with the help of 
Eqs.~(\ref{eq:fint_idealMHD_xA}) and (\ref{eq:fint_MHD-Euler_phi});
\beqn
&&
j^A \sqrt{-g} 
\,=\,
\left([\sqrt{-g}\Psi]''hu_\phi
+[\sqrt{-g}\Lambda_\phi]'
\right)\delta^{AB}B_B
\nonumber\\
&&\qquad\qquad
\,-\,[\sqrt{-g}\Psi]'
\delta^{AB}\omega_B,  
\label{eq:current_jA_Aphi}
\\[3mm]
&&
j^\phi\sqrt{-g}
\,+\,A'_t\, j^t\sqrt{-g}
\nonumber\\
&&\quad
\,=\,
\left([\sqrt{-g}\Psi]''hu_\phi 
+ [\sqrt{-g}\Lambda_\phi]'\right)B_\phi
\,-\,[\sqrt{-g}\Psi]'\omega_\phi
\nonumber\\
&&\quad
\,-\,\left(A''_t \,hu_\phi+\Lambda'\right)\rho u^t \sqrt{-g}
\,-\, s' T \rho \sqrt{-g} , 
\label{eq:current_jphijt_Aphi}
\eeqn
where $\delta^{AB}$ is the Kronecker delta, and 
$B_A$ and $\omega_A$ defined in Eqs.~(\ref{eq:F_Aphi}) 
and (\ref{eq:dhu_Aphi}) are substituted.

\subsubsection{Calculation of $j^t$ }

Among the four components of the current $j^\alpha$, there are three 
independent components; the $t$ and $\phi$ components appear to be 
a combination as in Eq.~(\ref{eq:current_jphijt_Aphi}).  
Therefore we propose to use the $t$-component of Maxwell's equations 
to determine the $t$-component of the current $j^t$.  
From Eq.~(\ref{eq:Maxwell_n}), using the relations 
$\rho_\Sigma = -j^\alpha n_\alpha = \alpha j^t$, and 
$D_a F^a 
=\psi^{-6}\zD_a(\psi^6 F^a)$, 
$j^t$ is written 
\beq
j^t \,=\, \frac1{4\pi\alpha\psi^6}\zD_a(\psi^6 F^a)
\,=\,
\frac1{4\pi\alpha\psi^6}\zD_a(\psi^2 \tgmabu F_b). 
\label{eq:jt}
\eeq
Then, we move the $j^t$ term in Eq.~(\ref{eq:current_jphijt_Aphi}) to 
the right-hand side to isolate $j^\phi$, and use this $j^\phi$ 
as source of the spatial components of Maxwell's equations 
(\ref{eq:Maxwell_vecpotA}) for evaluating $A_a$.  
This method works because $j^t$ is related to $A_t$ 
under a choice of $\Upsilon=A_\phi$ and hence $A_t$ is 
a prescribed function of $A_\phi$, $A_t=A_t(A_\phi)$ 
on the support of ideal MHD fluid, that is 
$F_a$ in Eq.~(\ref{eq:jt}) is related to $A_t$ through 
Eq.~(\ref{eq:Maxwell_n_tsym}) as 
\beq
F_a \,=\, \frac1{\alpha}\Lie_\beta A_a 
\,+\, \frac1{\alpha}D_a \big(-A_t(\Upsilon)+ A_a \beta^a\big) , 
\eeq
since $ A_t = -\alpha\Phi_\Sigma + A_a \beta^a$.  

In actual computations, we also solve the $n^\alpha$ component 
of Maxwell's equations (\ref{eq:Maxwell_n_poisson}) 
to cross-check the consistency of this method by comparing 
a solution and a prescribed function $A_t(A_\phi)$.  
It is also necessary to solve (\ref{eq:Maxwell_n_poisson}) 
in the electro-vacuum spacetime outside of the compact stars 
because the above argument is valid only on the support of 
ideal MHD fluid.  

\subsection{Elliptic PDE solver}
\label{sec:PDEsolver}

As discussed in a series of papers 
\cite{Uryu:2016dqr,Tsokaros:2018zlf,cocal}, one of the 
basic concepts of the \cocal\ code is to develop a simple and 
straightforward numerical method for computing 
data sets on a 3D slice $\Sigma$.  Our idea is to 
formulate vectorial or tensorial elliptic PDEs 
in terms of Cartesian components, and apply 
the same elliptic PDE solver as that for the 
elliptic PDE of a scalar function on spherical 
coordinates $(r,\theta,\phi)$.  Our scalar elliptic 
PDE solver, for example for Eq.~(\ref{eq:GRfieldeq}), 
uses a multipole expansion of Green's  
function
\beq
\Phi({\bf x}) \,=\, -\frac1{4\pi}\int_V 
\frac{{\cal S}({\bf x}')}{\left| {\bf x}-{\bf x}' \right|}d^3x' 
\,+\, \chi({\bf x}),
\label{eq:PDEsolver}
\eeq
where ${\bf x}$ and ${\bf x}'$ are points in $V$, 
${\bf x}, {\bf x}' \in V\subset\Sigma$, 
\beqn
\frac1{\left| {\bf x}-{\bf x}' \right|}
&=&
\sum_{\ell=0}^\infty 
\frac{r_<^\ell}{r_>^{\ell+1}} \sum_{m=0}^\ell \epsilon_m \,
\frac{(\ell-m)!}{(\ell+m)!}
\nonumber\\
&\times&
P_\ell^{~m}(\cos\theta)\,P_\ell^{~m}(\cos\theta')
\cos m(\varphi-\varphi'), \qquad
\label{eq:PDEGreenFn}
\eeqn
where 
$r_> := \sup\{r,r'\}$, $r_< := \inf\{r,r'\}$, 
$\epsilon_m = 1$ for $m=0$, 
and $\epsilon_m = 2$ for $m\ge 1$, and 
$P_\ell^{~m}(\cos\theta)$ are the associated Legendre polynomials.  
We will truncate the expansion in $\ell$ at a certain positive 
integer $L$ so that $0 \leq \ell \leq L$.\footnote{
Obviously, in the present aim for computing axisymmetric configurations, 
it is not necessary to expand in the azimuthal angle $\phi$.  
The Cartesian components of vector or tensor variables have trivial 
dependencies on $\phi$, which may be easily integrated analytically.  
We, however, keep $\phi$ dependencies in the formulation and 
the $\phi$ integrals as Eq.~(\ref{eq:PDEsolver}) 
in the numerical code for future extensions.  }  

The function $\chi({\bf x})$ in Eq.~(\ref{eq:PDEsolver}) is 
a homogeneous solution, $\zLap\chi({\bf x})=0$, 
to be used for imposing boundary conditions on $\Phi({\bf x})$.  
The function $\chi({\bf x})$ may be included in the Green's function, 
if the boundary is a concentric sphere on the spherical coordinate
\cite{cocal}.  
For this particular problem, that is, computations of compact stars 
that have flat asymptotics, among all elliptic PDEs, 
Eqs.~(\ref{eq:Ham})-(\ref{eq:trfreeG}), 
(\ref{eq:gauge_eq_xi})--(\ref{eq:gauge_source_xia}), 
(\ref{eq:Maxwell_n_poisson})-(\ref{eq:Maxwell_vecpotA_source}) 
and (\ref{eq:CoulombPDE}), all of them except for one can be integrated 
setting $\chi({\bf x})$ to be constant, since errors introduced 
to the potential is negligible if the boundary of the computational 
domain is taken far enough from the source.  An exception for the choice 
for $\chi({\bf x})$ is Eq.~(\ref{eq:Maxwell_n_poisson}) 
to determine $\alpha\Phi_\Sigma$.

As mentioned in the previous subsection, $\Phi_\Sigma$ is related to $A_t$, 
and $A_t$ is determined from the integrability condition $A_t=A_t(A_\phi)$ 
on the support of the ideal MHD fluids, and from Maxwell's equations 
on electro-vacuum spacetime outside of the fluids.  Because we also 
assume that $A_\phi$ as well as its derivatives are continuous 
across the stellar surface, while $A_t$ and hence $\alpha\Phi_\Sigma$ are 
continuous across the surface but their derivatives are not.  
Therefore, in solving $\alpha\Phi_\Sigma$, we impose a boundary 
condition not only at the boundary of the computational domain, but also 
at the stellar surface.  Our idea to impose the boundary condition 
at the stellar surface is essentially the same as the one described 
in Sec 3.1 of \cite{Bocquet:1995je}.  


We assume that the stellar surface is a single valued function of 
spherical coordinates $\theta$ and $\phi$ as $r=R(\theta,\phi)$, 
whose origin $r=0$ is placed inside of the star.  The homogenous 
solution $\chi({\bf x})$ outside of fluid support is regular 
at $r\rightarrow \infty$, that is, $\chi \propto r^{-\ell-1}$
for $r\geq R(\theta,\phi)$, and $\chi({\bf x})$ is determined 
so that Eq.~(\ref{eq:PDEsolver}) (in which $\Phi$ is replaced by 
$\alpha\Phi_\Sigma$) satisfies the boundary value 
\beq
(\alpha\Phi_\Sigma)^{\rm B}\,:=\, 
\left[\, A_t(A_\phi) \,+\,A_a\beta^a \,\right]\,\big|_{r=R(\theta,\phi)}\ .
\eeq
To achieve this, we expand $\chi({\bf x})$ with coefficients 
$(a_{\ell m},b_{\ell m})$ as, 
\beq
\chi({\bf x})\,=\,
\sum_{\ell=0}^{\infty}\sum_{m=0}^{\ell} 
\frac1{r^{\ell+1}}{\cal Y}_\ell^{~m}(\theta)
\left(a_{\ell m}\cos m\phi +
      b_{\ell m}\sin m\phi
\right), 
\label{eq:homosol}
\eeq
where ${\cal Y}_\ell^{~m}(\theta)$ is defined by 
\beq
{\cal Y}_\ell^{~m}(\theta)
\,=\,\sqrt{\frac{\epsilon_m (2\ell+1)(\ell-m)!}{4\pi(\ell+m)!}}
P_\ell^{~m}(\cos\theta), 
\eeq
and the expansion in $\ell$ is also truncated at $L$ here.
To determine $(a_{\ell m},b_{\ell m})$ from imposing boundary conditions, 
we apply the method of least squares.  Writing the boundary value of 
the volume integral term in Eq.~(\ref{eq:PDEsolver}), 
\beq
\left.
I^{\rm B} 
\,:=\, -\frac1{4\pi}\int_V 
\frac{{\cal S}({\bf x}')}{\left| {\bf x}-{\bf x}' \right|}
\right|_{r=R(\theta,\phi)}d^3x' , 
\eeq
and $\chi^{\rm B}\,=\,\chi({\bf x})\big|_{r=R(\theta,\phi)}$, 
we define the squared residuals,  
\beq
I \,:=\,\frac12\sum_{\theta_j,\phi_k} \left[\,(\alpha\Phi_\Sigma)^{\rm B}
\,-\,(I^{\rm B}+\chi^{\rm B}) \,\right]^2, 
\label{eq:LSI}
\eeq
and apply the method of least squares to minimize $I$, that is, 
we solve a linear system, 
\beq
\frac{\pa I}{\pa a_{\ell m}}\,=\,0 \quad \mbox{and} \quad
\frac{\pa I}{\pa\, b_{\ell m}}\,=\,0, 
\eeq
to determine a set of coefficients $(a_{\ell m},b_{\ell m})$.  
In the above Eq.~(\ref{eq:LSI}), a summation is taken 
over all grid points $(\theta_j,\phi_k)$ which will be 
introduce in later section, and $\ell$ is truncated also here 
$0\le\ell\le L$.

\subsection{Equations for the matter variables}
\label{sec:SCFmatter}

Finally, we introduce final forms of the first integrals and 
integrability conditions which are suitable for the self-consistent 
field (SCF) iteration scheme used in our numerical method.  

We assume the one-parameter EOS to have a single independent 
thermodynamic variable for simplicity, and assume homentropic 
flow, $s(\Upsilon)={\rm constant}$.  Because of these choices, 
we have 5 independent variables for the matter $\{h, u^\alpha \}$.  
In the following, the 4-velocity $u^\alpha$ may also be written 
in 3+1 form, 
\beq
u^\alpha 
= u^t (t^\alpha + v^\alpha)
= u^t (\alpha n^\alpha +\beta^\alpha + v^\alpha), 
\eeq
where $v^\alpha$ is the spatial component of the velocity that satisfies 
$v^\alpha\na_\alpha t=0$, and both expressions $u^\alpha$ and $v^\alpha$ 
(or $v^a$) are mixedly used.  
The meridional components $u^A$ are written $u^A = u^t v^A$.

Assuming a choice $\Upsilon=A_\phi$, a possible arrangement of 
equations for the SCF iteration of hydrodynamic variables becomes as follows.

For the meridional velocity $u^A$, the rest mass conservation (\ref{eq:merflowfn}) 
is used; 
\beqn
u^A 
&=& \frac1{\rho\sqrt{-g}}\epsilon^{AB}\pa_B[\sqrt{-g}\Psi](\Upsilon) 
\nonumber\\
&=& \frac1{\rho\alpha \psi^6\sqrt{\tgamma}}
[\sqrt{-g}\Psi]'\epsilon^{AB}\pa_B A_\phi.
\label{eq:uA}
\eeqn
%

For the $t$-component of the 4-velocity, $u^t$, the norm $u\cdot u=-1$ is used; 
\beq
u^t\,=\, \frac1{\left[\alpha^2-\psi^4\tgmabd
(v^a+\beta^a)(v^b+\beta^b)\right]^{1/2}}.
\label{eq:ut}
\eeq
%

For the $\phi$-component of the 4-velocity, $u^\phi$, $x^A$ component of ideal MHD 
condition (\ref{eq:fint_idealMHD_xA}) is used; 
\beqn
u^\phi
&=& \frac{[\sqrt{-g}\Psi]'B_\phi}{A'_\phi\,\rho\sqrt{-g}}\,-\,\frac{A'_t}{A'_\phi}u^t
\nonumber\\
&=& \frac{[\sqrt{-g}\Psi]'B_\phi}{\rho\alpha\psi^6\sqrt{\tgamma}}-A'_t u^t.
\label{eq:uphi}
\eeqn
%

For a thermodynamic variable, the enthalpy $h$, the combination of $t$ and 
$\phi$-components of MHD-Euler equations (\ref{eq:fint_MHD-Euler_tphi}) is used; 
\beq
h 
\,=\, \frac{\Lambda}{A'_\phi u_t - A'_t u_\phi}
\,=\, \frac{\Lambda}{u_t - A'_t u_\phi} .
\label{eq:h}
\eeq


As mentioned earlier, even in the case of no meridional flows 
(purely rotational flows), $u^A=0$ and 
$[\sqrt{-g}\Psi](\Upsilon)={\rm constant}$, the above set of 
equations for matter are valid, and Eq.~(\ref{eq:h}) can be used 
as the same form (see, Appendix \ref{sec:App:fint_MHD-Euler}).  

\subsection{Assumptions for arbitrary functions}
\label{sec:fnc}

To specify a model of a rotating star, a concrete 
form of each arbitrary function that appears in the integrability 
conditions (\ref{eq:def_fn_At_Psi}) and in the first integrals 
(\ref{eq:fint_MHD-Euler_t})-(\ref{eq:fint_MHD-Euler_tphi}) 
has to be prescribed.  
We partly follow a choice made in \cite{YYE06} for these functions, 
which are used in our previous paper \cite{Uryu:2014tda}, but we also 
introduce new functional forms below.  As mentioned in 
Sec.\ref{sec:current}, we choose $\Upsilon=A_\phi$ as 
the independent variable instead of the master potential $\Upsilon$.  

\subsubsection{A smoothed step function}
\label{sec:stepfnc}

We introduce a class of a two parameter sigmoid function $\Xi'(x;b,c)$ 
that varies from $0$ to $1$ in a region $x \in [0, 1]$ whose 
transition width is $b$ $(0<b<1)$, and transition center $c$ $(0 < c < 1)$, 
\beq
\Xi'(x;b,c) = \frac12\left[\,\tanh\left(\frac{x}{b}-c \right)+1\,\right],  
\label{eq:MHDfnc_dXi_x}
\eeq
and its integral $\Xi(x;b,c)$, 
\beq
\Xi(x;b,c) = \frac12\left[\,b\,\ln\cosh\left(\frac{x}{b}-c\right)+x \,\right]
+{\rm constant}. 
\label{eq:MHDfnc_Xi_x}
\eeq

We make use of these functions in a region where $A_\phi$ varies 
on the fluid support and its contour is closed as will be explained later.  
Functions $\Xi'(A_\phi)$ and $\Xi(A_\phi)$ are defined by
\beq
\Xi'(A_\phi)
= \frac12\left[\,\tanh\left(\frac1{b}
\frac{A_\phi-A_{\phi, \rm S}^{\rm max}}{A_\phi^{\rm max} - A_{\phi, \rm S}^{\rm max}}
-c\right)+1\,\right],  
\label{eq:MHDfnc_dXi}
\eeq
and
\beqn
\Xi(A_\phi)
&=& \frac12\left[\,b (A_\phi^{\rm max}-A_{\phi, \rm S}^{\rm max})\times 
\phantom{\frac1{b}}\right.
\nonumber\\
&&\left.
\ln\cosh\left(
\frac1{b}
\frac{A_\phi-A_{\phi, \rm S}^{\rm max}}{A_\phi^{\rm max} - A_{\phi, \rm S}^{\rm max}}
-c\right)+A_\phi\,\right].  \qquad
\label{eq:MHDfnc_Xi}
\eeqn
The smoothed step function $\Xi'(A_\phi)$ varies on a region 
$A_\phi \in [A_{\phi, \rm S}^{\rm max}, A_\phi^{\rm max}]$, where 
$A_\phi^{\rm max}$ and $A_{\phi, \rm S}^{\rm max}$ are the maximum values of 
$A_\phi$ on the fluid support, and that on the stellar surface, respectively, 
Here, $A_\phi^{\rm max} > A_{\phi, \rm S}^{\rm max}$ is assumed.  
Note that Eqs.~(\ref{eq:MHDfnc_dXi}) and (\ref{eq:MHDfnc_Xi}) are 
not a mere substitution of 
\beq
x=\frac{A_\phi - A_{\phi, \rm S}^{\rm max}}{A_\phi^{\rm max} - A_{\phi, \rm S}^{\rm max}}
\eeq
to Eqs.~(\ref{eq:MHDfnc_dXi_x}) and (\ref{eq:MHDfnc_Xi_x}), since 
the prime of Eq.~(\ref{eq:MHDfnc_dXi}) is with respect to $A_\phi$, not $x$, 
and a constant of integration in Eq.~(\ref{eq:MHDfnc_Xi}) is chosen appropriately.  

The parameters $(b,c)$ determine the width and position of the transition 
of $\Xi'$, and are set to be $(b,c)=(0.2,0.5)$ in the following applications.  
Other types of smooth step functions such as those made from Hermite 
interpolation polynomials could be used in the same manner.  


\subsubsection{Models}
\label{sec:models}


For the function $\Lambda(A_\phi)$, we choose 
\beq
\Lambda = - \Lambda_0 \Xi(A_\phi) - \Lambda_1 A_\phi - {\cal E},
\label{eq:MHDfnc_Lambda}
\eeq
where $\Lambda_0$, $\Lambda_1$, and ${\cal E}$ are constants.  
$\Lambda_0$ and $\Lambda_1$ are set by hand, while ${\cal E}$ is calculated 
from a condition to be imposed on a solution.  
In case of pure rotational flows without magnetic fields, 
the constant ${\cal E}$ agrees with the injection 
energy \cite{RNS}.

For the function $A_t(A_\phi)$, we choose 
\beq
A_t = - \Omegac A_\phi + C_e, 
\label{eq:MHDfnc_At}
\eeq
where $C_e$ is a constant that relates to the net electric 
charge of the star, and $\Omegac$ is a constant.  As discussed 
in \cite{YYE06}, the choice of the first term corresponds 
to the rigid rotation in the limits of 
$[\sqrt{-g}\Psi]' \rightarrow 0$ or $B_\phi \rightarrow 0$, 
because of relation (\ref{eq:rotlaw_noMC}).  

The current (\ref{eq:current_jA_Aphi}) and (\ref{eq:current_jphijt_Aphi}) 
involve terms with a derivative of function 
$[\sqrt{-g}\Lambda_\phi](A_\phi)$ coupled with the magnetic fields.  
Since we assume no electric current outside of the star, 
$[\sqrt{-g}\Lambda_\phi]'(A_\phi)$ has to vanish outside of 
the fluid support.  This is why we prepare 
a smooth function that varies between $[0,1]$ in a region 
$A_\phi \in [A_{\phi, \rm S}^{\rm max}, A_\phi^{\rm max}]$ as 
in Sec.~\ref{sec:stepfnc}.  
We choose a smooth function, 
\beqn
[\sqrt{-g}\Lambda_\phi] &=& \Lambda_{\phi 0}\, \Xi(A_\phi),
\label{eq:MHDfnc_Lambda_phi}
\\[1mm]
[\sqrt{-g}\Lambda_\phi]' &=& \Lambda_{\phi 0}\, \Xi'(A_\phi), 
\label{eq:MHDfnc_dLambda_phi}
\eeqn
where the parameter $\Lambda_{\phi 0}$ is the range of the function, 
$[\sqrt{-g}\Lambda_\phi]'(A_\phi)\in [0,\Lambda_{\phi 0}]$ set by hand.

In the later sections, we only present solutions without 
the meridional circulation flows, hence for $[\sqrt{-g}\Psi](A_\phi)$, 
we set 
\beqn
[\sqrt{-g}\Psi] &=& {\rm constant},
\\[1mm]
[\sqrt{-g}\Psi]' &=& 0.
\eeqn
We have also tested a few models for $[\sqrt{-g}\Psi](A_\phi)$, 
and successfully computed solutions with meridional flows, 
although so far we have calculated solutions whose meridional 
flows do not affect equilibrium of the stars.  
For example, we may choose the same form as 
Eqs.~(\ref{eq:MHDfnc_Lambda_phi}) and (\ref{eq:MHDfnc_dLambda_phi}),  
\beqn
[\sqrt{-g}\Psi] &=& a_\Psi\, \Xi(A_\phi),
\label{eq:MHDfnc_Psi}
\\[1mm]
[\sqrt{-g}\Psi]' &=& a_\Psi\, \Xi'(A_\phi), 
\label{eq:MHDfnc_dPsi}
\eeqn
where $a_\Psi$ is a parameter to be set by hand.  

\subsubsection{Differentially rotating models}

When magnetic fields and meridional flows exist inside of compact stars, 
Eq.~(\ref{eq:uphi}) implies that the stellar rotation $\Omega:=u^\phi/u^t$ 
becomes inevitably differential in general, because a combination 
$B_\phi/\rho\sqrt{-g}$ is not a function of $A_\phi$ or $\Upsilon$.  
When there is no meridional flow $u^A=0$, $[\sqrt{-g}\Psi]'=0$, on the 
other hand, the form of the function $A_t$ (\ref{eq:MHDfnc_At}) 
results in a uniform rotation as mentioned.  

It seems that the latter case with no meridional flows $[\sqrt{-g}\Psi]'=0$ 
is sometimes misinterpreted in the literature, as stated in 
\cite{Bocquet:1995je}, that only the uniform rotation 
$\Omega={\rm constant}$ is allowed in this case.  
This statement seems to have been made because a distribution of 
$A_\phi$ usually becomes toroidal and hence such a toroidal differential 
rotation $\Omega(A_\phi)$ was considered to be unnatural.  
Such differential rotation laws in which $\Omega$ depends on 
$\Upsilon$ (or $A_\phi$) are, however, allowed mathematically, 
and may not necessarily be too unrealistic to be rejected.  
For example, one can assume moderate, or weak, differential rotations, 
\beq
\frac{u^\phi}{u^t} = \frac{A_t'}{A'_\phi}:=\Omegac +\delta \Omega(\Upsilon),
\eeq
by setting $\max|\delta \Omega(\Upsilon)|$ to be a few tens of \%, or less, 
of $\Omegac$.  Various rotation laws can also be used for the case with 
meridional flow 
$u^A\neq 0$ (that is, $[\sqrt{-g}\Psi]'\neq 0$), 
but it is more likely that some kind of instability such as the magnetorotational 
instability may be induced.  
%

\subsubsection{Other models}

In our previous paper \cite{Uryu:2014tda}, the functional form for 
$A_t(A_\phi)$ was chosen the same as Eq.~(\ref{eq:MHDfnc_At}), and 
for the function $\Lambda(A_\phi)$, $\Lambda_0$ was set to be zero in 
Eq.~(\ref{eq:MHDfnc_Lambda}).  
Our previous choices for $[\sqrt{-g}\Lambda_\phi](A_\phi)$ 
and $[\sqrt{-g}\Psi](A_\phi)$ in \cite{Uryu:2014tda} were taken from those 
used in \cite{YYE06}. 
For $[\sqrt{-g}\Lambda_\phi]$, we have chosen 
\beqn
[\sqrt{-g}\Lambda_\phi]
&=& \frac{a}{k+1}(A_\phi - A_{\phi, \rm S}^{\rm max})^{k+1}
\Theta(A_\phi - A_{\phi, \rm S}^{\rm max}), \qquad
\label{eq:MHDfnc_Lambda_phi_bad}
\\[1mm]
[\sqrt{-g}\Lambda_\phi]' 
&=& a(A_\phi - A_{\phi, \rm S}^{\rm max})^k
\Theta(A_\phi - A_{\phi, \rm S}^{\rm max}), 
\label{eq:MHDfnc_dLambda_phi_bad}
\eeqn
where values of the constant coefficient $a$ and index $k$ 
are set by hand, and $\Theta(x)$ is the Heaviside function.  
In \cite{YYE06}, it was found that 
the solutions have comparable strength in poloidal and 
toroidal components of magnetic fields when 
the index is about $k = 0.1$.  
We replace Eqs.~(\ref{eq:MHDfnc_Lambda_phi_bad}) and (\ref{eq:MHDfnc_dLambda_phi_bad}) 
with Eqs.~(\ref{eq:MHDfnc_Lambda_phi}) and (\ref{eq:MHDfnc_dLambda_phi}) to bring 
a smoothness as well as better control in the behavior of the functional forms, 
although either choice of the function reproduce qualitatively 
the same set of solutions.  
Also in \cite{Uryu:2014tda}, for $[\sqrt{-g}\Psi](A_\phi)$, 
we have chosen
\beqn
[\sqrt{-g}\Psi] &=& \frac{a_\Psi}{p+1}(A_\phi - A_\phi^{\rm max})^{p+1}
\Theta(A_\phi - A_\phi^{\rm max}), 
\qquad
\label{eq:MHDfnc_Psi_bad}
\\[1mm]
[\sqrt{-g}\Psi]' &=& a_\Psi(A_\phi - A_\phi^{\rm max})^p
\Theta(A_\phi - A_\phi^{\rm max}), 
\label{eq:MHDfnc_dPsi_bad}
\eeqn
where the values of constant coefficient $a_\Psi$ and index $p$ 
are given by hand, for which we set $p=1$ following \cite{YYE06}.

\subsection{Alternative form of the first integral of 
MHD-Euler equations under pure rotational flows}

When the flow fields are pure rotational $u^A=0$, 
the 4-velocity $u^\alpha$ becomes
\beq
u^\alpha \,=\, u^t(t^\alpha + \Omega \phi^\alpha) \,=\, u^t k^\alpha.  
\label{eq:4ve_GS}
\eeq
For the stationary and axisymmetric perfect fluid spacetime 
{\it without} magnetic fields, the first integral of 
the relativistic Euler equations can be derived as 
a consequence of the Cartan identity (\ref{eq:Cartan}).  
For the simplest uniformly rotating case, $\Omega={\rm constant}$, 
the helical vector $k=t+\Omega \phi$ becomes a Killing vector.  
Then, the Euler equations are written, with the help of 
Eq.~(\ref{eq:Cartan}), 
\beqn
u\cdot d(\hu) &=& u^t\left[\Lie_k \hu \,-\, d(k\cdot\hu)\right]
\nonumber\\
&=&-u^t d(k\cdot\hu) \,=\,0, 
\eeqn
and hence the first integral is derived as $\hu\cdot k=\mbox{constant}$.
This relation is used for determining a thermodynamic 
variable, the enthalpy $h$ in this case, of uniformly rotating 
non-magnetized stars.  

In the presence of magnetic fields, the corresponding first integral 
(\ref{eq:fint_MHD-Euler_tphi}) (or (\ref{eq:fint_MHD-Euler_tphi_GS})) 
for determining the relativistic enthalpy $h$, in place of the above relation 
$\hu\cdot k=\mbox{constant}$, was not derived from the Cartan identity 
(\ref{eq:Cartan}) as discussed in previous sections.  We show 
an alternative derivation of the first integral of ideal MHD flow 
using the Cartan identity (\ref{eq:Cartan}), which is valid only 
for the case of pure rotational flows (\ref{eq:4ve_GS}).  

The canonical momentum, $\pi_\alpha=hu_\alpha$, respects 
the symmetries $\Lie_t \pi_\alpha=\Lie_\phi \pi_\alpha=0$.  
Although the angular velocity $\Omega$ is a certain function 
which also respects the symmetries $\Lie_t \Omega=\Lie_\phi \Omega=0$, 
$\Omega \phi^\alpha$ is not a Killing vector.  
Hence for a certain p-form $Q$, a relation, 
\beq
\Lie_{\Omega\phi}Q\,=\,\Omega\Lie_\phi Q \,+\, d\Omega\wedge(\phi \cdot Q) ,
\eeq
is satisfied.  
Then, the first term of the MHD-Euler equations (\ref{eq:MHD-Euler}) 
divided by the enthalpy $h$ becomes 
\beqn
\frac{u}{h}\cdot d(\hu)
&=& \frac{u^t}{h}k \cdot d(\hu)
\,=\,\frac{u^t}{h} \left[\, \Lie_k d(\hu) \,-\,  d(k \cdot\hu) \,\right]
\nonumber\\ 
&=& \frac{u^t}{h} \left[\, d\Omega\wedge(\phi \cdot\hu) \,-\,  d(k \cdot\hu) \,\right]
\nonumber\\ 
&=& u^t u_\phi d\Omega \,+\, \frac{u^t}{h} d\left(\frac{h}{u^t}\right) . 
\eeqn
Hence, Eq.~(\ref{eq:MHD-Euler}) is rewritten,
\beq
d\ln\frac{h}{u^t}
\,+\,u^t u_\phi d\Omega
\,-\,\frac{T}{h}ds
\,+\,\frac1{\rho h}j\cdot dA
\,=\,0 . 
\label{eq:MHD-Euler_GS}
\eeq

Substituting the current (\ref{eq:current}), the $t$ and $\phi$-components 
of Eq.~(\ref{eq:MHD-Euler_GS}) become, \\
$t$-component:
\beq
j\cdot dA\cdot t
\,=\, j\cdot \left[-\Lie_t A\,+\,d(t\cdot A)\right]
\,=\, j^A \pa_A A_t
\,=\,0 , 
\label{eq:MHD-Euler_GS_t}
\eeq
$\phi$-component:
\beq
j\cdot dA\cdot \phi 
\,=\, j\cdot \left[-\Lie_\phi A\,+\,d(\phi\cdot A)\right]
\,=\, j^A \pa_A A_\phi
\,=\,0.
\label{eq:MHD-Euler_GS_phi}
\eeq
Above we used $t\cdot dQ=0$ and $\phi\cdot dQ=0$ for a scalar $Q$. 
For the $x^A$-components, combining Eq.~(\ref{eq:MHD-Euler_GS}) 
dotted with the basis $e_A$
\beq
\left(d\ln\frac{h}{u^t}
\,+\,u^t u_\phi d\Omega
\,-\,\frac{T}{h} ds
\,+\,\frac1{\rho h}j\cdot dA\right) \cdot e_A
\,=\,0, 
\eeq
and
\beqn
&&
j\cdot dA\cdot e_A 
\nonumber\\
&&\ 
\,=\, \left\{\,j^t\left[\Lie_t A\,-\,d(t\cdot A)\right]
\,+\,      j^\phi\left[\Lie_\phi A\,-\,d(\phi\cdot A)\right]
\right.
\nonumber\\
&&\left.\qquad
\,+\, j^B e_B\cdot dA \,\right\}\cdot e_A
\nonumber\\
&&\ 
\,=\,-j^t e_A\cdot dA_t 
\,-\, j^\phi e_A\cdot dA_\phi
\,-\,j^B (dA)_{AB}
\,=\,0,\qquad
\eeqn
we have \\
$x^A$-component:
\beqn
&&
\pa_A\ln\frac{h}{u^t}
\,-\,\frac{T}{h} \pa_A s
\,+\,u^t u_\phi \pa_A\Omega
\nonumber\\
&&
\,-\,\frac1{\rho h}
\left[j^t \pa_A A_t 
\,+\, j^\phi \pa_A A_\phi
\,+\,j^B (dA)_{AB}\right]
\,=\,0, \quad
\label{eq:MHD-Euler_GS_xA}
\eeqn
where $e_A\cdot dQ=\pa_A Q$ for a scalar $Q$.

Substituting the $x^A$-components of Maxwell's equations 
(\ref{eq:mercurrent}), to the $t$ and $\phi$ components of MHD-Euler 
equations (\ref{eq:MHD-Euler_GS_t}) and (\ref{eq:MHD-Euler_GS_phi}), 
\beqn
&&
\epsilon^{AB}\pa_B(\sqrt{-g}B)\pa_A A_t\,=\,0,
\\[1mm] &&
\epsilon^{AB}\pa_B(\sqrt{-g}B)\pa_A A_\phi\,=\,0,
\eeqn
the integrability conditions, 
\beqn
&&
A_t=A_t(\Upsilon),
\quad\ \ 
A_\phi=A_\phi(\Upsilon),
\nonumber\\[1mm]
&&\!\!\!\!\!\!\!\!\!\!\!\!\!\!\!\!\!\!
\mbox{and}
\qquad\quad
\sqrt{-g}B=[\sqrt{-g}B](\Upsilon), 
\label{eq:def_fn_At_B}
\eeqn
are derived.  
Hence, using Eqs.~(\ref{eq:mercurrent}) and (\ref{eq:FABd})
the current term of $x^A$-components (\ref{eq:MHD-Euler_GS_xA}) 
becomes 
\beq
j^B(dA)_{AB}
\,=\,-\frac1{4\pi\sqrt{-g}}B_\phi[\sqrt{-g}B]'\pa_A\Upsilon.  
\label{eq:current_term_xA}
\eeq
Substituting Eq.~(\ref{eq:current_term_xA}), and the integrability 
conditions (\ref{eq:def_fn_At_B}) for $A_t$ and $A_\phi$ and 
(\ref{eq:rotlaw_noMC}) for $\Omega$ into 
Eq.~(\ref{eq:MHD-Euler_GS_xA}), we have \\
$x^A$-component: 
\beqn
&&
\pa_A\ln\frac{h}{u^t}
\,-\,\frac{T}{h} \pa_A s
\,+\,\left[
u^t u_\phi \Omega'
\phantom{\frac12}\right.
\nonumber\\
&&\left.
\,-\,\frac1{\rho h}
\left(\,A'_t j^t  \,+\, A'_\phi j^\phi 
\,-\,\frac1{4\pi\sqrt{-g}}B_\phi[\sqrt{-g}B]'
\,\right)\right]\pa_A\Upsilon
\,=\,0.
\nonumber\\
\label{eq:MHD-Euler_GS_xA_2}
\eeqn

To derive a first integral of this Eq.~(\ref{eq:MHD-Euler_GS_xA_2}), 
we have a few choices to reduce the first two terms: 
we may assume any of (i) $s=\mbox{constant}$, (ii) $T/h=[T/h](s)$, 
(iii) $\rho h = [\rho h](p)$ with $d h-Tds=dp/\rho$, or (iv) $s=s(\Upsilon)$.  
Then the following argument goes analogously.  Here we assume a homentropic fluid, 
$s=\mbox{constant}$, for simplicity, and rewrite Eq.~(\ref{eq:MHD-Euler_GS_xA_2}), 
\beq
d\ln\frac{h}{u^t}\,=\, \lambda d\Upsilon, 
\label{eq:MHD-Euler_GS_xA_3}
\eeq
or we separate the contribution of differential rotation by defining 
$j(\Omega):=u^t u_\phi$, and rewrite 
\beq
d\ln\frac{h}{u^t}\,+\, \left[\,j(\Omega) \Omega' \,-\, \lambda\,\right] d\Upsilon \,=\,0.
\label{eq:MHD-Euler_GS_xA_4}
\eeq
From the converse of the Poincar\'e lemma, the integrability condition becomes 
\beq
\lambda\,=\,\lambda(\Upsilon).
\eeq
For the latter Eq.~(\ref{eq:MHD-Euler_GS_xA_4}), the arbitrary function 
$\lambda(\Upsilon)$ is related to the current as 
\beq
\,A'_t j^t  \,+\, A'_\phi j^\phi
\,=\, \rho h \lambda
\,+\,\frac1{4\pi\sqrt{-g}}B_\phi[\sqrt{-g}B]'.
\eeq
and the first integral is written 
\beq
\ln\frac{h}{u^t}\,+\,\int j(\Omega) d\Omega(\Upsilon)
\,-\,\int\lambda(\Upsilon) d\Upsilon\,=\,{\cal E},
\eeq
where ${\cal E}$ is a constant.  

We have implemented this formulation, 
assuming $\Upsilon=A_\phi$, and $\Omega={\rm constant}$, 
\beq
\frac{h}{u^t}\,=\,{\cal E}e^{-\Lambda}, 
\quad \mbox{where}\quad 
\Lambda = \Lambda(A_\phi), 
\label{eq:h_alt}
\eeq
and, although we do not show the result in this paper, 
we have computed a few solutions which agree well 
with those calculated from Eq.~(\ref{eq:h}).  

\subsection{Remarks on numerical method}

\subsubsection{Finite difference and iteration}

Given the forms of functions presented in Sec \ref{sec:fnc}, 
a set of integral equations of the field equations (\ref{eq:PDEsolver}) 
and algebraic equations arranged from the first integrals and 
integrability conditions (\ref{eq:uA})-(\ref{eq:h}) 
derived in Sec.~\ref{sec:PDEsolver} and \ref{sec:SCFmatter} 
are a full system of equations for computing magnetized rotating 
compact stars.  These equations are discretized on spherical 
coordinates $(r,\theta,\phi)\in [r_a,r_b]\times[0,\pi]\times[0,2\pi]$ 
that cover a star and an asymptotic region, where the origin $r_a=0$ 
is located at the center of the star.  Then a self-consistent 
field iteration method is applied to calculate a converged solution.  
The numerical code is developed on \cocal\ code extending previously 
developed rotating compact star code in which the waveless 
formulation is used \cite{Shibata:2004qz,Uryu:2016dqr,cocal}.  
A numerical method used in the present code for magnetized compact 
stars, including setups for coordinate grid points and grid spacing, 
finite difference schemes for derivatives and integrals, and 
the self-consistent iteration scheme, are common with the above 
mentioned rotating compact star code on \cocal.  Readers who 
are interested in the details of the code are advised to consult 
papers \cite{Uryu:2016dqr,cocal}.

In Table \ref{tab:grids_param}, we reproduce a list of relevant 
parameters for the coordinate grids presented in previous papers 
\cite{Uryu:2016dqr,cocal}, and, in Table \ref{tab:RNS_grids}, 
present the numbers of grid points and other grid parameters used 
in actual computations shown in the later sections.  
An important difference from the previous calculations for 
non-magnetized rotating stars is the inclusion of higher multipoles 
and higher resolution in the $\theta$ direction.  As we will see 
below, stronger toroidal magnetic fields tend to concentrate near 
the equatorial plane, hence it is necessary to increase the number of 
terms in the multipole expansion in (\ref{eq:PDEGreenFn}) and (\ref{eq:homosol}) 
to as high as $L \agt 30$, and accordingly the grid points in the $\theta$ 
direction to $N_\theta \agt 144$.  

Typically, with the grid setup SE3 $L=40$ in Table \ref{tab:RNS_grids}, 
it takes 6 minutes per 1 iteration using 1 CPU thread of 
Xeon E5-2687W v3 3.1GHz, and for a convergence about 500-1000 
iterations are required.

\subsubsection{Parameters}

In our formulation, parameters to specify a magnetized rotating 
model appear in the integrability conditions shown in Sec.~\ref{sec:models}.
For the case without meridional flows, those are 
$\Lambda_0$, $\Lambda_1$, and $\cal E$ in Eq.~(\ref{eq:MHDfnc_Lambda}), 
$\Omegac$ and $C_e$ in Eq.~(\ref{eq:MHDfnc_At}), and $\Lambda_{\phi0}$ 
in Eq.~(\ref{eq:MHDfnc_Lambda_phi}).  A set of parameters $b$ and $c$ 
contained in smooth step functions $\Xi(A_\phi)$ in 
Eqs.~(\ref{eq:MHDfnc_Lambda}) and (\ref{eq:MHDfnc_Lambda_phi}) 
may be distinct in general but are set to have the same value in both 
equations.  In addition to these parameters, we augment the number 
of parameters by introducing an equatorial radius $R_0$ in coordinate 
length for rescaling the radial coordinate $r$ \cite{Uryu:1999uu}. 

Another set of parameters is introduced from the EOS, which is also 
one of the integrability conditions.  
In \cocal, a piecewise polytropic EOS and a variant of such 
piecewise EOS to model, for example, quark matter, are implemented 
\cite{CocalQuark}.  
In this paper, we simply use a (single segment) polytropic EOS 
\beq
p\,=\,K \rho^\Gamma, 
\label{eq:EOS}
\eeq
which introduces two parameters, $K$ and $\Gamma$, the adiabatic constant 
and the adiabatic index.  


The values of the parameters $\{ \Lambda_0, \Lambda_1, \Lambda_{\phi0}, b, c\}$ 
are prescribed, which control the strength of electromagnetic fields.  
As in the computations of non-magnetized rotating compact stars, 
three parameters, $\{{\cal E}, \Omegac, R_0 \}$, are determined 
from three conditions, which are a given value of the maximum 
rest mass density $\rhoc$, at (or near) the center of the star, 
the normalization of the equatorial radius, $r_{\rm eq}$, as 
$r_{\rm eq}/R_0=1$, and the given value for the deformation $r_p/r_{\rm eq}$ 
at the north pole, $r_p$.  These conditions are imposed on 
Eq.~(\ref{eq:h}) and the resulting set of three algebraic equations 
is simultaneously solved to determine $\{{\cal E}, \Omegac, R_0 \}$ 
during iteration.

Finally, the parameter $C_e$ in Eq.~(\ref{eq:MHDfnc_At}) is left 
to be determined.  We fix this value from the condition that 
the asymptotic (net) electric charge $Q$ vanishes:
\beq
Q\,=\, \frac1{4\pi}\int_\infty \Fabu dS_{\albe}, 
\label{eq:charge}
\eeq
where $\int_\infty$ is the surface integral over a sphere $S_r$ 
with radius $r$, in the limit, 
$\int_\infty:={\dis \lim_{r\rightarrow \infty}}\int_{S_r}$.  
This integral is evaluated at a large radius of our computational region, 
typically $r\sim 10^4 R_0$, at every 30 iterations, then the secant method is 
applied to find a solution of $C_e$ to have $Q=0$.  One can also compute 
a charged solution with setting a finite value to $Q$.  

%
\begin{table}
\caption{Summary of grid parameters. 
}
\label{tab:grids_param}
\begin{tabular}{lll}
\hline
$r_a$  &:& Radial coordinate where the radial grids start.       \\
$r_b$ &:& Radial coordinate where the radial grids end.     \\
$r_c$ &:& Radial coordinate between $r_a$ and $r_b$ where   \\
&\phantom{:}& the radial grid spacing changes.   \\
$N_{r}$ &:& Number of intervals $\Dl r_i$ in $r \in[r_a,r_{b}]$. \\
$\Nrf$ &:& Number of intervals $\Dl r_i$ in $r \in[r_a,1]$. \\
$\Nrm$ &:& Number of intervals $\Dl r_i$ in $r \in[r_a,r_{c}]$. \\
$N_{\theta}$ &:& Number of intervals $\Dl \theta_j$ in $\theta\in[0,\pi]$. \\
$N_{\phi}$ &:& Number of intervals $\Dl \phi_k$ in $\phi\in[0,2\pi]$. \\
$L$ &:& Order of included multipoles. \\
\hline
\end{tabular}  
\end{table}
\begin{table}
\caption{Grid parameters used for computing magnetized 
rotating compact stars.  Resolution types SD12-SD3 are used for 
computing model P2, and SE12-SE3 are for P1 and P3.
Normalized radial coordinates $r_a$, $r_b$, and $r_c$ are in the 
unit of equatorial radius $R_0$ (in coordinate length).
}
\label{tab:RNS_grids}
\begin{tabular}{lcccrrrrrc}
\hline
Type  & $r_a$ & $r_b$ & $r_c$ & $\Nrf$ & $\Nrm$ & $N_r$ & $N_\theta$ & $N_\phi$ & $L$ \\
\hline
SD12 & 0.0 & $10^6$ & 1.1  & 60  & 66   & 144  & 72   & 48 &  30 \\
SD2  & 0.0 & $10^6$ & 1.1  & 80  & 88   & 192  & 96   & 48 &  30 \\
SD23 & 0.0 & $10^6$ & 1.1  & 120 & 132  & 288  & 144  & 48 &  30 \\
SD3  & 0.0 & $10^6$ & 1.1  & 160 & 176  & 384  & 192  & 48 &  30 \\
\hline
SE12 & 0.0 & $10^6$ & 1.1  & 60  & 66   & 144  &  96  & 48 &  40 \\
SE2  & 0.0 & $10^6$ & 1.1  & 80  & 88   & 192  & 128  & 48 &  40 \\
SE23 & 0.0 & $10^6$ & 1.1  & 120 & 132  & 288  & 192  & 48 &  40 \\
SE3  & 0.0 & $10^6$ & 1.1  & 160 & 176  & 384  & 256  & 48 &  20,30,40 \\
SE3p & 0.0 & $10^6$ & 1.1  & 160 & 176  & 384  & 256  & 60 &  50 \\
SE3t & 0.0 & $10^6$ & 1.1  & 160 & 176  & 384  & 384  & 60 &  50 \\
SE3tp & 0.0 & $10^6$ & 1.1  & 160 & 176  & 384  & 384  & 72 &  60 \\
\hline
\end{tabular}
\end{table}

\begin{table}
\caption{Quantities of a TOV solution in $G=c=M_\odot=1$ units 
for the polytropic EOS $p=K\rho^\Gamma$ with $\Gamma=2$ and $3$.  
The values of maximum mass models of the corresponding EOS 
parameters are listed, where $p_c$ and $\rho_c$ are the pressure 
and the rest mass density at the center, 
$M_0$ is the rest mass, $M$ the gravitational mass, and 
$M/R$ the compactness (a ratio of the gravitational mass to 
the circumferential radius).  
The adiabatic constant $K$ is chosen so that the value of $M_0$ 
becomes $M_0=1.5$ at the compactness $M/R=0.2$.  
To convert a unit of $\rhoc$ to cgs, multiply the values by 
$\Msol(G\Msol/c^2)^{-3} \approx 6.176393\times 10^{17} {\rm g}\ {\rm cm}^{-3}$.}  
\label{tab:TOV_solutions}
\begin{tabular}{ccccccc}
\hline
$\Gamma$ & $\prhoc$ & $\rhoc$ & $M_0$ & $M$ & $M/R$  & Models \\
\hline
$2$ & $0.318244$ & $0.00448412$ & $1.51524$ & $1.37931$ & $0.214440$ & P1, P2\\
$3$ & $0.827497$ & $0.00415972$ & $2.24295$ & $1.84989$ & $0.316115$ & P3 \\
\hline
\end{tabular}
\end{table}

\begin{table}
\caption{Parameters of functions in the integrability conditions 
(\ref{eq:MHDfnc_Lambda}) and (\ref{eq:MHDfnc_Lambda_phi}), 
and of EOS (\ref{eq:EOS}), used in computing solutions presented in 
Fig.~\ref{fig:P123sol}, and Tables \ref{tab:MRNS_solutions} and 
\ref{tab:MRNS_solutions_EMF}.} 
\label{tab:para_functions}
\begin{tabular}{ccccccc}
\hline
Models & $\Lambda_0$ & $\Lambda_1$ & $\Lambda_{\phi0}$ & $b$ & $c$ & $\Gamma$ \\
\hline
P1 & $-3.0$ & $0.3$ & $2.3$ & $0.2$ & $0.5$ & $2$  \\
P2 & $-1.7$ & $0.1$ & $1.7$ & $0.2$ & $0.5$ & $2$  \\
P3 & $-0.2$ & $0.3$ & $1.0$ & $0.2$ & $0.5$ & $3$  \\
\hline
\end{tabular}
\end{table}

\begin{figure*}
\begin{center}
\includegraphics[height=48mm]{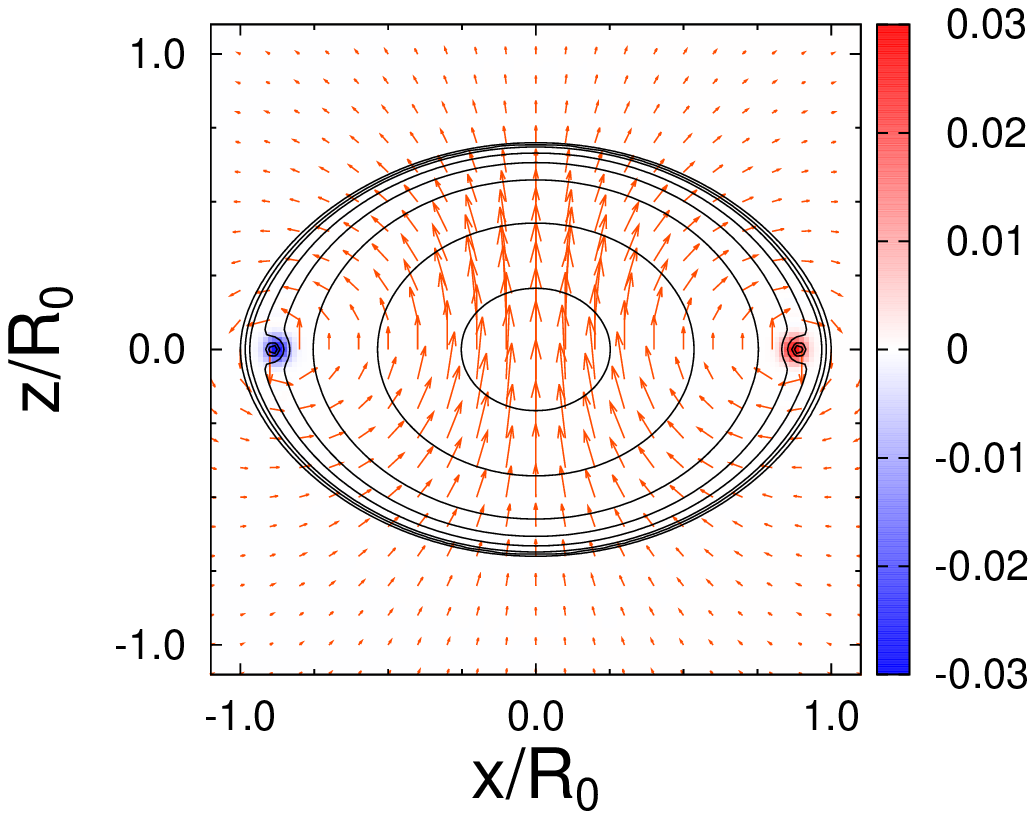}
\includegraphics[height=48mm]{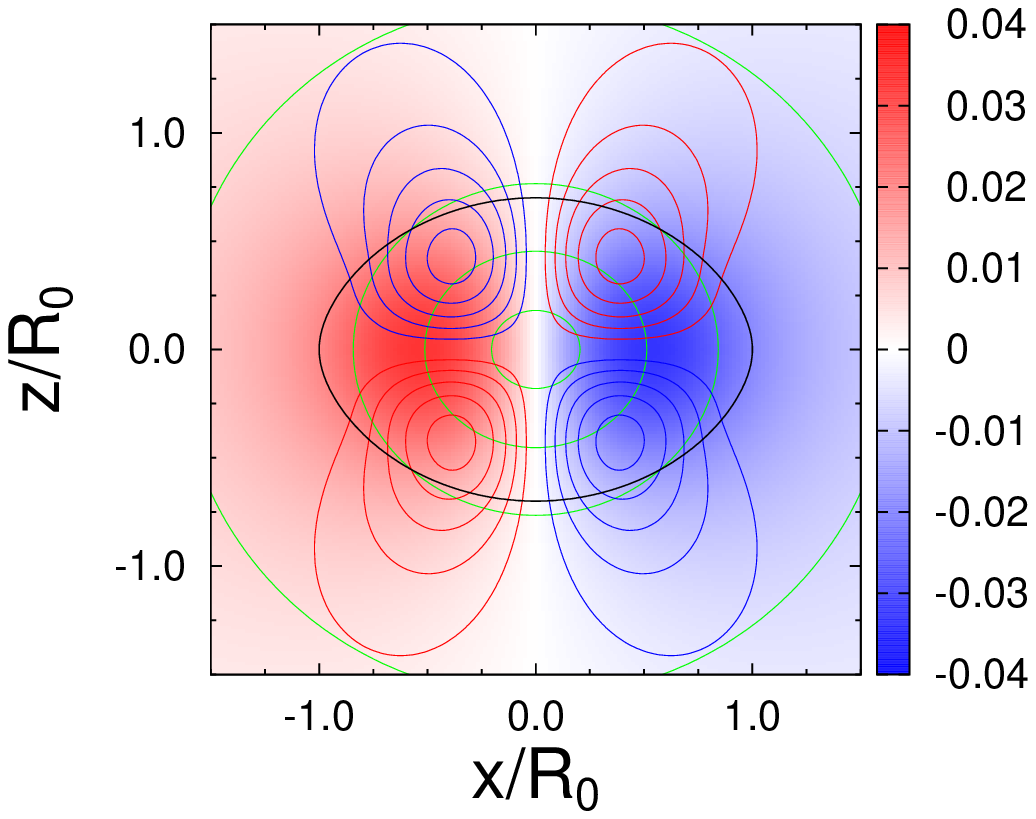}
\includegraphics[height=48mm]{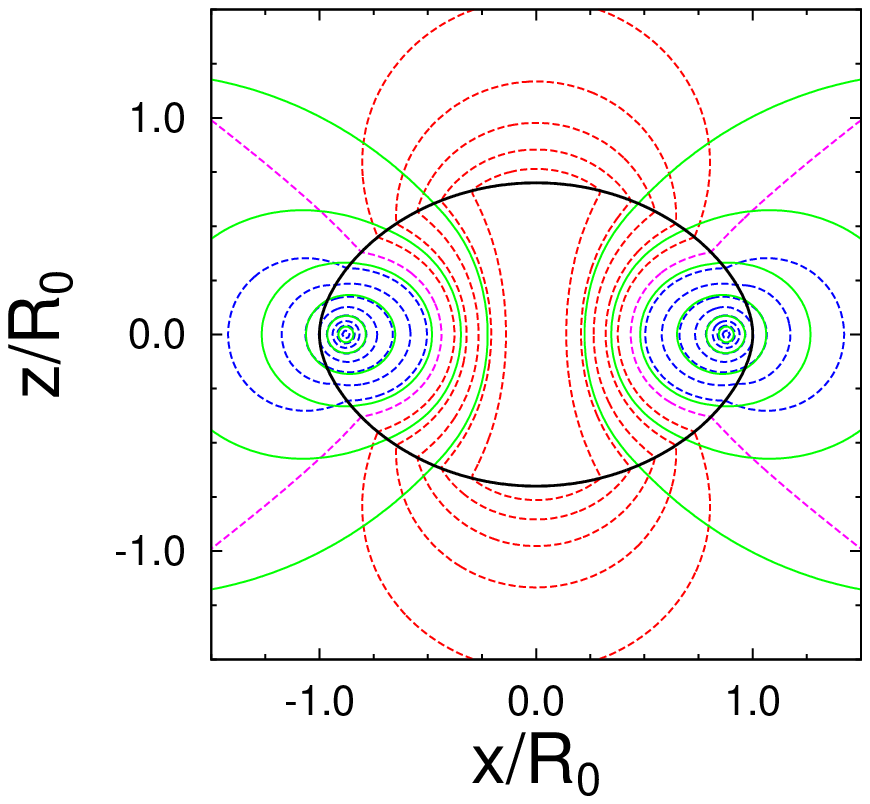}
\\
\includegraphics[height=48mm]{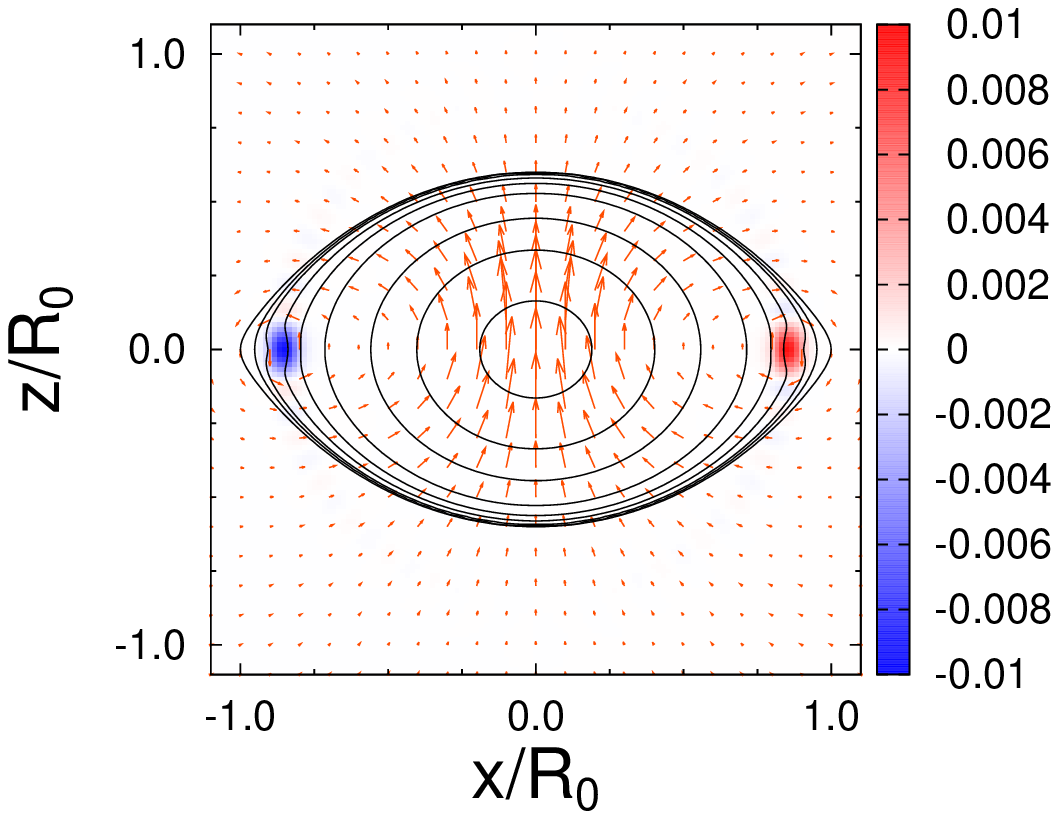}
\includegraphics[height=48mm]{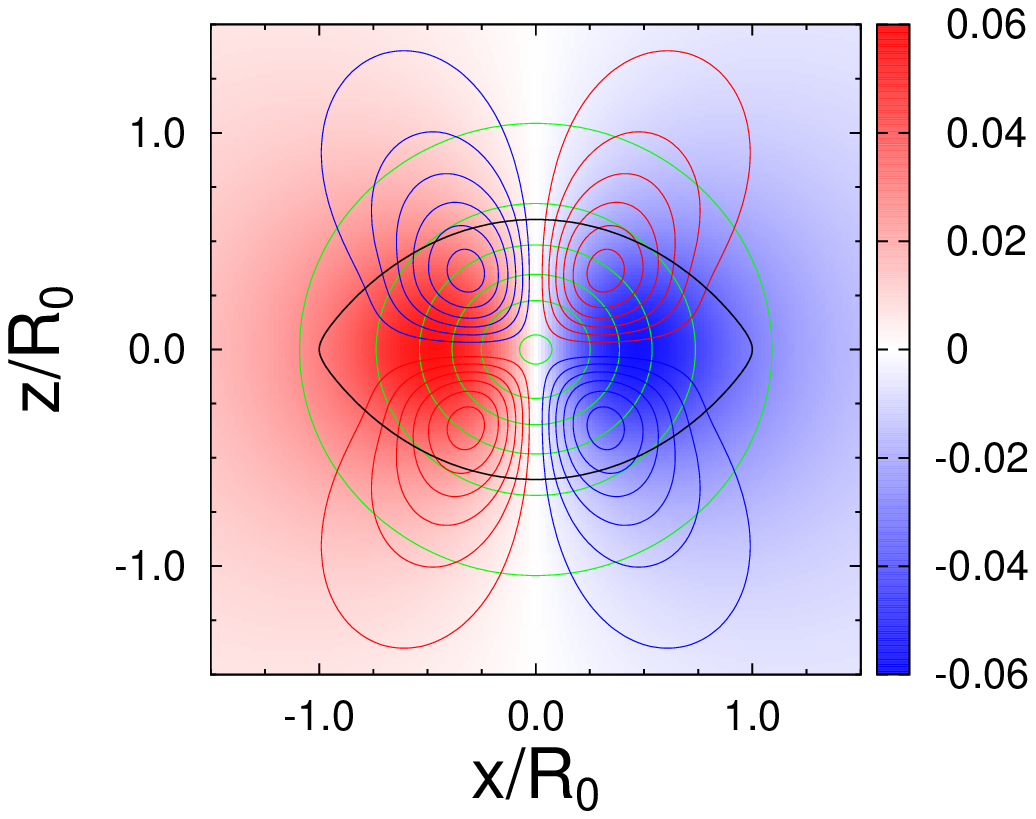}
\includegraphics[height=48mm]{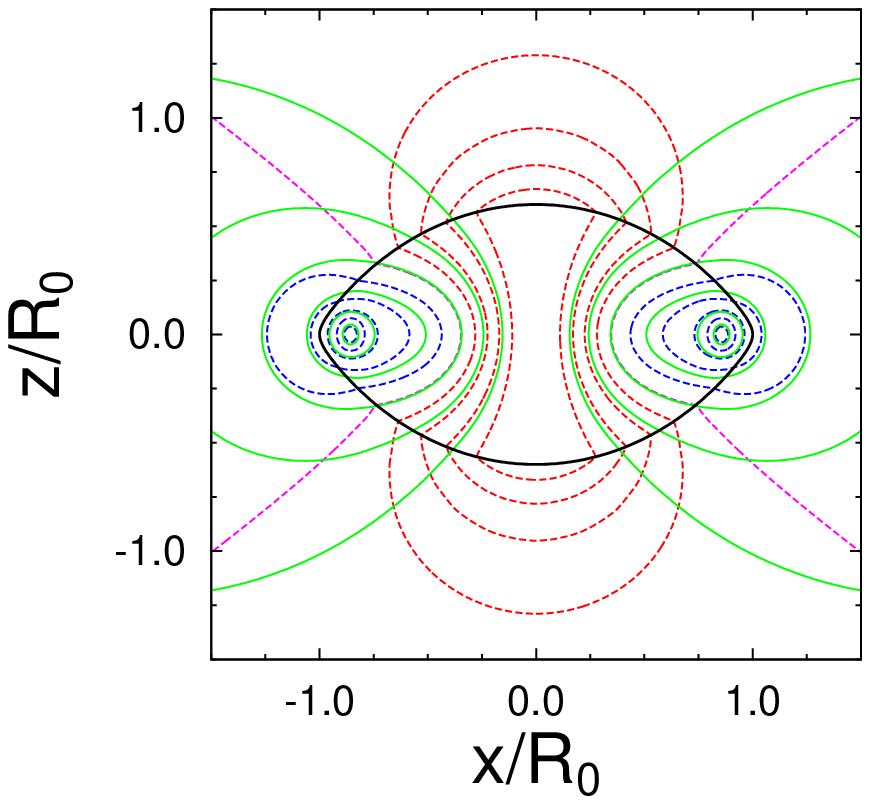}
\\
\includegraphics[height=48mm]{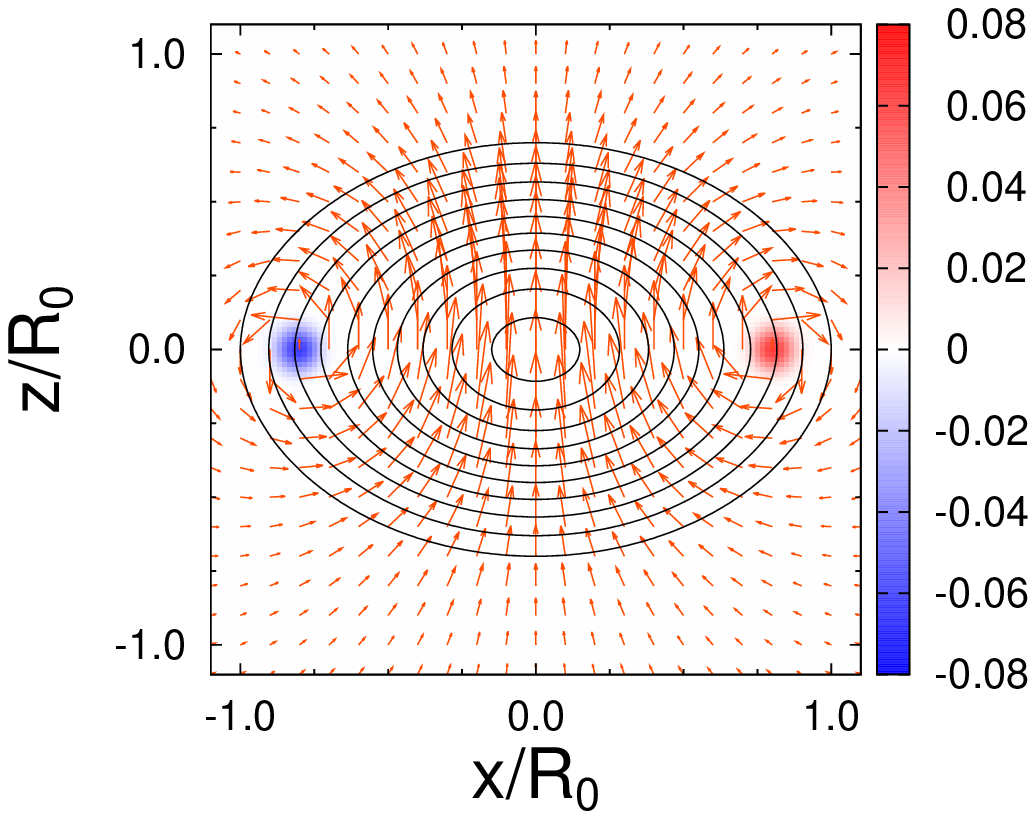}
\includegraphics[height=48mm]{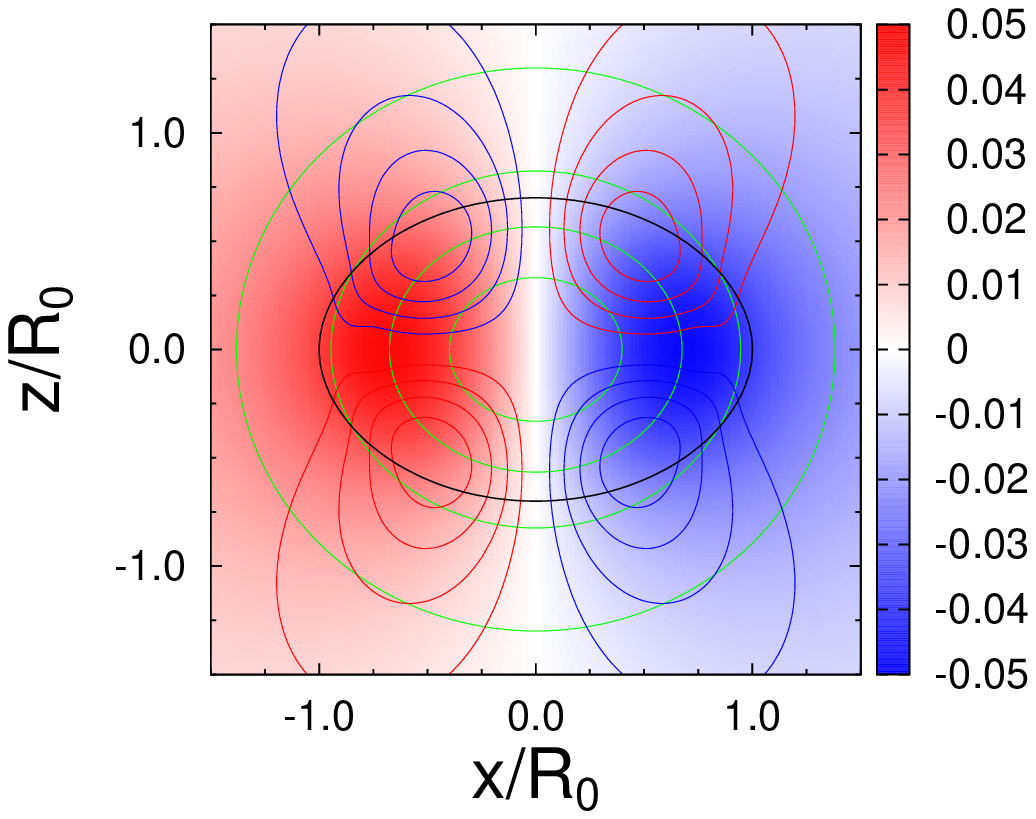}
\includegraphics[height=48mm]{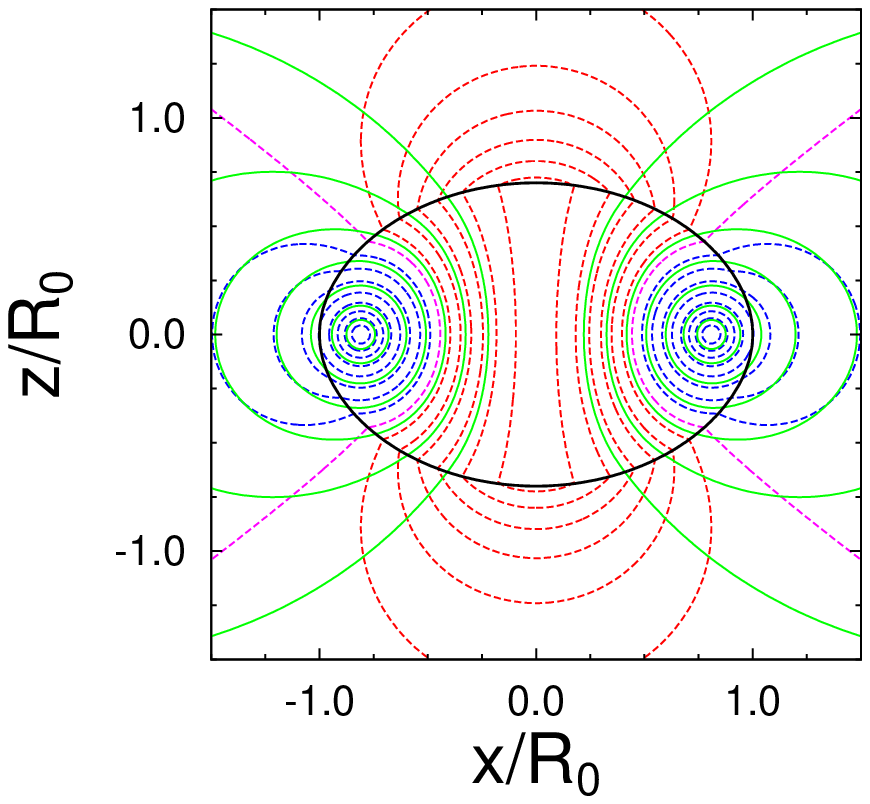}
\caption{Meridional sections of extremely magnetized solutions of rotating compact stars.  
Top, middle, and bottom rows correspond respectively to models 
P1 ($\Gamma=2$, normal mass), 
P2 ($\Gamma=2$, supramassive), 
and P3 ($\Gamma=3$, normal mass).  
Panels in 
the left column: the solid curves (in black) are contours of $p/\rho$, the arrows (in orange) 
correspond to the poloidal magnetic field and the color density maps (in red and blue) 
to the toroidal magnetic fields.  For the models P1 and P2, the contours of $p/\rho$ 
are drawn at $p/\rho=0.001,0.002,0.005,0.01,0.02,0.05,0.1$, and for P3, the contours 
are drawn linearly every $0.02$.  
Panels in the middle column: the metric potentials are 
shown.  Green curves correspond to equi-contours of $\psi$, 
the red and blue color density maps to $\tbeta_y$, and 
the red and blue curves to contours of $h_{xz}$.  
Panels in the right column: contours of $t$ and $\phi$ components of 
electromagnetic 1-forms $A_t$ and $A_\phi$.  Dashed red, purple, blue curves 
correspond to $A_t$, and solid green curves to $A_\phi$.  $A_t$ vanishes on 
the purple curves, and is positive (negative) on the red (blue) curves.  
A black curve in each panel in the middle and right columns 
represents the surface of the star.  The left and right panels 
correspond to the models P1 and P2, respectively.  
}  
\label{fig:P123sol}
\end{center}
\end{figure*}

\begin{figure}
\begin{center}
\includegraphics[height=55mm]{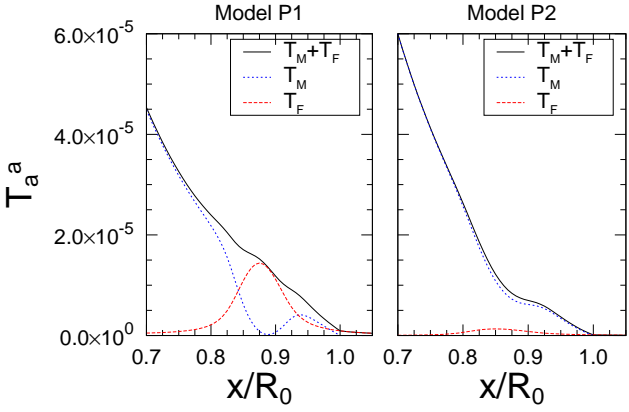}
\caption{
Plots for the spatial trace part of stress-energy tensor 
$T_a{}^a=\Tabu\gamma_{\albe}$ along 
the x-axis (the radial coordinate in the equatroial plane) near the surface.  
$T_M$ stands for the contribution from the matter $\TabMu$ (\ref{eq:TabMu}), and 
$T_F$ from the electromagnetic fields $\TabFu$ (\ref{eq:TabFu}).}  
\label{fig:SEM}
\end{center}
\end{figure}

\begin{table*}
\caption{Selected solutions for extremely magnetized rotating compact 
stars are presented.  Models P1 -- P3 are calculated using the corresponding 
parameters in Table \ref{tab:para_functions}.  
Listed quantities include 
the equatorial and polar radii in proper length $\bar{R}_0$ and $\bar{R}_z$, 
the ratio of the maximum values of the pressure to the rest mass density $\prhoc$, 
the angular velocity near the rotation axis $\Omega_{\rm c}$, 
the ADM mass $\Madm$, the rest mass $M_0$, 
the angular momentum $J$, 
the virial constant $I_{\rm vir}$, and a residual of the equality 
of the Komar mass $\MK$ and the ADM mass $\Madm$.  
The definitions of these quantities are found in Appendix.~\ref{sec:App:physq}
(see also, \cite{cocal}).  
To convert a unit of length from $G=c=\Msol=1$ to [km], multiply 
$G\Msol/c^2=1.477$[km].
}
\label{tab:MRNS_solutions}
\begin{tabular}{lccccccccccc}
\hline
Model &  
$\bar{R}_0$ &  $\bar{R}_z/\bar{R}_0$ &  $\prhoc$ & $\rhoc\ [\mbox{g/cm}^3]$ 
& $\Omegac$ & $\Madm$ & $M_0$ & 
$J/\Madm^2$ & 
$|1-\MK/\Madm|$ 
\\
\hline
\ \ P1 
\ \ ($\Gamma=2$, normal mass)
&
  11.0609 & 0.71996 & 0.12322 & $1.0717\times 10^{15}$ & 
  0.026384 & 1.35908 & 1.46223 & 
  0.52809 & 
$1.96\times 10^{-5}$
\\  
\ \ P2 
\ \ ($\Gamma=2$, supramassive)
&
  11.0787 & 0.64818 & 0.25582 & $2.2250\times 10^{15}$ & 
  0.043648 & 1.58645 & 1.74178 & 
  0.57113 & 
$4.06\times 10^{-5}$
\\
\ \ P3
\ \ ($\Gamma=3$, normal mass)
&
   8.8439 & 0.71839 & 0.18830 & $1.2248\times 10^{15}$ & 
   0.033124 & 1.59179 & 1.80318 & 
   0.51282 & 
$7.26\times 10^{-5}$
\\
\hline
\end{tabular}
\end{table*}

\begin{table*}
\caption{Continuing from Table \ref{tab:MRNS_solutions}, 
listed for the same solutions are the maximum values 
of poloidal and toroidal magnetic fields, $\Bpolmax$ and $\Btormax$, 
the ratios of poloidal and toroidal magnetic field energies, 
${\cal M}_{\rm pol}$ and ${\cal M}_{\rm tor}$, and electric field energy 
${\cal M}_{\rm ele}$ to the total electromagnetic field energy ${\cal M}$, 
the ratios of the kinetic, internal, and electromagnetic field energies 
to the gravitational energy, ${\cal T}/|{\cal W}|$, $\Pi/|{\cal W}|$, 
and ${\cal M}/|{\cal W}|$, respectively, and the virial constant 
$I_{\rm vir}$, and the electric charge contribution from the volume integral 
of the star $Q_M$.  
Details of the definitions are found in Appendix.~\ref{sec:App:physq}.  
}
\label{tab:MRNS_solutions_EMF}
\begin{tabular}{ccccccccccccc}
\hline
Model & 
$\Bpolmax$[G] & $\Btormax$[G] & 
${\cal M}_{\rm pol}/{\cal M}$ & ${\cal M}_{\rm tor}/{\cal M}$ &
${\cal M}_{\rm ele}/{\cal M}$ & ${\cal T}/|{\cal W}|$ & $\Pi/|{\cal W}|$ & 
${\cal M}/|{\cal W}|$ & $I_{\rm vir}/|{\cal W}|$ & $Q_M$ 
\\
\hline
P1 &
 $6.5382\times 10^{17}$ & $6.5133\times 10^{17}$ & 
 0.93381 & 0.043905 & 0.022284 & 
 0.063996 & 0.29637 & 0.019832 & 
 $3.2202\times 10^{-4}$ & 0.037041
 \\
P2 &
 $6.2207\times 10^{17}$ & $2.2065\times 10^{17}$ & 
 0.92609 & 0.033001 & 0.040905 & 
 0.083577 & 0.29918 & 0.001624 & 
 $2.4514\times 10^{-4}$ & 0.024655
 \\
P3 &
 $1.7797\times 10^{18}$ & $1.4487\times 10^{18}$ & 
 0.93967 & 0.040486 & 0.019840 & 
 0.068800 & 0.29176 & 0.043991 & 
 $1.3324\times 10^{-4}$ & 0.068080
\\
\hline
\end{tabular}
\end{table*}


%
%

\section{Results}
\label{sec:Results}

In \cite{Uryu:2014tda}, we have presented preliminary 
results for relativistic rotating star solutions associated with 
mixed poloidal and toroidal magnetic fields.  As mentioned 
in Sec.~\ref{sec:fnc}, we have modified the form of arbitrary 
functions from those used in \cite{Uryu:2014tda}.  We have also 
improved numerical codes to maintain expected accuracy; 
for example, the virial relation is satisfied in higher precision.  
The numerical computations presented below are performed using smaller to 
larger numbers of grid points and multipoles as shown in Tables 
\ref{tab:grids_param} and \ref{tab:RNS_grids} for studying 
the convergence of the solutions.  

In the following computations, we choose the adiabatic EOS (\ref{eq:EOS}) 
with indices $\Gamma=2$ or $3$ and the adiabatic constant $K$ 
so that the compactness of a spherical solution having 
rest mass $M_0=1.5\Msol$ becomes $M/R=0.2$.  
Reference quantities for the Tolman-Oppenheimer-Volkov (TOV) solutions 
for these EOSs are tabulated in Table \ref{tab:TOV_solutions}.  
For the model parameters to determine the strength of electromagnetic 
fields, we choose three sets listed in Table \ref{tab:para_functions}.  
To our knowledge, since our first paper \cite{Uryu:2014tda} was published, 
\cocal\ is the only code that can calculate fully relativistic 
rotating compact stars associated with mixed poloidal and toroidal 
magnetic fields without any approximation in the formulation other than 
assumptions of stationarity and axisymmetry.

\subsection{Extremely magnetized solutions}

We present three solutions of magnetized rotating 
compact stars in Fig.~\ref{fig:P123sol}, 
and corresponding physical quantities in Tables 
\ref{tab:MRNS_solutions} and \ref{tab:MRNS_solutions_EMF}.  
Definitions of these quantities are summarized in 
Appendix \ref{sec:App:physq}. 
The model parameter of each solution is P1, P2, and P3, 
respectively in Table \ref{tab:para_functions}, where 
the model P1 is a normal mass solution with $\Gamma=2$ EOS, 
P2 is a supramassive solution with $\Gamma=2$ and is rotaing 
near the Kepler limit, and P3 is a normal mass solution with 
$\Gamma=3$.  

As shown in Table \ref{tab:MRNS_solutions_EMF}, these solutions 
are associated with extremely strong poloidal and toroidal magnetic 
fields about an order of $\sim 10^{17}$--$10^{18}$[G], while 
the mass and radius of these compact stars are close to those of 
common neutron stars.  
For the models P1 and P3, the maximum values of the toroidal and poloidal 
components, $\Bpolmax$ and $\Btormax$ respectively, are comparable, 
and even for P2, $\Btormax$ is about $1/3$ of $\Bpolmax$.  
As reported also in other works, however, the bulk energy of 
the toroidal magnetic fields ${\cal M}_{\rm tor}$ is much smaller than 
that of the poloidal fields ${\cal M}_{\rm pol}$; as shown in Table 
\ref{tab:MRNS_solutions_EMF}, the energy of the poloidal fields 
accounts for more than 90\% of the total electromagnetic energy $\cal M$.  

In the top to bottom left panels of Fig.~\ref{fig:P123sol}, 
the contours of $p/\rho$ and the poloidal and toroidal magnetic fields 
are presented.  Although the toroidal magnetic field component 
$\Btor$ is not dominating in the whole electromagnetic energy, 
$\Btor$ is concentrated near the equatorial surface so that its 
maximum value is comparable to that of poloidal component $\Bpol$.  
This feature has been often observed in the other Newtonian 
\cite{MRNSNewtonian} or approximate calculations \cite{PolTorMS}.  

A new feature can be seen in these panels for models P1 and P2.  
When the toroidal field $\Btor$ is extremely strong, 
the magnetic energy density locally dominates over the mass energy 
density, and hence expel the matter from the region of extremely 
strong toroidal magnetic fields.  In the middle left panel for 
the model P2, we can observe that the $p/\rho$ contours are deformed 
around the $\Btormax$, and in the top left panel for the model P1, 
there are small closed circles of the density contours near 
the equatorial surface.  For the model P1, a profile of $p/\rho$ 
along the equatorial radius near the surface (and hence $\rho$ or 
$\varepsilon$) almost drops to zero.  Hence, we expect that, 
with a little stronger magnetic fields, which can be easily 
achieved by changing the parameters in Table \ref{tab:para_functions}, 
the matter will be completely expelled from this region, and hence 
a toroidal electro-vacuum tunnel will be formed inside of 
the compact star.  

Roughly speaking, this happens because the pressure/energy density 
of the electromagnetic fields dominates over those of the matter in 
this toroidal region near the surface.  To see this, in Fig.~\ref{fig:SEM}, 
we show the plots of spatial trace part of stress-energy tensor 
$T_a{}^a = \Tabu\gamma_{\albe}=(\TabMu+\TabFu)\gamma_{\albe}$ separating 
contributions from the matter $\TabMu$ (\ref{eq:TabMu}) 
($T_a{}^a = \TabMu\gamma_{\albe}$) and the electromagnetic fields 
$\TabFu$ (\ref{eq:TabFu}) ($T_a{}^a = \TabFu\gamma_{\albe}$).  
As can be seen in the left panel for the model P1, the dominance of 
the matter to the electromagnetic fields exchange in this region.  
In the right panel for model P2, there is a sizable amount of contribution 
from $\TabFu$ but it does not dominate over $\TabMu$.  

In the middle panel of each row of Fig.~\ref{fig:P123sol}, contours 
of metric potentials around the compact stars are plotted.  Using 
the waveless formulation, we are able to compute non-conformal flat 
components of the metric such as $h_{xz}$ as shown in these panels.

In the right panel of each row of Fig.~\ref{fig:P123sol}, contours 
of $t$ and $\phi$ components of the electromagnetic 1-form 
$A_\alpha$ are shown.  As mentioned in Sec.~\ref{sec:fnc}, 
integrability of ideal MHD equations require $A_t$ to be a 
function of a master potential for which can be seen correctly 
imposed on the fluid support of compact stars.  Since we 
assume electro-vacuum spacetime outside of the star, the $A_t$ 
component is continuously but not smoothly connected at the 
stellar surface.  Since we also assume that the net charge 
at infinity $Q$ (evaluated at a larger radius from the source in 
actual computations) vanishes, the contours of $A_t$ become positive 
(red dashed curves) near the poles, and negative (blue dashed curves) 
near the equator.  These panels show that the method of solving for 
$A_t$ as described in Sec.~\ref{sec:PDEsolver} is working consistently 
in these computations.  

\begin{figure}
\begin{center}
\includegraphics[height=60mm]{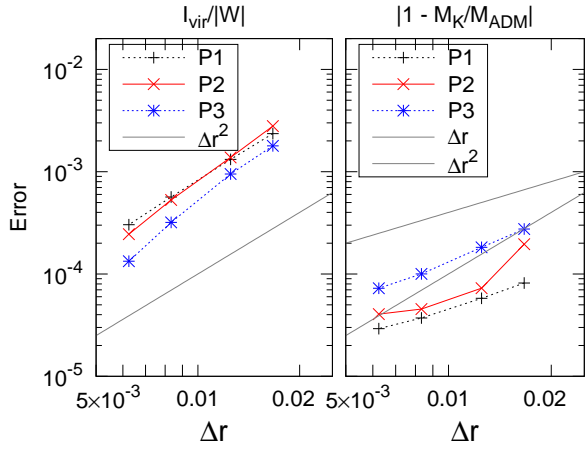}
\caption{Left: Convergence of the virial identity 
is plotted for the models P1, P2, and P3.  
Left panel: a convergence of $I_{\rm vir}/|{\cal W}|$.  
Right panel: a convergence of $\left| 1-\MK/\Madm \right|$.
}  
\label{fig:virial}
\end{center}
\end{figure}
\begin{figure}
\begin{center}
\includegraphics[height=55mm]{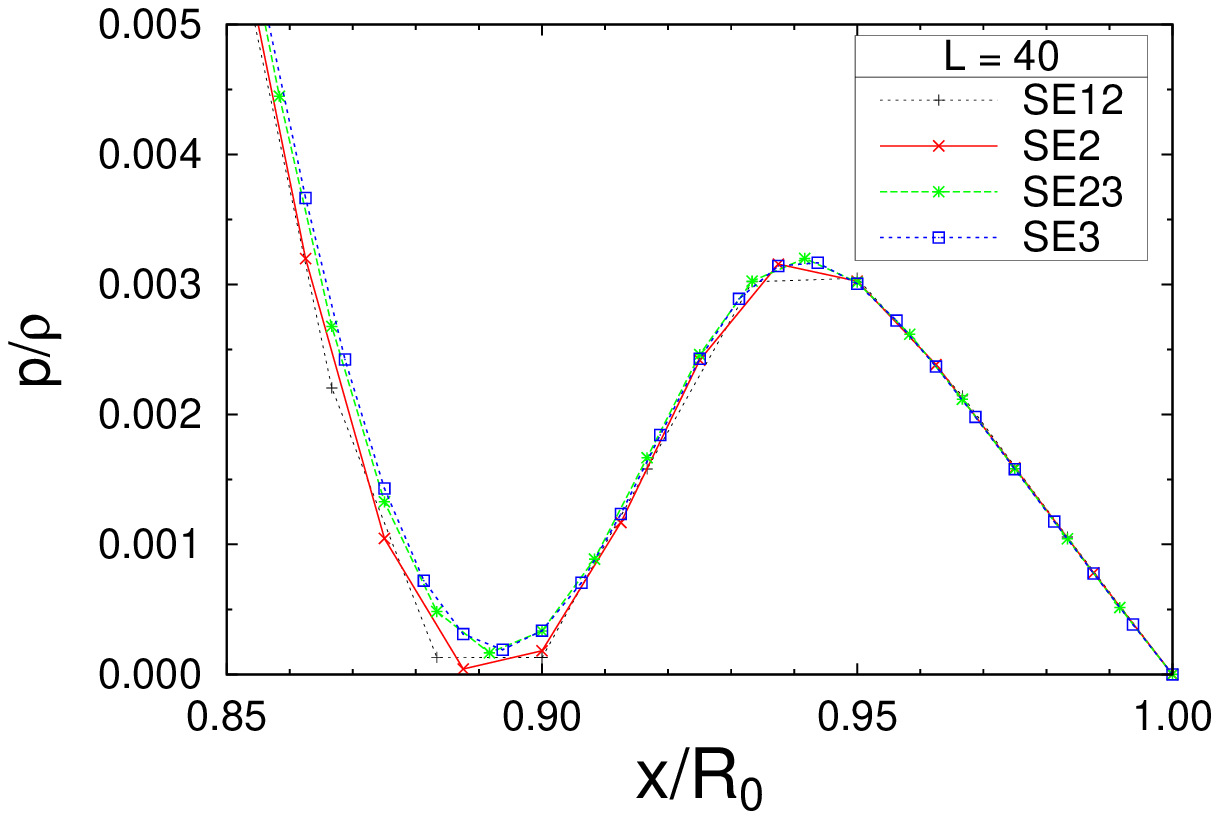}
\\
\includegraphics[height=55mm]{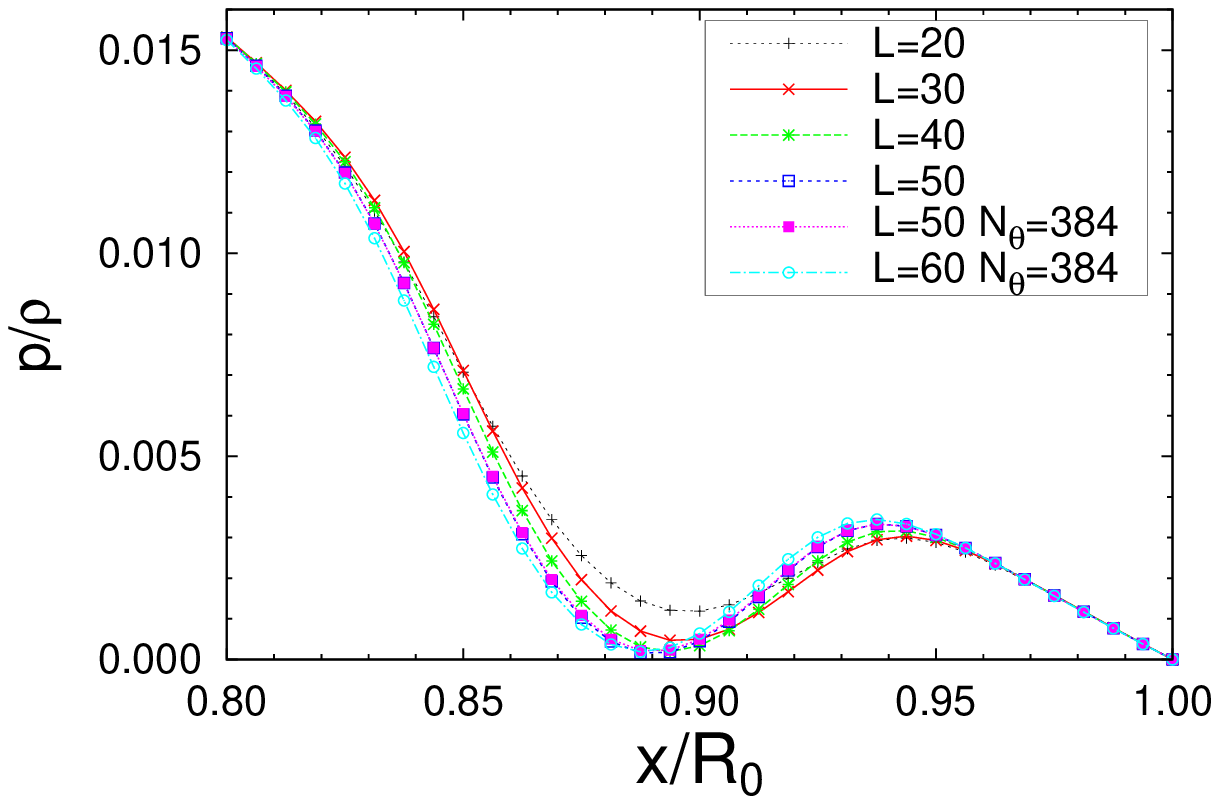}
\caption{Close up of the density profile $\rho$ 
of Model P1 along 
the x-axis (the radial coordinate in the equatorial plane)
near the surface.  Top: Convergence 
of $\rho$ with respect to increasing resolutions from SE12 to SE3.  
Bottom: Convergence with respect to increasing multipoles from 
$\ell=20$ to $60$.   
}  
\label{fig:rho}
\end{center}
\end{figure}

\subsection{Convergence test}

In Fig.~\ref{fig:virial}, the convergence of integrated quantities 
are plotted for the models P1, P2, and P3. Those are 
the convergence of $I_{\rm vir}/|{\cal W}|$ in the left, and 
that of $|1-\MK/\Madm|$ in the right panels.  The relativistic 
virial relation $I_{\rm vir}$ is defined as the volume integral of 
the spatial trace of Einstein's equations as (\ref{eq:App:traceG}) 
in the Appendix, and its residuals shown in the left panel 
decrease as $\Od(\Delta r^2)$ as expected.  
Strictly speaking, the numerically evaluated volume integral $I_{\rm vir}$ 
does not approach to zero as the resolution goes much higher,  
because it is evaluated on a large but finite computational 
domain $r\in[0,10^6 R_0]$, and also evaluated from a solution 
in which a large but a finite number of multipoles are used for approximation.  
Hence what we can conclude here is the fact that the actual value of 
$I_{\rm vir}/|{\cal W}|$ in our setup is smaller than the finite difference 
error and hence can not be probed with the present highest resolutions 
such as SD3 or SE3.  

The asymptotic Komar 
and ADM mass, $\MK$ and $\Madm$ are known to agree in the framework 
of the waveless formulation under the gauge choice Eq.~(\ref{eq:gauge}) 
\cite{Shibata:2004qz}.  
The residual $|1-\MK/\Madm|$ of each model decrease slower 
than $\Od(\Delta r^2)$ as shown in the right panel, which is the 
same behavior as that of non-magnetized rotating compact stars \cite{Uryu:2016dqr}.  
Hence we may conclude that the strongly magnetized solutions are calculated 
with comparable precision as the non-magnetized solutions in the \cocal\ code.  

In Fig.~\ref{fig:rho}, we show the profile of $p/\rho$ along 
the x-axis (the radial coordinate in the equatorial plane)
near the surface for the model P1.  As mentioned in the above, extremely 
strong toroidal magnetic fields expel the matter from the toroidal region.  
To resolve such a relatively small scale toroidal structure, it is necessary to 
increase the numbers of grid points and multipoles.  We performed convergence 
tests to examine the profile of $p/\rho$ with systematically increasing 
the resolution under a fixed number of multipoles, and also increasing 
the number of multipoles under a fixed resolution.  In the top panel of 
Fig.~\ref{fig:rho}, the number of multipoles is fixed to $L=40$, and 
the resolution is increased from SE12 to SE3 in Table \ref{tab:RNS_grids}.  
It can be observed that the largest error appears at the bottom of the $p/\rho$ 
profile around $x=0.89 R_0$, and that the profile converges at the levels of 
resolutions around SE23 to SE3.

In the bottom panel of Fig.~\ref{fig:rho}, plotted are a convergence of 
$p/\rho$ profiles near the equatorial surface of the model P1 with 
respect to the number of multipoles used in the elliptic solver (\ref{eq:PDEsolver}).  
In this test, resolution SE3 and modified resolutions of it (SE3p, SE3t, 
SE3tp in Table \ref{tab:RNS_grids}) are used.  It is confirmed that 
those modified resolutions do not affect the profile.  For example, 
we compute the case with $L=50$ with SE3p and SE3t with increasing 
$N_\theta$, the number of grid points in $\theta$ coordinate, but those 
profiles overlap as seen in the plot.  The profiles are also the same 
when the number of $N_\phi$ is increased, which is not related to 
an accuracy of a solution but is necessary for computating solutions 
with larger $L$.  As seen in the panel, the profiles 
gradually change as increasing $L$ from 20 to 60.  Because of the limit 
of computational resources, we do not perform computations higher than 
$L=60$ with a resolution SE3tp.  The solutions still slightly changes 
from $L=50$ with SE3t resolution to the $L=60$ with SE3tp resolution, 
but the overall difference of the profile is getting smaller with 
increasing $L$.

\section{Discussion}
\label{sec:Discussion}

In this paper, we have presented the full details of a formulation 
and a numerical method for computing stationary and axisymmetric 
equilibriums of fully relativistic rotating compact stars associated 
with mixed poloidal and toroidal magnetic fields.  
We have successfully calculated solutions associated with extremely 
strong poloidal and toroidal magnetic fields, and found solutions 
whose mass energy density is expelled by the energy density of toroidal 
magnetic fields.  As presented in Fig.~\ref{fig:rho}, the structure of 
this toroidal low density region can be calculated accurately using a large 
number of multipoles in our Poisson solver (Sec.~\ref{sec:PDEsolver}).  
From the top left panel of Fig.~\ref{fig:P123sol}, the size of the toroidal 
region in the $\theta$ direction is around $\sim 0.03$ radian and hence 
requires a resolution $N_\theta > \pi/0.03 \sim 100$.  The number 
$N_\theta=384$ of the resolution SE3tp is sufficient to resolve this structure.  
On the other hand, a Legendre polynomial $P_{60}^0(\cos \theta)$ has only 
60 nodes, which may resolve roughly $\pi/60\sim 0.05$ radians in the $\theta$ 
direction, and hence this is a reason for slow convergence in 
the number of multipoles $L$.  

One of the new features of our method is to solve all components of 
Maxwell's equations to determine all components of the electromagnetic 
potential 1-form $A_\alpha$.  This allows us to compute electromagnetic 
configurations under various circumstances.  In this paper, 
we assume an electro-vacuum spacetime outside of the compact star, 
which results in a surface charge distribution when we compute 
an asymptotically charge neutral solution.  This is the same 
assumption used in the first relativistic computation of magnetized 
rotating equilibrium by the Meudon group \cite{Bocquet:1995je}.  
It would be also possible to compute a solution which 
continues to force free magnetosphere outside of the star.  
The \cocal\ code is also available for computing black holes.  
Hence an extension to a black hole associated with 
electromagnetic fields, that is, a black hole magnetosphere, 
is also a part of our future project.

\acknowledgments
This work was supported by 
JSPS Grant-in-Aid for Scientific Research(C) 18K03624, 15K05085, 
18K03606, 17K05447, NSF Grant PHY-1662211, NASA Grant 80NSSC17K0070, 
and the Marie Sklodowska-Curie grant agreement No.753115.
%

\appendix

\section{Notation and relations}

\subsection{Orthogonal curvilinear coordinates and basis}
\label{secApp:basis}

When a system of coordinate functions $x^A$ is introduced for a set of 
points $x^\alpha$ in a space, the basis and dual basis are respectively 
$\{e_A^\alpha \}$ and $\{\na_\alpha x^A \}$, where 
$e_A^\alpha:=\frac{\pa x^\alpha}{\pa x^A}$.  The coordinates are called 
orthogonal when 
\beq
e_A^\alpha \na_\alpha x^B \,=\,\Lie_{e_A}x^B=\dl_A{}^B
\eeq 
for any pair of indices $A$ and $B$.  Derivatives of this expression give
\beq
\na_\alpha(e_A^\beta \na_\beta x^B)
\,=\,\na_\alpha\Lie_{e_A}x^B
\,=\,\Lie_{e_A}\na_\alpha x^B
\,=\,0, 
\eeq
since $\na_\alpha\dl_A{}^B=0$.  Projecting these to the basis $e_B^\alpha$ 
we have 
\beqn
e_B^\alpha\,\Lie_{e_A}\na_\alpha x^C
&=&\Lie_{e_A}(e_B^\alpha \na_\alpha x^C)-\na_\alpha x^C \,\Lie_{e_A}e_B^\alpha 
\nonumber\\
&=&-\na_\alpha x^C \,\Lie_{e_A}e_B^\alpha 
\,=\,0, 
\eeqn
for any dual basis $\na_\alpha x^C$.  Therefore, we have 
\beq
\Lie_{e_A}e_B^\alpha \,=\, [e_A^\alpha,e_B^\beta]\,=\,0 .
\eeq

\subsection{Index free notation for differential forms and vectors}
\label{secApp:forms}

In some manipulations of equations in Sec.~\ref{sec:Formulation} 
and \ref{sec:Numerical}, we use index free notations for differential 
forms and vectors for convenience.  We summarize correspondences between 
index and index free notations in this subsection.  For more details 
see e.g. \cite{Gourgoulhon:2011gz,Gourg2012} 

\subsubsection{Differential forms}

For a 1-form $w$, we write an exterior derivative $dw$ which corresponds to 
the index notation as 
\beq
dw = (dw)_\albe = \na_\alpha w_\beta - \na_\beta w_\alpha.
\eeq
This notation is often used for an electromagnetic potential 1-form $A$, 
and for a canonical momentum 1-form $\hu$, where $\underbar{u}$ is a 
dual 1-form of the 4-velocity vector $u$ (which is $u^\alpha$ in the 
abstract index notation).  Faraday 2-form $F=dA$ is written, 
\beq
F_\albe = \na_\alpha A_\beta - \na_\beta A_\alpha.  
\eeq
$F$ is a closed 2-form, $dF = 0$, which is written in index notation, 
\beq
dF = 3\,\na_{[\alpha} F_{\beta\gamma]} 
= \na_\alpha F_{\beta\gamma}
+ \na_\beta F_{\gamma\alpha} 
+ \na_\gamma  F_{\alpha\beta}=0.
\eeq

\subsubsection{Inner product and Cartan identity}

Inner product of a $p$-form $\omega$ and a vector $u$ is denoted with a dot 
in index free notation, 
\beq
u\cdot \omega \,=\, u^\gamma \omega_{\gamma\alpha\dots\beta}.  
\eeq
Using this, the Cartan identity for a $p$-form $\omega$ is written 
\beq
\Lie_u \omega \,=\, u\cdot d\omega \,+\, d(u\cdot \omega), 
\eeq
in index free notation.  As a rule for the inner product 
between a vector and a $p$-form, we assume that when the vector is 
operated to $p$-form from left (right), the vector is contracted with 
the left (right) most index of the $p$-form, 
for example, $u\cdot F = - F\cdot u$ for a 2-form $F$.

\subsubsection{Ideal MHD condition}
\label{secApp:idealMHD}

Ideal MHD condition is written $F\cdot u = 0$.  This implies $\Lie_u F = 0$.  
This is shown using $dF =0$ as 
\beqn
\Lie_u F_\albe
&=& u^\gamma \na_\gamma F_\albe
+ F_{\gamma\beta} \na_\alpha u^\gamma 
+ F_{\alpha\gamma} \na_\beta u^\gamma 
\nonumber\\
&=& u^\gamma(\na_\gamma F_\albe
+ \na_\alpha F_{\beta\gamma} 
+ \na_\beta  F_{\gamma\alpha} )
\,=\,0.
\eeqn
Or, using a potential 1-form $A$, ($F = dA$)
\beq
\Lie_u F = \Lie_u dA = d\Lie_u A = d(u\cdot dA + d(A\cdot u)) =0, 
\eeq
where $u\cdot dA = u\cdot F = 0$, $d^2=0$ and Cartan identity are used.

Also for any vector proportional to $u$, that is, with an arbitrary 
scalar function $\lambda$, $F\cdot (\lambda u) = 0$ holds , which 
implies $ \Lie_{\lambda u} F = 0 $.  This guarantees that a flux of 
$F$ over any surface along a given family of flow lines is 
conserved \cite{Carter79}.

\subsubsection{Integrability condition}

When smooth functions $A$ and $f$ satisfy $fdA=C={\rm const}$, 
a relation is derived:
\beq
d(fdA) \,=\, df \wedge dA + f d^2A \,=\, df \wedge dA \,=\, 0.
\eeq
Hence $f = f(A)$.  As $f(A) dA = dF(A) = C={\rm const}$, 
$dF(A)/dA = f $.  If the constant $C=0$, then $f=dF/dA = 0$.

\subsection{4-velocity}

We decompose the 4-velocity $u^\alpha$ with respect to 
$t^\alpha$ as 
\beq
u^\alpha = u^t(t^\alpha + v^\alpha)
\eeq
where $v^\alpha$ is spatial vector $v^\alpha \nabla_\alpha t = 0$.  
In the first integrals and in the currents, $t$ and $\phi$ components 
$u_\alpha$ appears, which are calculated as follows; 
\beqn
u_t &=& u_\alpha t^\alpha \,=\, 
u^t (\alpha n_\alpha + \beta_\alpha + v_\alpha)(\alpha n^\alpha + \beta^\alpha)
\nonumber \\
&=& u^t \left[-\alpha^2 + \beta_a(\beta^a + v^a)\right] 
\nonumber \\
&=& u^t \left[-\alpha^2 + \psi^4\tbeta_a(\tbeta^a + \tilde v^a) \right], 
\eeqn
\beqn
u_\phi &=& u_\alpha \phi^\alpha \,=\, 
u^t (\alpha n_\alpha + \beta_\alpha + V_\alpha)\phi^\alpha
\nonumber \\
&=& u^t (\beta_a + V_a)\phi^a 
\,=\, u^t \psi^4(\tbeta_a + \tilde V_a)\tphi^a, 
\eeqn
where $t^\alpha = \alpha n^\alpha + \beta^\alpha$ is used.  

The coordinate basis for vector $\phi^\alpha$ is related to 
the Cartesian basis $\hat{x}^\alpha$ and $\hat{y}^\alpha$ as 
\beq
\phi^\alpha \,=\, \tphi^\alpha 
\,=\, -y \hat{x}^\alpha +  x \hat{y}^\alpha
\eeq
and the basis for 1-form 
\beqn
\nabla_\alpha \phi 
&=& 
-\frac{y}{x^2+y^2}\na_\alpha x + \frac{x}{x^2+y^2}\na_\alpha y 
\nonumber \\
&=& 
-\frac{\sin\phi}{r\sin\theta}\na_\alpha x + \frac{\cos\phi}{r\sin\theta}\na_\alpha y .
\eeqn
Hence, the $\phi$-components of the 4-velocity are related to those of Cartesian 
coordinates as 
\beqn
&& u_\phi \,=\, u_\alpha \phi^\alpha \,=\, -y u_x + x u_y,
\\
&& u^\phi \,=\, u^\alpha \nabla_\alpha\phi 
\,=\, -\frac{y}{x^2+y^2} u^x + \frac{x}{x^2+y^2} u^y,
\eeqn
and these becomes on the $\phi = 0$ (meridional) plane, 
\beqn
&& u_\phi|_{\phi=0} \,=\, x u_y,
\\
&& u^\phi|_{\phi=0} \,=\, \frac1{x} u^y.  
\eeqn
The same relations between $\phi$-components 
and the Cartesian components are used for 
the electric currents $j^\alpha$ ($j^\phi$ and $j_\phi$).

\section{3+1 decomposition of Faraday tensor}
\label{sec:App:Fab}

In this Appendix, we derive the 3+1 form of Faraday tensor and its divergence.  
The spatial projection of $\Fabd = (dA)_\albe$ can be derived explicitly as follows; 
For $\bar{F}_\alpha$, we use the Cartan identity 
$n\cdot dA = \Lie_n A - d(n\cdot A)$, 
\beqn
\bar{F}_\alpha 
&=& \gamma_\alpha{}^\beta F_{\beta\gamma} n^\gamma
\,=\, -\,\gamma_\alpha{}^\beta [\Lie_n A_\beta\,-\,\na_\beta(n_\gamma A^\gamma)]
\nonumber\\
&=& -\,\gamma_\alpha{}^\beta\Lie_n(\Phi_\Sigma\, n_\beta + \bar{A}_\beta)
\,-\, D_\alpha\Phi_\Sigma
\nonumber\\
&=& -\,\gamma_\alpha{}^\beta\Lie_n \bar{A}_\beta
\,-\, \frac1{\alpha}D_\alpha(\alpha\Phi_\Sigma), 
\eeqn
where the relation, $\Lie_n n_\alpha = D_\alpha \ln\alpha$, is used.  
For $\bar{F}_\albe$, 
since $\Fabd = (dA)_\albe$ is independent of the geometry of the manifold, 
its spatial projection becomes its spatial part, 
\beq
\bar{F}_\albe \,=\, D_\alpha  \bar{A}_\beta \,-\,D_\beta  \bar{A}_\alpha,
\eeq
which can be shown more explicitly as 
\beqn
\bar{F}_\albe &=& \gamma_\alpha{}^\gamma\gamma_\beta{}^\delta \Fcdd, 
\nonumber\\
&=& \gamma_\alpha{}^\gamma\gamma_\beta{}^\delta 
[(\na_\gamma\bar{A}_\delta
\,-\,\na_\delta\bar{A}_\gamma)
\,+\,\Phi_\Sigma\,(\na_\gamma n_\delta 
\,-\, \na_\delta n_\gamma)]
\nonumber\\
&=& 
D_\alpha\bar{A}_\beta
\,-\,D_\beta\bar{A}_\alpha
\,-\,\Phi_\Sigma\, (K_{\alpha\beta} \,-\, K_{\beta\alpha}), 
\nonumber\\
&=& 
D_\alpha\bar{A}_\beta
\,-\,D_\beta\bar{A}_\alpha, 
\eeqn
as $K_\albe$ is a symmetric tensor.
The divergence $\na_\beta \Fabu$ is also decomposed 
with respect to $\Sigma_t$.  
The projection of $\na_\beta \Fabu$ to the hypersurface normal 
$n_\alpha$ becomes 
\beqn
n_\alpha \na_\beta \Fabu 
&=& 
\na_\beta (\Fabu n_\alpha) - \Fabu \na_\beta n_\alpha
\nonumber\\
&=& - \na_\alpha \bar{F}^\alpha
\,+\, \Fabu(K_{\albe} + n_\beta D_\alpha \ln\alpha)
\nonumber\\
&=& - \frac1{\alpha} D_\alpha (\alpha \bar{F}^\alpha)
\,+\, \bar{F}^\alpha \frac1{\alpha} D_\alpha \alpha
\nonumber\\
&=& -D_\alpha \bar{F}^\alpha.
\eeqn
The projection of $\na_\beta \Fabu$ to the hypersurface $\Sigma_t$ 
becomes 
\beqn
\gamma^\alpha{}_\gamma \na_\beta F^{\gamma\beta}
&=&
\gamma^\alpha{}_\gamma \na_\beta 
(\bar{F}^{\gamma\beta} + n^\gamma \bar{F}^\beta - n^\beta \bar{F}^\gamma)
\nonumber\\
\,=\,
D_\beta \bar{F}^\albe
&+&\bar{F}^{\alpha\gamma} n^\beta \na_\beta n_\gamma
\,-\, \gamma^\alpha{}_\beta \Lie_n \bar{F}^\beta
\,+\, K \bar{F}^\alpha
\nonumber\\
&=&
\frac1{\alpha}D_\beta (\alpha \bar{F}^\albe)
\,-\, \Lie_n \bar{F}^\alpha
\,+\, K \bar{F}^\alpha.  
\eeqn
Hence, on $\Sigma_t$, we have 
\beqn
&& \ \ 
F_a
\,=\,
-\,\Lie_n A_a
\,-\, \frac1{\alpha}D_a(\alpha\Phi_\Sigma), 
\label{eq:App:Fabd}
\\[1mm]
&& \ \ 
F_{ab}
\,=\,
D_a A_b \,-\, D_b A_a, 
\label{eq:App:Fabd}
\\[1mm]
&&
n_\alpha \na_\beta \Fabu 
\,=\, -D_a F^a, 
\label{eq:App:projn_divF}
\\[1mm]
&&
\gamma^a{}_\alpha \na_\beta F^{\alpha\beta}
\,=\,
\frac1{\alpha}D_b (\alpha {F}^{ab})
\,-\, \Lie_n {F}^a
\,+\, K {F}^a.  
\label{eq:App:projSigma_divF}
\eeqn

\section{Derivation of equations for electromagnetic potentials}
\label{sec:App:Maxwell}

In this Appendix, we derive the final form of Maxwell's equations 
implemented in the \cocal\ code for computing electromagnetic potentials.  
Since we introduce a conformal decomposition of 
the spatial metric Eq.~(\ref{eq:Conformal}) as in Sec.\ref{sec:3+1}, 
the divergence with respect to the conformal metric $\tgmabd$ 
is simplified to that of flat metric $\fabd$, $\tD_aA^a=\zD_a A^a$. 

Projecting along $n^\alpha$, Eq.~(\ref{eq:Maxwell_n_tsym}) becomes 
\beqn
&&
(\na_\beta\Fabu - 4\pi j^\alpha)n_\alpha 
\,=\, -D_a F^a +4\pi\rho_\Sigma 
\nonumber\\
&&\quad
\,=\, -\frac1{\psi^6}\tD_a (\psi^2 \tgmabu F_b) +4\pi\rho_\Sigma 
\nonumber\\
&&\quad
\,=\, \frac1{\psi^6}\tD_a \left\{\frac{\psi^2}{\alpha} \tgmabu 
\left[\tD_b(\alpha\Phi_\Sigma)-\Lie_\beta A_b\right]\right\}
+4\pi\rho_\Sigma 
\nonumber\\
&&\quad
\,=\, \frac1{\alpha\psi^4}\left\{\tD_a\tD^a(\alpha\Phi_\Sigma)
\phantom{\frac12}\right.
\nonumber\\
&& \phantom{\frac12}\qquad\qquad
\,+\, \tgmabu \frac{\alpha}{\psi^2}
\tD_a \Big(\frac{\psi^2}{\alpha}\Big)
\left[\tD_b(\alpha \Phi_\Sigma)- \Lie_\beta A_b\right]
\nonumber\\
&&
\left.\phantom{\frac12} \qquad\qquad
\,-\, \tgmabu \tD_a \Lie_\beta A_b
+4\pi\alpha\psi^4 \rho_\Sigma\right\}\,=\,0 .
\label{eq:Maxwell_n_cfdcmp}
\eeqn
Separating the flat Laplacian from the first term, 
\beqn
\tD_a \tD^a(\alpha\Phi_\Sigma)
&=&
\frac1{\sqrt{\tgamma}}\zD_a 
\left[\,\sqrt{\tgamma}\tD^a(\alpha\Phi_\Sigma)\, \right]
\nonumber\\
&=&
\zD_a \left[\,\tgmabu\zD_b(\alpha\Phi_\Sigma)\,\right]
\nonumber\\
&=&
\zD_a \zD^a (\alpha \Phi_\Sigma)
\,+\,
h^{ab} \zD_a \zD_b(\alpha\Phi_\Sigma)
\nonumber\\
&&
\,+\,
\zD_a \tgmabu \zD_b(\alpha\Phi_\Sigma), 
\eeqn
an elliptic equation for $\alpha \Phi_\Sigma$ is derived 
\beq
\zLap(\alpha \Phi_\Sigma) \,=\, S, 
\label{eq:App:Maxwell_n_poisson}
\eeq
where the source $S$ is written
\beqn
S &=& 
-\,
h^{ab} \zD_a \zD_b(\alpha\Phi_\Sigma)
\,-\,
\zD_a \tgmabu \zD_b(\alpha\Phi_\Sigma)
\nonumber\\
&+& \tgmabu \frac{\alpha}{\psi^2}
\tD_a \Big(\frac{\psi^2}{\alpha}\Big)\,\alpha{ F}_b
\,+\,\zD_a \tgmabu\, \Lie_\beta A_b 
\nonumber\\
&+& \tgmabu \zD_a \Lie_\beta A_b
-4\pi\alpha\psi^4 \rho_\Sigma. \qquad 
\label{eq:App:Maxwell_n_source}
\eeqn
The second term of the source (\ref{eq:App:Maxwell_n_source}) vanishes 
under the Dirac gauge condition $\zD_a \tgmabu=0$.  Also note that, 
the fourth and fifth terms are derived as below since 
$\tilde \gamma = f$ is satisfied, 
\beqn
\tgmabu \tD_a \Lie_\beta A_b
&=& \zD_a (\tgmabu \Lie_\beta A_b) 
\nonumber\\
&=& \zD_a \tgmabu\, \Lie_\beta A_b 
\,+\,\tgmabu \zD_a \Lie_\beta A_b . 
\eeqn

Projecting to $\Sigma_t$, Eq.~(\ref{eq:Maxwell_Sigma_tsym}) becomes 
\beqn
&&
(\na_\beta\Fabu - 4\pi j^\alpha)\gamma_{a \alpha}
\nonumber\\
&&
\,=\, \frac1{\alpha}D_b (\alpha F_a{}^b)
\,+\, \frac1{\alpha}\Lie_\beta F_a 
\,-\, 2A_a{}^b F_b 
\,+\, \frac13 K F_a 
-4\pi j^\Sigma_a 
\nonumber\\
&&
\,=\, 0.  
\label{eq:Maxwell_Sigma_cfdcmp}
\eeqn
The first term, from which an elliptic operator is separated as below, 
is rewritten,  
\beqn
\frac1{\alpha}D_b (\alpha F_a{}^b)
&=&
\frac1{\alpha\psi^6}\gamma_{ac}\tD_b (\alpha \psi^6 F^{cb})
\nonumber\\
&=&
\frac1{\alpha\psi^2}\tD_b \Big(\frac{\alpha}{\psi^2} F_{ac}\Big)\tgamma^{bc}
\nonumber\\ 
&=&
\frac1{\psi^4}\tgamma^{bc}\tD_b F_{ac}
\,+\,\frac1{\alpha \psi^2}\tgamma^{bc}
\tD_b\Big(\frac{\alpha}{\psi^2}\Big) F_{ac}.
\nonumber\\ 
\eeqn
Using an identity, 
\beq
\ttR_{ab} {\tilde v}^b \,=\, (\tD_b \tD_a - \tD_a \tD_b){\tilde v}^b, 
\eeq
where ${\tilde v}^a = \tgmabu v_b$, we have 
\beqn
\tgamma^{bc}\tD_b F_{ac} 
&=&\tgamma^{bc}\tD_b(\tD_a \tA_c - \tD_c \tA_a) 
\nonumber\\
&=&- \tD_b \tD^b \tA_a \,+\, \tD_a \tD_b \tA^b \,+\, \ttR_{ab} \tA^b. \ \ 
\eeqn
Hence, 
\beqn 
\frac1{\alpha}D_b (\alpha F_a{}^b)
&=&
\frac1{\psi^4}\left[\,
- \tD_b \tD^b \tA_a  
\,+\, \tD_a \tD_b \tA^b 
\phantom{\frac{\alpha}{\psi^2}}\right.
\nonumber\\
&&\left.
\,+\, \ttR_{ab} \tA^b
\,+\,\tgamma^{bc}\frac{\psi^2}{\alpha}
\tD_b\Big(\frac{\alpha}{\psi^2}\Big) F_{ac}
 \,\right]. \qquad
\label{eq:dF_Lap_cfdcmp}
\eeqn 

From the first term of the right hand side of Eq.~(\ref{eq:dF_Lap_cfdcmp}), 
the flat Laplacian  
$\dis - \zLap \tA_a$ is isolated, 
\beqn
-\tD_b\tD^b\tA_a 
&=&
-\zLap\tA_a 
\,-\, h^{bc}\zD_b\zD_c\tA_a
\,+\,\tgamma^{bc}\zD_b(C^d_{ca}\tA_d) 
\nonumber\\
&&
\,+\,\tgamma^{bc}C^d_{bc}\tD_d\tA_a 
\,+\,\tgamma^{bc}C^d_{ba}\tD_c\tA_d
\label{eq:App:Aterms}
\eeqn
We keep $\tD_a$ instead of replacing it by $\zD_a$ and a connection $C^c_{ab}$
in a couple of terms in the Eq.~(\ref{eq:App:Aterms}), to shorten the 
equation.
Then, a set of elliptic equations for $A_a$ is derived,  
\beq
\zLap A_a \,=\, S_a, 
\label{eq:App:Maxwell_vecpotA}
\eeq
where the source $S_a$ is written 
\beqn
S_a  
&:=&
\,-\,h^{bc}\zD_b\zD_c\tA_a
\,+\,\tgamma^{bc}\zD_b(C^d_{ca}\tA_d)
\,+\,\tgamma^{bc}C^d_{bc}\tD_d\tA_a 
\nonumber\\
&&
\,+\,\tgamma^{bc}C^d_{ba}\tD_c\tA_d
\,+\, \tD_a \tD_b \tA^b
\,+\,\ttR_{ab}\tA^b
\nonumber\\
&&
\,+\,{\tilde F}_a\!{}^b \frac{\psi^2}{\alpha}
\tD_b \left(\frac{\alpha}{\psi^2}\right)
\,+\, \frac{\psi^4}{\alpha}\Lie_\beta F_a 
\,-\, 2\psi^4 A_a{}^b F_b 
\nonumber\\
&&
\,+\, \frac13 \psi^4 K F_a 
-4\pi \psi^4 j^\Sigma_a , 
\label{eq:App:Maxwell_vecpotA_source}
\eeqn


\section{Derivation of first integrals of MHD-Euler equations}
\label{sec:App:fint_MHD-Euler}

In this Appendix, we derive a set of integrability conditions 
and first integrals 
(\ref{eq:App:fint_MHD-Euler_t})-(\ref{eq:App:fint_MHD-Euler_xA})
of the relativistic MHD-Euler equations (\ref{eq:MHD-Euler}).  

For the $t$ and $\phi$ components of the MHD-Euler equations 
(\ref{eq:MHD-Euler_t}) and (\ref{eq:MHD-Euler_phi}), 
substituting Eq.~(\ref{eq:merflowfn}) to $u^A$ and 
Eq.~(\ref{eq:mercurrent}) to that of the current $j^A$ 
in the above set of equations, and multiplying by $\rho\sqrt{-g}$, 
we have
\beqn
&&
\epsilon^{AB}\pa_B(\sqrt{-g}\Psi)\pa_A(hu_t)
+\frac1{4\pi}\epsilon^{AB}\pa_B(B\sqrt{-g})\pa_A A_t
\,=\,0, 
\nonumber\\
\\
&&
\epsilon^{AB}\pa_B(\sqrt{-g}\Psi)\pa_A(hu_\phi)
+\frac1{4\pi}\epsilon^{AB}\pa_B(B\sqrt{-g})\pa_A A_\phi
\,=\,0. 
\nonumber\\
\eeqn
Substituting the integrability conditions (\ref{eq:def_fn_At_Psi}), 
these are rewritten 
\beqn
&&
\epsilon^{AB}\left\{-[\sqrt{-g}\Psi]' \,\pa_B(hu_t)
+\frac1{4\pi}A'_t \,\pa_B(B\sqrt{-g})\right\}\pa_A \Upsilon
\,=\,0, 
\nonumber\\
\\
&&
\epsilon^{AB}\left\{-[\sqrt{-g}\Psi]' \,\pa_B(hu_\phi)
+\frac1{4\pi}A'_\phi\pa_B(B\sqrt{-g})\right\}\pa_A  \Upsilon
\,=\,0.
\nonumber\\
\eeqn
These relations imply that the terms in parenthesis are 
a function of $\Upsilon$.  
%
Hence, introducing the densitized scalars 
$[\sqrt{-g}\Lambda_t](\Upsilon)$, and 
$[\sqrt{-g}\Lambda_\phi](\Upsilon)$, for each component, 
the sufficient conditions for the $t$ and $\phi$ components are written, \\ 
$t$-component:
\beq
-[\sqrt{-g}\Psi]' \, hu_t
\,+\,\frac1{4\pi}A'_t \,B\sqrt{-g}
\,=\,[\sqrt{-g}\Lambda_t](\Upsilon), 
\label{eq:App:fint_MHD-Euler_t}
\eeq
$\phi$-component:
\beq
-[\sqrt{-g}\Psi]' \, hu_\phi
\,+\,\frac1{4\pi}A'_\phi \,B\sqrt{-g}
\,=\,[\sqrt{-g}\Lambda_\phi](\Upsilon).
\label{eq:App:fint_MHD-Euler_phi}
\eeq
These are combined and written using another function of $\Upsilon$, 
$\Lambda(\Upsilon)$, 
\beq
A'_\phi\, hu_t \,-\, A'_t \, hu_\phi
\,=\, \Lambda(\Upsilon)
\,:=\,
\frac{A'_t[\sqrt{-g}\Lambda_\phi]\,-\,A'_\phi[\sqrt{-g}\Lambda_t]}
{[\sqrt{-g}\Psi]'}. 
\label{eq:App:fint_MHD-Euler_tphi}
\eeq

For the $x^A$-component (\ref{eq:MHD-Euler_xA_formA}), 
multiplying $\rho\sqrt{-g}$, 
and, substituting Eq.~(\ref{eq:merflowfn}) and 
Eq.~(\ref{eq:mercurrent}) as well as definitions 
$(dA)_{AB} = \epsilon_{AB} B_\phi$ and 
$d(\hu)_{AB} = -\epsilon_{AB} \omega_\phi$, 
we have  
\beqn
&&
\rho u^t \sqrt{-g}\pa_A (hu_t)
\,+\,\rho u^\phi \sqrt{-g} \pa_A (hu_\phi)
\,+\, \omega_\phi\pa_A[\sqrt{-g}\Psi]
\nonumber\\
&&
\,+\,j^t \sqrt{-g}\pa_A A_t
\,+\,j^\phi \sqrt{-g}\pa_A A_\phi
\,-\,B_\phi \pa_A\Big(\frac1{4\pi}B\sqrt{-g}\Big)
\nonumber\\
&&
\,+\,\rho T \sqrt{-g}\pa_A s
\,=\,0,  \qquad
\label{eq:MHD-Euler_xA_formD}
\eeqn
In this Eq.~(\ref{eq:MHD-Euler_xA_formD}), the first two terms 
and the sixth term multiplied by $A'_t A'_\phi$ 
become as follows; substituting the integrals of $t$ and $\phi$ components 
of MHD-Euler equations (\ref{eq:App:fint_MHD-Euler_t}) and 
(\ref{eq:App:fint_MHD-Euler_phi}), 
\begin{widetext}
\beqn
&&
\frac12 \rho u^t\sqrt{-g}\, A'_t 
\left[\pa_A(A'_t\,hu_\phi+\Lambda)- hu_t \pa_A A'_\phi\right]
\,+\,\frac12 \rho u^t\sqrt{-g}\, A'_t A'_\phi\pa_A(hu_t)
\nonumber\\
&+&
\frac12 \rho u^\phi\sqrt{-g}\, A'_\phi 
\left[\pa_A(A'_\phi\,hu_t-\Lambda)- hu_\phi \pa_A A'_t\right]
\,+\,\frac12 \rho u^\phi\sqrt{-g}\, A'_t A'_\phi\pa_A(hu_\phi)
\nonumber\\
&-&
\frac12 B_\phi\,\pa_A\left(
A'_\phi[\sqrt{-g}\Psi]'hu_t + A'_\phi[\sqrt{-g}\Lambda_t]
+A'_t[\sqrt{-g}\Psi]'hu_\phi + A'_t[\sqrt{-g}\Lambda_\phi]
\right)
\,+\, B_\phi \frac1{4\pi}B\sqrt{-g}\,\pa_A (A'_t A'_\phi)
\nonumber\\
&=&
\frac12 \left(A'_t \rho u^t\sqrt{-g}
\,+\, A'_\phi \rho u^\phi\sqrt{-g}
\,-\,[\sqrt{-g}\Psi]'B_{\phi}\right)
\left[A'_t \pa_A(hu_\phi) + A'_\phi \pa_A(hu_t)\right]
\nonumber\\
&+&
\frac12(\pa_A A'_t\,hu_\phi-\pa_A A'_\phi\,hu_t+\pa_A\Lambda)
(A'_t u^t - A'_\phi u^\phi)\rho\sqrt{-g}
\nonumber\\
&-&
\frac12\left\{
A'_\phi\left(\pa_A[\sqrt{-g}\Psi]'hu_t + \pa_A[\sqrt{-g}\Lambda_t] \right)
\,-\,\pa_A A'_\phi \left([\sqrt{-g}\Psi]'hu_t + [\sqrt{-g}\Lambda_t] \right)
\right.
\nonumber\\
&& 
\left.
\,+\, A'_t\left(\pa_A[\sqrt{-g}\Psi]'hu_\phi + \pa_A[\sqrt{-g}\Lambda_\phi] \right)
\,-\,\pa_A A'_t \left([\sqrt{-g}\Psi]'hu_\phi + [\sqrt{-g}\Lambda_\phi] \right)
\right\}B_\phi
\label{eq:MHD-Euler_xA_terms}
\eeqn
\end{widetext}
The terms in the first parenthesis of r.h.s.~vanish because of 
the first integral of the $x^A$-components of the ideal 
MHD condition (\ref{eq:fint_idealMHD_xA}), 
and all other terms are proportional to $\pa_A \Upsilon$, 
as $A_t$, $A_\phi$, $\sqrt{-g}\Psi$, $\sqrt{-g}\Lambda_t$, 
$\sqrt{-g}\Lambda_\phi$, and $\Lambda$, are functions of $\Upsilon$. 
Hence, 
with an assumption to the thermodynamic variable, 
that is, the entropy $s$ to be a function of $\Upsilon$, 
$s=s(\Upsilon)$, 
Eq.~(\ref{eq:MHD-Euler_xA_formD}) multiplied by $A'_t A'_\phi$ 
is rewritten 
\beqn
&&\left\{
-\,\frac12\left[
A'_\phi\left([\sqrt{-g}\Psi]''hu_t + [\sqrt{-g}\Lambda_t]' \right)
\right.\right.
\nonumber\\
&&\quad\quad 
\,-\,A''_\phi \left([\sqrt{-g}\Psi]'hu_t + [\sqrt{-g}\Lambda_t] \right)
\nonumber\\
&&\quad\quad 
\,+\, A'_t\left([\sqrt{-g}\Psi]'' hu_\phi + [\sqrt{-g}\Lambda_\phi]' \right)
\nonumber\\
&&
\left.\quad\quad 
\,-\, A''_t \left([\sqrt{-g}\Psi]' hu_\phi + [\sqrt{-g}\Lambda_\phi] \right)
\right]B_\phi
\nonumber\\
&&\ 
\,+\,\frac12(A''_t\,hu_\phi - A''_\phi\,hu_t+\Lambda')
(A'_t u^t - A'_\phi u^\phi)\rho\sqrt{-g}
\nonumber\\
&&\ 
\,+\, A'_t A'_\phi[\sqrt{-g}\Psi]'\omega_\phi
\,+\,A'_t  A'_\phi  s'   T   \rho \sqrt{-g}
\nonumber\\
&&\!\!\!\!\!
\left.\phantom{\frac11}
\,+\, (A'_t)^2 A'_\phi \, j^t\sqrt{-g}
\,+\,  A'_t (A'_\phi)^2\, j^\phi\sqrt{-g}
\right\}\pa_A \Upsilon
\,=\,0.
\nonumber\\
\eeqn
Therefore, we obtain the first integral of the $x^A$-components of 
the MHD-Euler equations (\ref{eq:MHD-Euler_xA_formD}) as 
\beqn
&&
-\,\frac12\left[
A'_\phi\left([\sqrt{-g}\Psi]''hu_t + [\sqrt{-g}\Lambda_t]' \right)
\right.
\nonumber\\
&&\quad
\,-\,A''_\phi \left([\sqrt{-g}\Psi]'hu_t + [\sqrt{-g}\Lambda_t] \right)
\nonumber\\
&&
\quad
\,+\, A'_t\left([\sqrt{-g}\Psi]'' hu_\phi + [\sqrt{-g}\Lambda_\phi]' \right)
\nonumber\\
&&
\left.\quad
\,-\, A''_t \left([\sqrt{-g}\Psi]' hu_\phi + [\sqrt{-g}\Lambda_\phi] \right)
\right]B_\phi
\nonumber\\
&&
\,+\,\frac12(A''_t\,hu_\phi - A''_\phi\,hu_t+\Lambda')
(A'_t u^t - A'_\phi u^\phi)\rho\sqrt{-g}
\nonumber\\
&&
\,+\, A'_t A'_\phi[\sqrt{-g}\Psi]'\omega_\phi
\,+\,A'_t  A'_\phi  s'   T   \rho \sqrt{-g}
\nonumber\\
&&
\,+\, (A'_t)^2 A'_\phi \, j^t\sqrt{-g}
\,+\,  A'_t (A'_\phi)^2\, j^\phi\sqrt{-g}
\,=\,0.
\label{eq:App:fint_MHD-Euler_xA}
\eeqn

\section{First integral and integrability conditions for the case 
of pure rotational flow}
\label{sec:App:purerotational}

For the case without meridional flow fields, $u^A=0$, the system of 
first integrals and Maxwell's equations under stationarity and 
axisymmetry can be recast into a single equation to be solved for
a single independent variable which may be called the Grad-Shafranov 
equation with a toroidal flow fields.  
However as we have noted, we do not reduce the number of variables 
in our formulation, but rather solve the hydrostationary equation and
Maxwell's equations simultaneously.  Although the derivation 
presented in Sec.~\ref{sec:fint} can be applied for the case with pure rotational 
flow, we repeat the derivation below for clarity.

\subsection{Ideal MHD condition for purely rotational flow}

We assume the 4-velocity $u^\alpha$ of the flow field in the absence of 
meridional flow $u^A=0$ as 
\beq
u^\alpha \,=\, u^t(t^\alpha + \Omega \phi^\alpha) \,=\, u^t k^\alpha.
\label{eq:App:4ve_GS}
\eeq
The ideal MHD condition $\Fabd u^\beta=0$ in this case becomes \\
$t$-component
\beq
t\cdot F\cdot u \ (\,=\,-u^A\pa_A A_t)\,\equiv\, 0
\eeq
$\phi$-component
\beq
\phi\cdot F\cdot u \ (\,=\,-u^A\pa_A A_\phi)\,\equiv\, 0
\eeq
$x^A$-component
\beq
e_A\cdot F\cdot u 
\,=\, u^t\pa_A A_t \,+\, u^\phi\pa_A A_\phi \,=\,0
\label{eq:idealMHD_GS_xA}
\eeq
For the case with meridional flow, integrability conditions 
can be found in the above ideal MHD condition alone.  The absence 
of the meridional stream function in this case trivializes the 
$t$ and $\phi$ components of ideal MHD conditions and hence 
the integrability conditions are not derived from these equations.

\subsection{MHD-Euler equations for pure rotational flow}

Substituting $u^A=0$ and 
$j^\alpha = j^t t^\alpha + j^\phi \phi^\alpha + j^A e_A^\alpha$, 
the MHD-Euler equations becomes \\
$t$-component:
\beq
\frac1{\rho} j^A \pa_A A_t \,=\, 0
\label{eq:MHD_Euler_GS_t}
\eeq
$\phi$-component:
\beq
\frac1{\rho} j^A \pa_A A_\phi \,=\, 0
\label{eq:MHD_Euler_GS_phi}
\eeq
$x^A$-component:
\beqn
&& u^t\pa_A (hu_t) 
\,+\, u^\phi\pa_A (hu_\phi) 
\nonumber\\
&&
\,+\, \frac1{\rho} j^t \pa_A A_t 
\,+\, \frac1{\rho} j^\phi \pa_A A_\phi 
\,+\, \frac1{\rho} j^B (d A)_{AB} 
\,+\,T\pa_A s
\,=\,0.
\nonumber\\
\label{eq:MHD_Euler_GS_xA}
\eeqn

\subsection{Integrability conditions for the case of purely rotational flow}

Substituting the $x^A$-components of Maxwell's equations (\ref{eq:mercurrent}) 
to the meridional current $j^A$ appearing in the $t$ and $\phi$ 
components of the MHD-Euler equations (\ref{eq:MHD_Euler_GS_t}) and 
(\ref{eq:MHD_Euler_GS_phi}), we have 
\beqn
&& 
\frac1{4\pi\rho\sqrt{-g}}\epsilon^{AB}\pa_B(\sqrt{-g}B)\,\pa_A A_t\,=\,0, 
\\
&& 
\frac1{4\pi\rho\sqrt{-g}}\epsilon^{AB}\pa_B(\sqrt{-g}B)\,\pa_A A_\phi\,=\,0.  
\eeqn
These relations require integrability conditions for consistency, 
namely, a master potential $\Upsilon$ is introduced as follows, 
\beqn
A_t\,=\,A_t(\Upsilon), 
&\quad &
A_\phi\,=\,A_\phi(\Upsilon), 
\nonumber\\[1mm]
\mbox{and} \qquad 
\sqrt{-g}B \,&=&\, [\sqrt{-g}B](\Upsilon).  
\label{eq:def_fn_At_Aphi_B}
\eeqn

The $x^A$-components of ideal MHD conditions (\ref{eq:idealMHD_GS_xA}), and 
the above integrability conditions for $A_t$ and $A_\phi$ imply 
\beq
u^t\pa_A A_t \,+\, u^\phi\pa_A A_\phi 
\,=\, (u^t A'_t \,+\, u^\phi A'_\phi)\pa_A \Upsilon
\,=\,0, 
\eeq
and hence, 
\beq
u^t A'_t \,+\, u^\phi A'_\phi\,=\,0, 
\label{eq:MHDfnc_MHD-Euler_xA_GS}
\eeq
or, introducing the angular velocity $\Omega$, 
\beq
\frac{u^\phi}{u^t}\,=\,\Omega\,=\,-\frac{A'_t}{A'_\phi}.
\label{eq:MHDfnc_MHD-Euler_Omega_xA_GS}
\eeq
Therefore, $\Omega$ should be a function of 
$\Upsilon$ as well, 
\beq
\Omega=\Omega(\Upsilon).
\label{eq:def_fn_Omega}
\eeq

\subsection{First integral of meridional components of MHD-Euler equations}

Derivation of the integrability of $x^A$-component of
MHD-Euler equations proceeds analogous to the case with
non-zero meridional flow.  A difference is an absence
of the stream function $\sqrt{-g} \Psi$.  In 
Eq.~(\ref{eq:fint_MHD-Euler_tphi}), the stream function
$[\sqrt{-g}\Psi]'(\Upsilon)$ appears in the denominator of the 
definition of an arbitrary function $[\sqrt{-g}\Lambda](\Upsilon)$.
In paper \cite{Gourgoulhon:2011gz}, we have proved that such 
a combination always become finite, and hence the relations derived
in previous sections are valid also for the case of pure rotational 
flow under simply taking a limit 
$[\sqrt{-g}\Psi](\Upsilon) \rightarrow {\rm constant}$.  
In this section, we prove this fact by repeating the derivation of the 
previous section, and derive (\ref{eq:fint_MHD-Euler_tphi}) directly 
as a part of integrability conditions.  

We recast the $x^A$-components of MHD-Euler equations in the
same way as the case for generic flow in the previous section.
To proceed, we use a relation,  
\beq
A'_\phi u_t \,-\, A'_t u_\phi
\,=\, \frac1{2u^t u^\phi}(A'_t u^t \,-\, A'_\phi u^\phi), 
\eeq
derived from normalization
of the 4-velocity $u\cdot u=-1$, and the integrability condition
(\ref{eq:MHDfnc_MHD-Euler_xA_GS}).  
Multiplying by the factor $\frac{2A'_t A'_\phi}{A'_t u^t-A'_\phi u^\phi}$,
the kinetic term of the $x^A$-components of MHD-Euler equations 
for purely rotational flow (\ref{eq:MHD_Euler_GS_xA}), is rewritten  
\beqn
&&
\frac{2A'_t A'_\phi}{A'_t u^t-A'_\phi u^\phi}
\left[\,u^t\pa_A (hu_t) \,+\, u^\phi\pa_A (hu_\phi) \,\right]
\nonumber\\
&=& \pa_A (A'_\phi hu_t - A'_t hu_\phi) 
\,+\, (hu_\phi \pa_A A'_t - hu_t \pa_A A'_\phi), 
\qquad
\eeqn
where a consistency Eq.~(\ref{eq:MHDfnc_MHD-Euler_xA_GS}) is used.  
%
Also, the Lorenz force term of Eq.~(\ref{eq:MHD_Euler_GS_xA}) becomes 
\beqn
&&
\frac1{\rho} \left[\,j^t \pa_A A_t 
\,+\, j^\phi \pa_A A_\phi 
\,+\, j^B (d A)_{AB} \,\right]
\,=\,
\nonumber\\ &&
\frac1{\rho} \left[j^t \pa_A A_t 
\,+\, j^\phi \pa_A A_\phi 
\,+\, \frac1{4\pi\sqrt{-g}}B_\phi
\pa_A \left(B\sqrt{-g}\right)\right], 
\qquad
\eeqn
with Eqs.~(\ref{eq:FABd}) and (\ref{eq:mercurrent}).  

Because $A_t$, $A_\phi$ and $B\sqrt{-g}$ are 
functions of the master potential $\Upsilon$ as shown in 
Eq.~(\ref{eq:def_fn_At_Aphi_B}), 
the $x^A$-components of MHD-Euler equations (\ref{eq:MHD_Euler_GS_xA})
multiplied by the factor 
$\frac{2A'_t A'_\phi}{A'_t u^t-A'_\phi u^\phi}$,
is rewritten, with an assumption of $s=s(\Upsilon)$, 
\beqn
&&
\pa_A (A'_\phi hu_t - A'_t hu_\phi) 
\nonumber\\
&+& \left\{ (hu_\phi \pa_A A'_t - hu_t \pa_A A'_\phi) 
\,+\, \frac{2A'_t A'_\phi}{A'_t u^t-A'_\phi u^\phi}\times 
\right.
\nonumber\\&&
\left[
\frac1{\rho} \left(\,j^t \pa_A A_t 
\,+\, j^\phi \pa_A A_\phi 
\,+\, \frac1{4\pi\sqrt{-g}}B_\phi
\left[B\sqrt{-g}\right]'\,\right)
\right.
\nonumber\\&&
\left.
\left.\phantom{\frac12}
\,+\,T s'\right]\right\}\pa_A \Upsilon
\,=\,0.
\label{eq:MHD_Euler_GS_int_xA}
\eeqn
The above relation suggests that because of the converse 
of the Poincar\'e lemma, the first term is a function of $\Upsilon$, 
\beq
A'_\phi hu_t \,-\, A'_t hu_\phi \,=\, \Lambda(\Upsilon).
\label{eq:fint_MHD-Euler_tphi_GS}
\eeq
This is compared with Eq.~(\ref{eq:fint_MHD-Euler_tphi}).

Since $\sqrt{-g}B=[\sqrt{-g}B](\Upsilon)$, we introduce 
the following functions of $\Upsilon$, 
\beqn
\frac1{4\pi}A'_t \sqrt{-g}B &=& [\sqrt{-g}\Lambda_t](\Upsilon),
\label{eq:fint_MHD-Euler_t_GS}
\\
\frac1{4\pi}A'_\phi \sqrt{-g}B &=& [\sqrt{-g}\Lambda_\phi](\Upsilon).
\label{eq:fint_MHD-Euler_phi_GS}
\eeqn
Taking derivatives of these with respect of $\Upsilon$, 
and combining them, we have a relation, 
\beqn
\frac1{4\pi}A'_t A'_\phi [\sqrt{-g}B]' 
&=& 
\frac12\left(
A'_\phi [\sqrt{-g}\Lambda_t]' 
- A''_t [\sqrt{-g}\Lambda_\phi] \right.
\nonumber\\&&
\left.
\,+\, A'_t [\sqrt{-g}\Lambda_\phi]' 
- A''_\phi [\sqrt{-g}\Lambda_t]\right). \qquad\  
\label{eq:MHDfnc_Bprime_GS}
\eeqn
Finally, substituting Eqs.~(\ref{eq:fint_MHD-Euler_tphi_GS}) and 
(\ref{eq:MHDfnc_Bprime_GS}) to (\ref{eq:MHD_Euler_GS_int_xA}), 
the consistency of the $x^A$ components yields 
\beqn
&&
-\frac12\left(
A'_\phi [\sqrt{-g}\Lambda_t]' 
- A''_\phi [\sqrt{-g}\Lambda_t] 
\right.
\nonumber\\
&& \left.\quad
\,+\, A'_t [\sqrt{-g}\Lambda_\phi]' 
- A''_t [\sqrt{-g}\Lambda_\phi]\right)B_\phi 
\nonumber\\ &&
\,+\,\frac12\left( A''_t hu_\phi - A''_\phi hu_t + \Lambda'\right)
(A'_t hu^t \,-\, A'_\phi hu^\phi)\rho\sqrt{-g}
\nonumber\\ &&
\,+\, A'_t A'_\phi s' T \rho\sqrt{-g}
\nonumber\\
&&
\,+\, (A'_t)^2 A'_\phi j^t \sqrt{-g}
\,+\, A'_t (A'_\phi)^2 j^\phi \sqrt{-g} \,=\,0.
\label{eq:fint_MHD-Euler_xA_GS}
\eeqn
This relation is compared with the result for the generic 
flow (\ref{eq:fint_MHD-Euler_xA}); Eq.~(\ref{eq:fint_MHD-Euler_xA_GS}) 
agrees with Eq.~(\ref{eq:fint_MHD-Euler_xA}) in the limit  
$[\sqrt{-g}\Psi](\Upsilon) \rightarrow \mbox{constant}$.


\section{Definitions of mass, angular momentum, and virial relation}
\label{sec:App:physq}

For the reader's convenience, we summarize definitions of tabulated 
quantities in Tables \ref{tab:MRNS_solutions} and 
\ref{tab:MRNS_solutions_EMF}  (see also e.g., \cite{RNS,Gourg2012}).  
Those include the rest mass $M_0$, ADM mass $\Madm$, Komar mass $\MK$, 
total angular momentum $J$, 
the virial relation $I_{\rm vir}$ and other related quantities 
including electromagnetic energy $\cal M$ and its decomposition.

In Table \ref{tab:MRNS_solutions}, $\bar{R}_0$ and $\bar{R}_z$ are 
the equatorial and polar radius of a compact star in the proper length, respectively.  
The proper equatorial radus $\bar{R}_0$ is defined by 
\beq
\bar{R}_0\,:=\,\int_0^{R_0} \psi^2 \sqrt{\tgamma_{xx}} \, dx, 
\label{eq:proper_radi}
\eeq
and for $\bar{R}_p$, the integral is taken along $z$-axis.
The unbarred $R_0$ and $R_p$ are the equatorial 
and polar radii in coordinate length.  

The rest mass $M_0$ is written 
\beq
M_{0}
\,:=\, \int_\Sigma \rho u^\alpha dS_\alpha 
\,=\, \int_\Sigma \rho u^t \alpha \psi^6\sqrt{\tgamma}d^3x, 
\label{eq:rest_mass}
\eeq
where $dS_\alpha = \na_\alpha t \sqrt{-g} d^3x$, 
and $d^3x=r^2 \sin\theta dr d\theta d\phi$ on the 
spherical coordinates.  This is conserved irrespective of 
the choice of a slice $\Sigma$ for the rest mass conservation 
Eq.~(\ref{eq:restmass}).  The above volume integral over the 
hypersurface $\Sigma$ is non zero only on the fluid support.

The ADM mass $\Madm$ is defined and calculated by
\beqn
\Madm
&:=& \frac1{16\pi}\int_\infty 
\left(f^{ac}f^{bd}-f^{ab}f^{cd}\right)
\zD_b\gamma_{cd}\, dS_a
\nonumber\\
&=& 
- \frac1{2\pi}\int_\infty \tD{}^a\psi \,d\tS_a 
\label{eq:ADM_surf}
\\
&=& \frac1{2\pi}\int_\Sigma \left[\,
\,-\,\frac{\psi}{8}\,\ttR
\,+\,\frac18\psi^5\left(\tA_{ab}\tA^{ab}-\frac23 K^2\right)\right.
\nonumber\\
&&\left.\phantom{\frac11}
\,+\,2\pi\psi^5\rhoH \,\right]\sqrt{\tgamma}d^3x.
\label{eq:ADM_vol}
\eeqn
where $\rho_H:=\Tabd n^\alpha n^\beta$ is a component of 
the stress energy tensor $\Tabd$ normal to the hypersurface $\Sigma$.  
To check the consistency of the solution, both of the surface integral 
(\ref{eq:ADM_surf}) and the volume integral (\ref{eq:ADM_vol}) 
are evaluated.  
Relative errors between those values are $\sim 0.01\%$ 
for the solutions presented in Sec.~\ref{sec:Results}.  
The surface integral is calculated on 
a sphere with radius around $r\sim 10^4 R_0$, while 
the volume integral is taken within a computational domain 
within a radius around $r\sim 10^6 R_0$.  
For the surface integral (\ref{eq:ADM_surf}), we replace 
$\tgamma^{ab}\rightarrow f^{ab}$, $\tD_a \rightarrow \zD_a$, and 
$d\tS_a = \na_a r \sqrt{\tgamma}d^2x = \na_a r \sqrt{f}d^2x = dS_a$.  
These are exact at spatial infinity, and they introduce 
only a negligible numerical error at the above radius where 
the surface integral (\ref{eq:ADM_surf}) is evaluated.

The Komar mass $\MK$ associated with the global 
timelike Killing field $t^\alpha$ is defined by
\beqn
\MK
&:=& -\frac1{4\pi}\int_\infty \na^\alpha\, t^\beta \,dS_{\albe}
\\
&=&  -\int_\Sigma \left(\,2\Tab - T \gab \,\right)\,t^\beta \,dS_\alpha 
\nonumber\\
&=& \int_\Sigma\left[\, \alpha \left(\rhoH+S\right) 
-2 j_a \beta^a\,\right] \psi^6\sqrt{\tgamma}d^3x , 
\label{eq:Komar_vol}
\eeqn
and the asymptotic Komar mass whose $t^\alpha$ is a 
symmetry of an asymptotically flat spacetime by
\beqn
\MK
&:=& -\frac1{4\pi}\int_\infty \na^\alpha\, t^\beta \,dS_{\albe}
\nonumber\\
&=& \frac1{4\pi}\int_\infty D^a \alpha\,dS_a
\label{eq:Komar_surf}
\\
&=& \frac1{4\pi}\int_\Sigma\left[\, 
 \tA_{ab}\tA^{ab}+\frac13 K^2
- \Lie_n K \right.
\nonumber\\
&&\left.\phantom{\frac11}
+4\pi\left(\rhoH+S\right) \,\right]\alpha \psi^6\sqrt{\tgamma}d^3x , 
\label{eq:Komar_vol_nc}
\eeqn
where the source terms $j_a:= -\Tabd \gmaa n^\beta$, and 
$S:= \Tabd \gamma^\albe$ are the components of 
the 3+1 decomposed stress-energy tensor $\Tabd$.  
In deriving (\ref{eq:Komar_vol_nc}), a relation, 
$(\Gabd - 8\pi \Tabd)(\gamma^{\albe}+n^\alpha n^\beta)=0$ 
is used.  
Since $\Tabd=\TabMd+\TabFd$ contains electromagnetic contributions, 
the support of the volume integral (\ref{eq:Komar_vol}) is non-compact.  
All integrals (\ref{eq:Komar_vol}), (\ref{eq:Komar_surf}), and 
(\ref{eq:Komar_vol_nc}) should reproduce the same value, when the waveless 
condition (\ref{eq:waveless}) and the coordinate conditions (\ref{eq:gauge}) 
are imposed at least asymptotically  \cite{Shibata:2004qz}.  
We computed Eqs.~(\ref{eq:Komar_vol}), (\ref{eq:Komar_surf}), and 
(\ref{eq:Komar_vol_nc}) to check the consistency of the solutions, 
and found that they agree in the same order as $\Madm$ mentioned above.

For the total angular momentum $J$, the surface and volume 
integrals are evaluated, 
\beqn
J
&:=& \frac1{8\pi}\int_\infty K^a{}_b\phi^b \,dS_a
\label{eq:J_surf}
\\
&=&  \frac1{8\pi}\int_\Sigma D_a (K^a{}_b \phi^b)\,dV
\nonumber\\
&=& \frac1{8\pi}\int_\Sigma 
\left(8\pi j_a \phi^a + A^a{}_b \tD_a\phi^b - \frac4{\psi}K\phi^a\zD_a\psi\right)\,
\nonumber\\
&&\ \times \psi^6\sqrt{\tgamma}d^3x, 
\label{eq:J_vol}
\eeqn
and the difference between the values from Eqs.~(\ref{eq:J_surf}) 
and from (\ref{eq:J_vol}) is typically $\Od(0.1)\%$.  
Also for $J$, a term including $j_a$ in the volume integral 
contains contributions from fluid as well as electromagnetic fields, 
and hence it is integrated over a non-compact support.  
The values of $\Madm$, $J$, and  $|1-\MK/\Madm|$ listed in Table 
\ref{tab:MRNS_solutions} are those of the volume integrals, (\ref{eq:ADM_vol}), 
(\ref{eq:Komar_vol}) and (\ref{eq:J_vol}).\footnote{
In previous papers for non-magnetized rotating stars \cite{Uryu:2016dqr,RNScocal},  
the ratio of the kinetic energy and the gravitational energy, $T/|W|$ 
was defined following \cite{RNS}, 
\beqn
W &:=& \Madm - M_{\rm P} - T, 
\\
T &:=& 
\textstyle
\frac12 \int_\Sigma \Omega \,dJ.  
\label{eq:ToverW}
\eeqn
where the proper mass $\MP$ was defined by
\beq
\textstyle
M_{\rm P}
\,:=\, \int_\Sigma \Tab u^\beta dS_\alpha \,=\, \int_\Sigma \epsilon u^\alpha \dSa .
\label{eq:proper_mass}
\eeq
In the solutions presented in Sec.~\ref{sec:Results}, $\Tab$ includes 
the electromagnetic $\TabFd$, while $u^\alpha$ is defined only on the 
fluid support.  Likewise $\Omega$ in (\ref{eq:ToverW}) is undefined 
outside of a star, although $dJ$ has a non zero electromagnetic 
contribution there.  Because of this ambiguity, we do not calculate 
the values of $\MP$ and $T/|W|$.  
}
%
%

The relativistic virial theorem for an Einstein-Maxwell spacetime coupled with 
charged and magnetized perfect fluid \cite{1994CQGra..11..443G} is computed 
to determine the accuracy of solutions.  
It is a vanishing integral of the spatial trace of Einstein's equations 
over a hypersurface $\Sigma$, 
\beqn
&&
\int_\Sigma \left(T_a{}^a - {\small \frac1{8\pi}} G_a{}^a \right)dV
\nonumber\\
&&
\,=\,
2{\cal T} \,+\, 3\Pi \,+\, {\cal M} +\, {\cal W} \,+\, \Madm \,-\, \MK \,=\, 0.  \qquad
\label{eq:App:traceG}
\eeqn
The equality of the ADM mass and the Komar mass $\Madm = \MK$ 
has been proved for stationary spacetimes \cite{1978PhLA...69..153B}, 
and for the waveless formulation \cite{Shibata:2004qz}.  
The integrals $\cal T$, $\Pi$, ${\cal M}$ and $\cal W$ are defined by 
\beqn
{\cal T} &=& \frac12\int_\Sigma (\epsilon+p)u_a u^a dV, 
\label{eq:App:energyKin}
\\[1mm]
\Pi &=& \int_\Sigma p\, dV, 
\label{eq:App:energyInt}
\\[1mm]
{\cal M} &=& \frac1{16\pi}\int_\Sigma (2 F_a F^a + F_{ab} F^{ab})dV
\label{eq:App:energyEM}
\\[1mm]
{\cal W} &=& \frac1{4\pi}\int_\Sigma [\psi^{-4}(2\tD^a \ln\psi \tD_a \ln\psi - \tD^a \ln\alpha \tD_a \ln\alpha)
\nonumber\\
&+&\frac34(\Aabd \Aabu -\frac23 K^2)+\frac1{\alpha}K\beta^a\tD_a\ln\alpha+\frac14 \ttR \psi^{-4}]dV, 
\label{eq:App:energyGR}
\nonumber\\
\eeqn
which become the kinetic, internal, electromagnetic, and gravitational energies, 
in the Newtonian limit.  In the integrand of $\cal W$ (\ref{eq:App:energyGR}), 
$\ttR$ is a scalar curvature of a conformally related spacelike hypersurface 
associated with a conformal 3-metric $\tgmabd$.  
We define a virial integral $I_{\rm vir}$ as 
\beq
I_{\rm vir}\,=\, \left|\,2 {\cal T}\,+\, 3 \Pi \,+\, {\cal M}\,+\, {\cal W} \,\right|, 
\label{eq:App:virial}
\eeq
whose values for the selected solutions are presented in Table 
\ref{tab:MRNS_solutions_EMF}.  
The magnetic energy term $\cal M$ (\ref{eq:App:energyEM}) is decomposed into 
contributions from the electric fields as well as the poloidal and toroidal 
magnetic fields, for which we define, respectively, 
\beqn 
{\cal M}_{\rm ele}&=&\frac1{8\pi}\int_\Sigma  F_a F^a dV, 
\\[1mm]
{\cal M}_{\rm pol}&=&\frac1{16\pi}\int_\Sigma F_{AB} F^{AB}dV, 
\\[1mm]
{\cal M}_{\rm tor}&=&\frac1{8\pi}\int_\Sigma F_{A\phi}F^{A\phi}dV,  
\eeqn
which are also listed in Table \ref{tab:MRNS_solutions_EMF}.  

Finally, the electric charge $Q$ defined in Eq.~(\ref{eq:charge}) 
becomes 
\beq
Q\,=\, \frac1{4\pi}\int_\infty \Fabu dS_{\albe}
\,=\,\frac1{4\pi}\int_\infty F^a dS_a
\label{eq:App:charge}
\eeq
where $\dSab = \frac12(\na_\alpha t \na_\beta r -\na_\alpha r \na_\beta t )\sqrt{-g}\,d^2x$, 
and $dS_a = \na_a r \sqrt{\gamma}\,d^2x$, which 
is evaluated on a large sphere $S$ in the asymptotics of $\Sigma$.  
Rewriting the charge $Q$ in the form of volume integral, 
\beqn
Q&=&\frac1{4\pi}\int_\Sigma \na_\beta \Fabu \dSa
\,=\,\int_\Sigma j^\alpha \dSa
\nonumber\\[1mm]
&=&
Q_M\,+\,Q_S, 
\label{eq:App:charge_vol}
\eeqn
the volume integral over the MHD fluid support $Q_M$ and 
the surface charge $Q_S$ at the stellar surface should 
contribute to the total charge $Q$.  In our formulation, 
the form of $Q_S$ is not given, and the values of $Q_M$ 
are listed in Table \ref{tab:MRNS_solutions_EMF}.

\section{Imposition of symmetry of the electromagnetic 
vector potential
}
\label{secApp:gauge}

When an exact 2-form $F = dA$ respects the symmetry 
$\Lie_t F = 0$, a gauge potential $f$ exists such that 
$A$ transformed by $A \rightarrow A+df$ satisfies $\Lie_t (A+df)=0$.  

\noindent
proof)  
The Cartan identity $t \cdot dF = \Lie_t F - d(t\cdot F)$ implies 
$d(t\cdot F)=0$ when $\Lie_t F=0$.  
Hence, because of the Poincar\'e lemma, 
a function $\Phi$ exists such that $t\cdot F = d\Phi$ on 
a simply connected manifold.  
This implies 
\beq
\Lie_t A 
\,=\, t \cdot dA \,+\, d(t\cdot A) 
\,=\, t\cdot F \,+\, d(t\cdot A)
\,=\, d(\Phi \,+\, t\cdot A).
\eeq
Under a gauge transformation with a potential $f$, 
$A \rightarrow A+df$, $(F\rightarrow F)$, $\Lie_t A$ 
is transformed as 
\beq
\Lie_t (A+df) = \Lie_t A + d\Lie_t f = d(\Lie_t f + \Phi + t\cdot A).
\eeq
Hence for an $f$ that satisfies 
\beq
\Lie_t f + \Phi + t\cdot A = {\rm const}, 
\eeq
$A+df$ satisfies the symmetry $\Lie_t (A+df)  =0$.  

We have freedom to choose another gauge potential 
${\widehat f}$ that respects the symmetry 
$\Lie_t \widehat f = 0$.  This gauge potential does not 
affect the above transformation to impose the symmetry 
on the potential $A$, namely 
$A\rightarrow A + df + d\widehat{f}$ respects the symmetry.  
With this gauge freedom we may, for example, impose Coulomb gauge 
(vanishing spatial divergence) $\zD^a A_a = 0$ 
where $a$ is a spatial 3D index.


\begin{thebibliography}{99}

\bibitem{Magnetar}
  R.~C.~Duncan and C.~Thompson,
  Astrophys.\ J.\  {\bf 392}, L9 (1992);
For a review, see e.g., 
  V.~M.~Kaspi and A.~Beloborodov,
  Ann.\ Rev.\ Astron.\ Astrophys.\  {\bf 55}, 261 (2017)
  R.~Turolla, S.~Zane and A.~Watts,
  Rept.\ Prog.\ Phys.\  {\bf 78}, no. 11, 116901 (2015)


\bibitem{Oron:2002gs} 
  A.~Oron,
  Phys.\ Rev.\ D {\bf 66}, 023006 (2002);
\bibitem{Konno:1999}
  K.~Konno, T.~Obata and Y.~Kojima,
  Astron.\ Astrophys.\ {\bf 352}, 211 (1999).
%
\bibitem{PolTorMS}
  K.~Ioka and M.~Sasaki,
  Phys.\ Rev.\ D {\bf 67}, 124026 (2003);
  K.~Ioka and M.~Sasaki,
  Astrophys.\ J.\  {\bf 600}, 296 (2004)
  R.~Ciolfi, V.~Ferrari, L.~Gualtieri and J.~A.~Pons,
  Mon.\ Not.\ Roy.\ Astron.\ Soc.\  {\bf 397}, 913 (2009); 
  R.~Ciolfi, V.~Ferrari and L.~Gualtieri,
  Mon.\ Not.\ Roy.\ Astron.\ Soc.\  {\bf 406}, 2540 (2010)
\bibitem{SSMS}
  S.~Yoshida, K.~Kiuchi and M.~Shibata,
  Phys.\ Rev.\ D {\bf 86}, 044012 (2012);
  S.~Yoshida,
   arXiv:1811.11564 [gr-qc]. 

\bibitem{RNS}
For reviews of relativistic rotating stars, see e.g.,
J.~L.~Friedman and N.~Stergioulas, 
\emph{Rotating Relativistic Stars}, 
Cambridge University Press, Cambridge, UK, 2013.
%
N.~Straumann, \emph{General Relativity}, Springer 
Science+Business Media Dordrecht, 2013;
%
  V.~Paschalidis and N.~Stergioulas,
  Living Rev.\ Rel.\  {\bf 20}, no. 1, 7 (2017)


\bibitem{Bocquet:1995je} 
  M.~Bocquet, S.~Bonazzola, E.~Gourgoulhon and J.~Novak,
  Astron.\ Astrophys.\  {\bf 301}, 757 (1995)

\bibitem{Kiuchi:2008ch} 
  K.~Kiuchi and S.~Yoshida,
  Phys.\ Rev.\ D {\bf 78}, 044045 (2008)

\bibitem{Frieben:2012dz} 
  J.~Frieben and L.~Rezzolla,
  Mon.\ Not.\ Roy.\ Astron.\ Soc.\  {\bf 427}, 3406 (2012)

\bibitem{Firenze}
        A.~G.~Pili, N.~Bucciantini and L.~Del Zanna,
        Mon.\ Not.\ Roy.\ Astron.\ Soc.,\ {\bf 439}, 3541 (2014);
%
  A.~G.~Pili, N.~Bucciantini and L.~Del Zanna,
  Mon.\ Not.\ Roy.\ Astron.\ Soc.\  {\bf 470}, no. 2, 2469 (2017)

\bibitem{Uryu:2014tda} 
  K.~Uryu, E.~Gourgoulhon, C.~Markakis, K.~Fujisawa, A.~Tsokaros and Y.~Eriguchi,
  Phys.\ Rev.\ D {\bf 90}, no. 10, 101501(R) (2014)


\bibitem{circular}
see e.g., 
B.~Carter, in \emph{Black holes --- Les Houches 1972}, edited by C.~DeWitt
\& B.S.~DeWitt, Gordon and Breach, New York (1973), p.~57; 
%

\bibitem{Shibata:2004qz} 
  M.~Shibata, K.~Uryu and J.~L.~Friedman,
  Phys.\ Rev.\ D {\bf 70}, 044044 (2004)
  [Erratum-ibid.\ D {\bf 70}, 129901 (2004)]

\bibitem{MeudonWL}
  S.~Bonazzola, E.~Gourgoulhon, P.~Grandclement and J.~Novak,
  Phys.\ Rev.\ D {\bf 70}, 104007 (2004);
%

\bibitem{WLBNS}
  K.~Uryu, F.~Limousin, J.~L.~Friedman, E.~Gourgoulhon and M.~Shibata,
  Phys.\ Rev.\ Lett.\  {\bf 97}, 171101 (2006);
  K.~Uryu, F.~Limousin, J.~L.~Friedman, E.~Gourgoulhon and M.~Shibata,
  Phys.\ Rev.\ D {\bf 80}, 124004 (2009).

\bibitem{Gourgoulhon:2011gz} 
  E.~Gourgoulhon, C.~Markakis, K.~Uryu and Y.~Eriguchi,
  Phys.\ Rev.\ D {\bf 83}, 104007 (2011)

\bibitem{NR}
M.~Alcubierre, ``Introduction to 3+1 Numerical Relativity''
Oxford University Press, New York (2008); 
M. Shibata, \emph{Numerical Relativity}, World Scientific, Singapore (2016)

\bibitem{NRBS}
T.~W.~Baumgarte and S.~L.~Shapiro, 
``Numerical Relativity: Solving Einstein's Equations on the Computer''
Cambridge University Press, New York (2010);

\bibitem{Gourg2012}
E.~Gourgoulhon : \emph{3+1 Formalism in General Relativity; Bases of Numerical Relativity},
Springer (Berlin) (2012) ;



\bibitem{Uryu:2016dqr} 
  K.~Uryu, A.~Tsokaros, F.~Galeazzi, H.~Hotta, M.~Sugimura, K.~Taniguchi and S.~Yoshida,
  Axisymmetric and triaxial rotating stars,''
  Phys.\ Rev.\ D {\bf 93}, no. 4, 044056 (2016).

\bibitem{York:1998hy} 
  J.~W.~York, Jr.,
  Phys.\ Rev.\ Lett.\  {\bf 82}, 1350 (1999)

\bibitem{Cook:2000vr}
  G.~B.~Cook,
  Living Rev.\ Rel.\  {\bf 3}, 5 (2000);

\bibitem{BD92D94}
  J.~K.~Blackburn and S.~Detweiler, Phys.\ Rev.\ D {\bf 46}, 2318 (1992); 
S.~Detweiler, Phys.\ Rev.\ D {\bf 50}, 4929 (1994).

\bibitem{Friedman:2001pf}
  J.~L.~Friedman, K.~Uryu and M.~Shibata, 
  Phys.\ Rev.\  D {\bf 65}, 064035 (2002) 
  [Erratum-ibid.\  D {\bf 70}, 129904 (2004)]

\bibitem{Klein:2004be}
  C.~Klein,
  Phys.\ Rev.\  D {\bf 70}, 124026 (2004)

\bibitem{PSWconsortium}
  J.~T.~Whelan, C.~Beetle, W.~Landry and R.~H.~Price,
  Class.\ Quant.\ Grav.\  {\bf 19}, 1285 (2002);
%
  C.~Beetle, B.~Bromley, N.~Hernandez and R.~H.~Price,
  Phys.\ Rev.\  D {\bf 76}, 084016 (2007);
%
  N.~Hernandez and R.~H.~Price,
  Phys.\ Rev.\  D {\bf 79}, 064008 (2009).

\bibitem{YBRUF06}
  S.~Yoshida, B.~C.~Bromley, J.~S.~Read, K.~Uryu and J.~L.~Friedman,
  Class.\ Quant.\ Grav.\  {\bf 23}, S599 (2006).

\bibitem{Tsokaros:2018zlf} 
  A.~Tsokaros, K.~Uryu and S.~L.~Shapiro,
  Phys.\ Rev.\ D {\bf 99}, no. 4, 041501(R) (2019)

\bibitem{Knapp:2002fm} 
  A.~M.~Knapp, E.~J.~Walker and T.~W.~Baumgarte,
  Phys.\ Rev.\ D {\bf 65}, 064031 (2002)

\bibitem{cocal} 
  X.~Huang, C.~Markakis, N.~Sugiyama and K.~Uryu,
  Phys.\ Rev.\ D {\bf 78}, 124023 (2008); 
  K.~Uryu and A.~Tsokaros,
  Phys.\ Rev.\ D {\bf 85}, 064014 (2012);  
  K.~Uryu, A.~Tsokaros and P.~Grandclement,
  Phys.\ Rev.\ D {\bf 86}, 104001 (2012);
%
  A.~Tsokaros, K.~Uryu and L.~Rezzolla,
  Phys.\ Rev.\ D {\bf 91}, no. 10, 104030 (2015)


\bibitem{Ferraro(1937)} V.~C.~A.~Ferraro, 
Mon.\ Not.\ Roy.\ Astron.\ Soc.\  {\bf 97}, 458 (1937)

\bibitem{YYE06}
S.~Yoshida, and Y.~Eriguchi, ApJS, 164, 156 (2006); 
S.~Yoshida, S.~Yoshida, and Y.~Eriguchi, \apj, 651, 462 (2006)


\bibitem{Uryu:1999uu} 
  K.~Uryu and Y.~Eriguchi,
  Phys.\ Rev.\ D {\bf 61}, 124023 (2000).

\bibitem{CocalQuark}
  E.~Zhou, A.~Tsokaros, L.~Rezzolla, R.~Xu and K.~Uryu,
  Phys.\ Rev.\ D {\bf 97}, no. 2, 023013 (2018);
%
  E.~Zhou, A.~Tsokaros, L.~Rezzolla, R.~Xu and K.~Uryu,
  Universe {\bf 4}, no. 3, 48 (2018);
%
  E.~Zhou, A.~Tsokaros, K.~Uryu, R.~Xu and M.~Shibata,
  arXiv:1902.09361 [astro-ph.HE].
  
  

\bibitem{MRNSNewtonian} 
  K.~Fujisawa and Y.~Eriguchi,
  Mon.\ Not.\ Roy.\ Astron.\ Soc.\  {\bf 432}, 1245 (2013);
%
C.~Armaza, A.~Reisenegger, and J.~Alejandro Valdivia, 
Astrophys.\ J.\  {\bf 802}, 121 (2015).


\bibitem{RNScocal} 
  K.~Uryu, A.~Tsokaros, L.~Baiotti, F.~Galeazzi, N.~Sugiyama, K.~Taniguchi and S.~Yoshida,
  Phys.\ Rev.\ D {\bf 94}, no. 10, 101302(R) (2016);
%
  K.~Uryu, A.~Tsokaros, L.~Baiotti, F.~Galeazzi, K.~Taniguchi and S.~Yoshida,
  Phys.\ Rev.\ D {\bf 96}, no. 10, 103011 (2017);
%
  E.~Zhou, A.~Tsokaros, L.~Rezzolla, R.~Xu and K.~Uryu,
  Phys.\ Rev.\ D {\bf 97}, no. 2, 023013 (2018)



\bibitem[Gourgoulhon \& Bonazzola(1994)]{1994CQGra..11..443G} 
E.~Gourgoulhon, and S.~Bonazzola, Classical and Quantum Gravity, 11, 443 (1994)

\bibitem[Beig(1978)]{1978PhLA...69..153B} 
R.~Beig, Physics Letters A, 69, 153 (1978)


\bibitem{Carter79}
B. Carter, 
in {\em Active Galactic Nuclei},
edited by C.~Hazard and S.~Mitton,
(Cambridge University Press, Cambridge, 1979), p.~273.


\end{thebibliography}
\end{document}